%% file: ms_ref_preprint.tex
\documentclass[preprint]{aastex}
\shortauthors{Sokal et al.}
\shorttitle{The Prevalence of WR Stars in Emerging Clusters}

\usepackage{url}
\usepackage{amsmath}

\usepackage{natbib}  
\begin{document}

\title{The Prevalence and Impact of Wolf-Rayet Stars in Emerging Massive Star Clusters}

\author{Kimberly R. Sokal\altaffilmark{1, 2}, Kelsey E. Johnson\altaffilmark{1}, R\'{e}my Indebetouw\altaffilmark{1}, and Philip Massey\altaffilmark{2,3}}

\altaffiltext{1}{Department of Astronomy, University of Virginia, P.O. Box 3818, Charlottesville, VA 22903, USA; krs9tb@virginia.edu}
\altaffiltext{2}{Visiting astronomer, Kitt Peak National Observatory, National Optical Astronomy Observatory, which is operated by the Association of Universities for Research in Astronomy (AURA) under a cooperative agreement with the National Science Foundation.}
\altaffiltext{3}{Lowell Observatory, 1400 W Mars Hill Road, Flagstaff, AZ 86001, USA}

\begin{abstract}

We investigate Wolf-Rayet (WR) stars as a source of feedback contributing to the removal of natal material in the early evolution of massive star clusters. Despite previous work suggesting that massive star clusters clear out their natal material before the massive stars evolve into the WR phase, WR stars have been detected in several emerging massive star clusters. These detections suggest that the timescale for clusters to emerge can be at least as long as the time required to produce WR stars (a few million years), and could also indicate that WR stars may be providing the tipping point in the combined feedback processes that drive a massive star cluster to emerge. We explore the potential overlap between the emerging phase and the WR phase with an observational survey to search for WR stars in emerging massive star clusters hosting WR stars. We select candidate emerging massive star clusters from known radio continuum sources with thermal emission and obtain optical spectra with the 4m Mayall Telescope at Kitt Peak National Observatory and the 6.5m MMT\footnote[4]{Observations reported here were obtained at the MMT Observatory, a joint facility of the University of Arizona and the Smithsonian Institution.}. We identify 21 sources with significantly detected WR signatures, which we term ``emerging WR clusters.'' WR features are detected in $\sim$50\% of the radio-selected sample, and thus we find that WR stars are commonly present in massive star clusters currently emerging. The observed extinctions and ages suggest that clusters without WR detections remain embedded for longer periods of time, and may indicate that WR stars can aid, and therefore accelerate, the emergence process.

\end{abstract}

\keywords{galaxies: star formation --- galaxies: star clusters: general --- HII regions ---  stars: Wolf-Rayet}

\noindent

\section{Introduction}

One of the most stunning early discoveries of the {\em Hubble Space Telescope} was of star clusters with masses of 10$^4$ M$_\odot$ - 10$^7$ M$_\odot$, which are quite different than the typical young star clusters in our own Galaxy. These clusters came to be known by a variety of names, including massive star clusters and Super Star Clusters or together they can be also called Young Massive Clusters. Not only can these star clusters dominate the appearance of their host galaxies, but their formation and early evolution can have a significant impact on their environment \citep{krum14}.  

 A simple version of the stages of early evolution of a massive star cluster as a scaled up version of a massive star was proposed by \citet{john02}, which starts with the collapse of a massive ($\gtrsim 10^6$ M$_\odot$) molecular cloud, subsequently forming hundreds to thousands of massive stars \citep[e.g.,][]{bbz98,evans99}. When star formation begins, the cluster is still heavily obscured and the presence of massive stars is first observable through radio free-free emission arising from ionized gas \citep[e.g.,][]{kj99, turn00}. The massive stars eventually clear the natal material away from the cluster, enabling the brilliant star cluster to be studied with optical and near-infrared wavelengths. 

However, closer inspection of the details in this simplified scheme reveals that both the timescales and the mechanisms at play are not well-constrained. For instance, \citet{sok15} identified a massive star cluster in NGC 4449 that suggests that the evolution cannot be so cleanly  divided into individual stages. This massive star cluster, called S26, was originally identified as radio continuum source with a thermal emission component \citep{rei08}; yet, surprisingly, was also discovered to host Wolf-Rayet stars \citep{rei10}. Wolf-Rayet (WR) stars are the evolved descendants of O-stars ($\gtrsim 25$ M$_\odot$) that have stripped off their outer layers via high mass loss rates \citep{con83}. As only massive stars will become WR stars, the WR phase for a cluster is short, typically occurring at a cluster age of $\sim$ 3 Myr and lasting only a few Myr depending on the stellar populations present. Due to the short WR phase and the feedback nature of the WR stars themselves, the simultaneous presence of the thermal radio emission and the WR stars in S26 in NGC 4449 has important implications for the evolution of young massive star clusters \citep{sok15}. 

One implication is that the lifetime of the thermal radio emission, which typically is observed while a cluster is still embedded in its natal cocoon, may not be as simple and short-lived as expected. Previous work has shown that some of the most massive star clusters can be observed without obscuring natal material or thermal counterparts within a few Myr \citep[e.g.,][]{wz02, bas14,hol15}, which is in agreement with predictions from population studies of embedded thermal sources that suggested a timescale of $\sim 1-2$ Myr for emergence \citep{kj99}. If the timescales for emergence are indeed as short as a few million years, this would require massive star clusters to typically clear out the embedding natal material before the first supernova has exploded, which for instance should occur at the end of a 60 M$_\sun$ star's life at 4 Myr \citep{groh14} and at even younger ages for more massive stars.  In line with this expectation, WR stars have been observed in a number of optically visible clusters \citep[e.g.,][]{bas14}, suggesting the timescale for cluster emergence can be less than $\sim$ 3 Myr.  On the other hand,  some clusters also in NGC 4449 are observed to stay embedded in natal material for up to 5 Myr \citep{rei08}, which could be past the onset of the WR phase. However, S26 in NGC 4449 has provided the first example that, for some massive clusters, there may be a period during which the thermal emission and the WR phase have overlapping timeframes, showing that at least some clusters are not fully emerged by the time the WR stars appear.

Perhaps more importantly, \citet{sok15} suggest that the WR stars in S26 might be contributing to the evolution of the cluster by helping to clear out embedding material. S26 is in the act of emerging from its natal molecular cloud, with additional evidence for ongoing feedback seen in the infrared spectral energy distribution and optical nebular morphology.  However, thus far, the contribution of the additional feedback from WR stars has not been addressed. There are good reasons to suspect that WR stars may be important in how massive star clusters emerge.  In general, the relative importance of different feedback mechanisms on the natal material in young massive star clusters and driving the evolution of the H \textrm{II} region is not yet understood \citep[e.g.,][]{lop13}. The expansion of the ionized gas likely dominates H \textrm{II} regions driven by single massive stars to massive star clusters, for instance in analytic studies \citep[e.g.,][]{mat02}, simulations \citep[e.g.,][]{dale05}, and observational studies \citep[e.g., in the Large Magellanic Cloud (LMC) by][]{lop13}. However, radiation pressure may also be the dominant feedback mechanism early on \citep{km09, lop11}. Stellar winds are thought to be less important energetically in the dynamics of H \textrm{II} regions, as the output energy equivalent to that of a supernova can leak and escape \citep[e.g.,][]{rp13}. Yet, winds are more efficient than the H \textrm{II} region pressure in removing extremely dense material and in determining the morphology \citep{dale14}. In addition, the impact of later supernova is increased  by up to a factor of two, if winds have cleared molecular material \citep{wn15}. These details demonstrate that the feedback phase of massive star clusters is complicated, with different mechanisms contributing at different times and interplay between them.  It is becoming increasingly clear that stellar winds should not be ignored \citep{calura15,frey03}. 

The feedback from many different physical processes may be increased during the WR phase, including these two leading candidates for the dominant feedback process (photoionization and radiation pressure). There might be a slight increase in the luminosity, which would produce more radiation pressure than that from O-stars, during the WR phase for stars with certain initial masses or properties, as follows.  The Geneva models at solar metallicity \citep{eks12} show that non-rotating stars have a higher luminosity during the WNL phase than during their O-star phase \citep[see Figure 1 in][]{geo12}. If rotating, this only follows for stars with lower initial stellar masses ($\sim$25$M_\sun$).  At a low metallicity of z$=$0.002 \citep{geo13}, where a higher initial mass is required make a WR star through single star evolutionary paths, the stars with initial stellar masses $>85$M$_\sun$ exhibit higher luminosities during the WR phase compared to the earlier stages. Observationally, this is hard to test and the findings may not be in agreement with the predictions; as shown in HR diagrams, WC stars in the MW appear below the previously predicted tracks \citep{sander12}, and WN stars in the LMC and M31 fall across and below evolutionary tracks \citep{hai14,sander14}. Regardless of the luminosity, the WR winds can cause chemical enrichment \citep{kpv13}, which can lead to greater opacities and a corresponding increase in radiation pressure.

Additionally, the ionizing flux for stars with an initial mass of 60 $M_\sun$ is increased by an order of magnitude, and more for stars with initial masses less than 60 $M_\sun$, when they have evolved into the WR phase \citep[at solar metallicity; see Figure 6 in][]{ts15}. Higher ionizing photon rates from WR stars can result in more photoionization in comparison to that produced by O-stars, and thus result in higher ionized gas pressure.

Lastly, WR winds will input roughly ten times the instantaneous energy than O-star winds. But, it is the short lifespan of the WR stars, which is roughly a tenth of the O-stars' lifetime, that has likely led to somewhat ignoring the WR star contributions to the feedback: after all, the integrated energy input by the winds is similar over the lifetime of the O-stars and the WR stars. Yet if there is an increase in feedback during the WR phase, the influence on the environment may be different than that of the previous O-star phase particularly because of the carved out cavities due to the winds. In the same way that the impact of supernova is increased if the region was previously cleared out, the impact of the WR feedback should be increased after the O-star feedback has (very slowly) done its work. Clearly, there is much work to be done to understand the interplay amongst the different feedback mechanisms  \citep{calura15}, including the WR contributions.  

In addition to these possibilities, an inspection of the literature reveals additional massive star clusters other than S26 in NGC 4449 may simultaneously exhibit thermal radio emission and WR signatures. For example, the irregular blue compact dwarf galaxy NGC 3125 is dominated by two bright star formation regions. The eastern lobe hosts a compact thermal radio source \citep{av11}, and also shows WR signatures \citep[e.g.,][]{had06}. The unusual supernebulae in NGC 5253, a site of dust enrichment and high star-formation efficiency, may be coincident with known WR features as well \citep{turn15}. Therefore, these objects may be currently undergoing the same short-lived evolutionary transition as S26 in NGC 4449, and suggest that S26 is not an unique massive star cluster, but in fact may be common.

Here, we follow up the in-depth investigation of S26 with an observational survey to search for WR stars in emerging massive star clusters to assess the role of WR stars in this evolutionary process.  WR stars produce high ionization stellar emission lines that are relatively unique to WR stars. Large populations of WR stars produce a broad, integrated spectral features in the optical spectrum known as WR ``bumps''. The ``blue bump'' includes He \textrm{II} at $\lambda$ 4686 \AA\ and N \textrm{III}/C \textrm{III} at $\lambda$ 4640/4650 \AA, and the ``red bump'' is due to C \textrm{IV} at $\lambda$ 5808 \AA. We used a novel method to identify emerging massive star clusters, explained in Section \ref{section-sample} and obtained optical spectra to look for the WR bumps. We present the observations in Section \ref{section-observations}. The sources are classified by presence or absence of a WR bump observed in the optical spectra in Section \ref{section-wremission}, and the environments are characterized in Section \ref{section-environment}, including nebular properties and massive star populations.  The frequency of WR detections in our sample and other results are discussed Section \ref{section-results}. Finally, a discussion of timescales and an indication that WR stars may accelerate the emergence of massive star clusters is presented in Section \ref{section-conclusions}.

\begin{deluxetable}{lllll}
\tabletypesize{\scriptsize}
\tablewidth{0pt} 
\tablecaption{\label{table-gals} Target Host Galaxies}
\tablehead{
	 \colhead{Galaxy}			&
	 \colhead{Morphological Type\tablenotemark{a}}				&
	 \colhead{SFR}			&
	 \colhead{Distance }		&		
	 \colhead{Distance Reference}			\\
	 \colhead{}			&
	 \colhead{}				&
	 \colhead{[M$_\sun$ yr$^{-1}$]}			&
	 \colhead{[Mpc] }			&
	 \colhead{}			\\

}
\startdata

NGC 2366	&	 IB(s)m &   0.11   &   3.34	&   \citet{tcd13} \\
NGC 4214	&    IAB(s)m &   0.16  &   2.93   &    \citet{tcd13} \\
NGC 4449	&    IBm   &  0.66  &   3.82   &    \citet{ann08} \\
NGC 6946	&	 SAB(rs)cd  &  5.65 	&   5.5     &   \citet{tul88} \\
M 51	    &    SAbc   &    4.48   &   7.62     &  \citet{ciar02} \\

\enddata

\tablecomments{An overview of the basic properties of the host galaxies in our sample. The given distances are used throughout this work. The star formation rates (SFR) were estimated using H$\alpha$ and are from \citet{lee09}.
}
\tablenotetext{a}{From NED}

\end{deluxetable}

\section{\label{section-sample} The Sample Selection: Emerging Massive Star Clusters}

In order to ascertain if there are many other massive star clusters that are emerging from their natal material and display WR features, similar to NGC 4449's S26  \citep{sok15}, we selected a sample of 45 radio sources in six nearby galaxies known to be
starbursts or have high star-formation rates.  Such galaxies are likely hosts to massive star clusters, and our sample included both dwarfs and spirals.  These galaxies all had previously known radio continuum detections with thermal emission,  as indicated by a radio spectral index $\alpha \geq -0.1$.  Optically thin thermal free-free emission results in $\alpha = -0.1$, which increases with density (n$_\text{e}$) and can become positive in dense enough conditions. We used a cutoff of $\alpha>-0.3$ including the uncertainties \citep[e.g.,][]{mad07}, which allowed for the unavoidable large uncertainties in the measured values for $\alpha$ in extragalactic sources; this cut-off will include mixed sources, where thermal free-free emission is contributing but is not necessarily the sole emission mechanism.  It is also important to note that while WR stars are known to produce free-free emission in their winds \citep{wb75}, this emission is not significant in comparison to the H \textrm{II} region. For example \citet{sok15} shows in the case of S26 with $\sim 20$ WR stars, any WR contributions to the observed radio emission are negligible.

The properties of the host galaxies, consisting of NGC 2366, NGC 4214, NGC 4449, NGC 6946, and M 51,  are listed in Table \ref{table-gals}. The targets in NGC 2366 and NGC 4214 are selected from \citet{cw09}, a study originally intended to identify supernova remnants. Using the Very Large Array (VLA), \citet{cw09} produce maps at 20cm, 6cm, and 3.6cm with synthesized beams of 3\farcs7$\times$3\farcs7 ($\sim$ 60 $\times$ 60 pc) for NGC 2366 and 1\farcs35$\times$1\farcs35 ($\sim$ 19 $\times$ 19 pc) for NGC 4214. All discrete radio sources with flux measurements greater than 3 $\sigma$ at 20 cm are identified and catalogued. In total, we include 3 sources in NGC 2366 and 5 sources in NGC 4214  in our sample. Targets in NGC 4449 are from the work of \citet{rei08}. \citet{rei08} mapped NGC 4449 with the VLA at 1.3cm, 3.6cm, and 6.0cm and convolved all data to a  synthesized beam 1\farcs3$\times$1\farcs3 ($\sim$ 24 $\times$ 24 pc). Radio sources are identified using requirement of a 3$\sigma$ detection (local rms) minimum in the 3.6cm image. Of these radio sources, 7 are included in our sample. The studies of  \citet{hym00}  and \citet{mad07}  provide our targets in NGC 6946 and M 51, respectively. \citet{hym00} re-evaluate  high resolution (2\farcs, roughly 53pc) observations of NGC 6946 at 6cm and 20cm with the VLA by \citet{ldg97}. These radio continuum maps reach sensitivities of 16 and 20 $\mu$Jy beam$^{-1}$ at 6cm and 20cm. \citet{mad07} produced high resolution images of M 51 with the VLA at 6cm and 20cm. The 6cm map, with an rms $=$ 11.7 $\mu$Jy beam$^{-1}$ has a deconvolved beam of 1\farcs47 $\times$ 1\farcs13 ($\sim$ 54 $\times$ 42); the 20cm map has a deconvolved beam measuring 1\farcs50 $\times$ 1\farcs21 ($\sim$ 55 $\times$ 45) and an rms of 22.5 $\mu$Jy beam$^{-1}$. From these observations, compact radio sources are identified via a detection algorithm and visual checks. We include 6 sources in NGC 6946 and 24 sources in M 51 in our sample.  Although we utilized a heterogeneous dataset composed of different catalogues to select our targets, all selected studies included a measurement at 6cm. From here on, we adopt a naming convention based on the original numbered identifications from these published radio catalogues and use Galaxy - Object ``ID number'' for all sources.

\section{\label{section-observations} Observations}

We obtained optical spectra of the targets with the 4m Mayall Telescope at the Kitt Peak National Observatory (National Optical Astronomical Observatory) and the 6.5m MMT at the Fred Lawrence Whipple Observatory to search for WR emission as well as characterize their environments. Target selection was heavily constrained in the observing process by weather, airmass, slit angle, and fiber placement. We include in our sample only the sources whose signal-to-noise ratio (SNR) in the optical continuum is observed to be $\geq$ 15 per pixel. An example of the optical spectra of several sources is shown in Figure \ref{fig-fullspec1}.
A summary of the spectral observations is presented in the Appendix in Table \ref{table-obs}, as well as the spectra of the rest of the sources in the sample that are shown in Figures \ref{fig-fullspec2} - \ref{fig-candidatespec}.

Additionally, we use archival imaging to provide the total V-band flux from each source. Not all observing was photometric while obtaining the optical spectra, but we correct the spectral flux in Section \ref{section-lines} using the photometric V-band fluxes.

\begin{figure}[t!]
\hspace{-10pt}
\includegraphics[width=0.55\textwidth,angle=0]{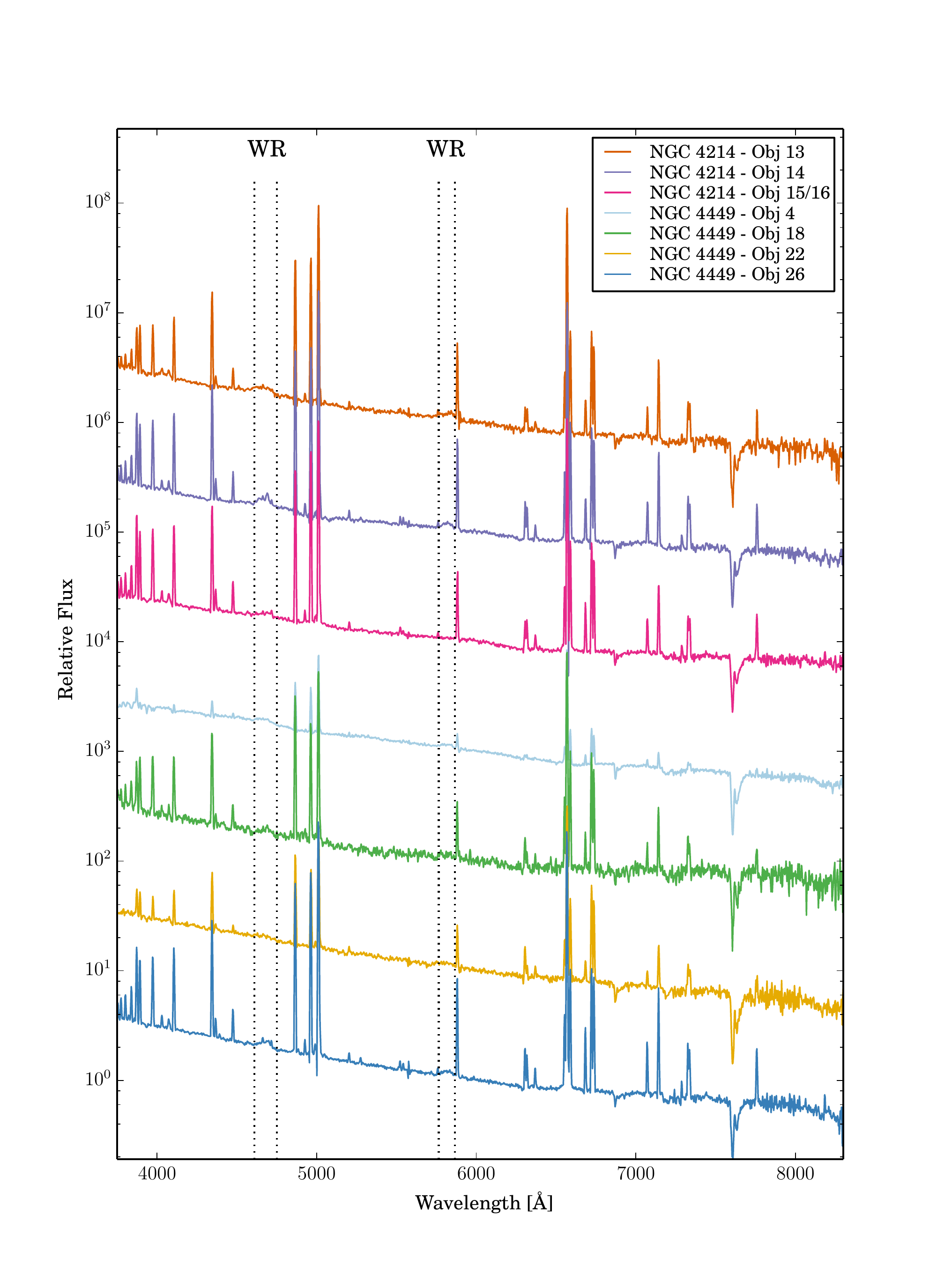}
\hspace{-25pt}
\includegraphics[width=0.55\textwidth,angle=0]{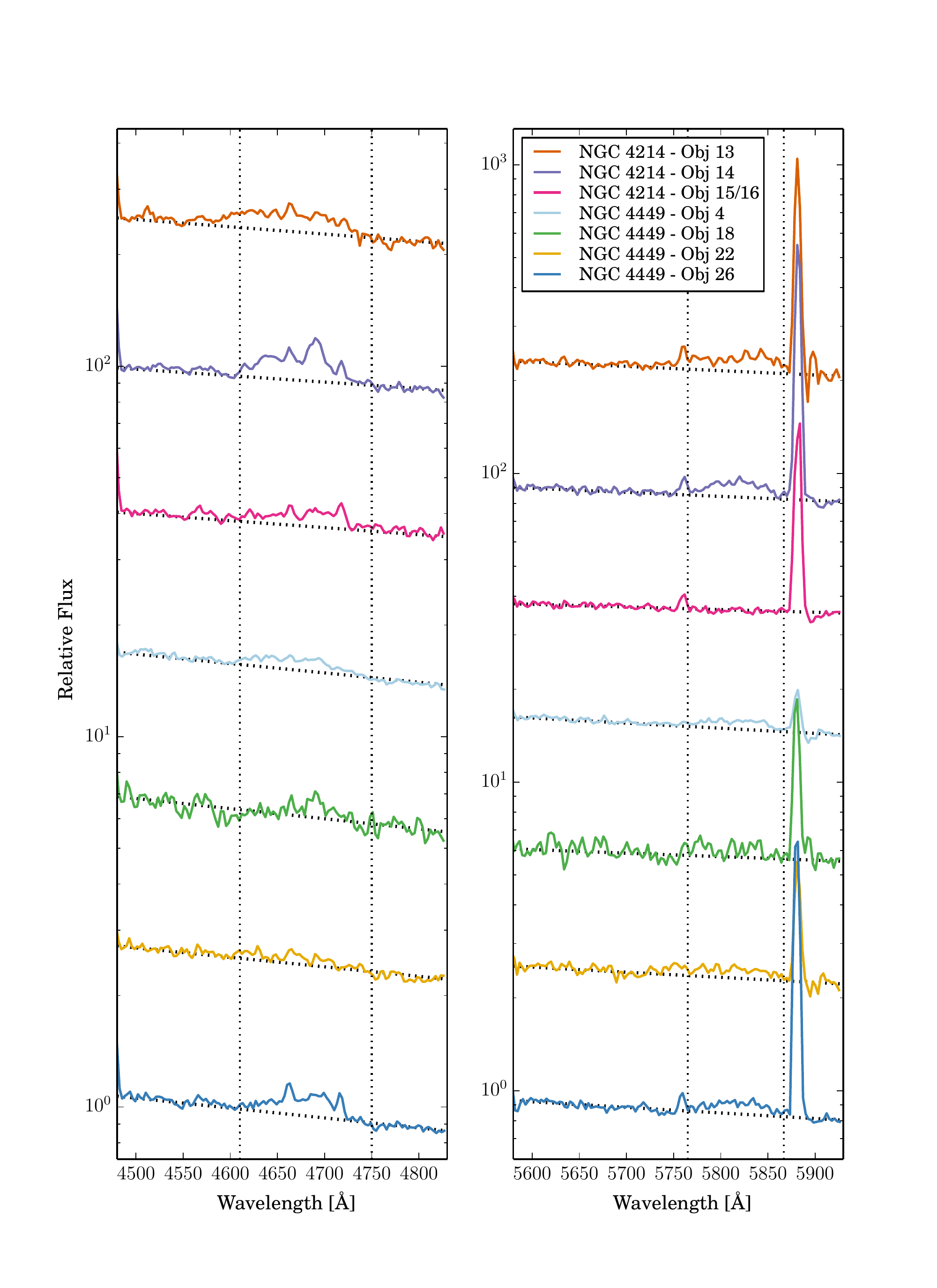}
\caption{\label{fig-fullspec1} Optical spectra observed with the 4m Mayall Telescope at KPNO of a subset of the WR clusters. Dashed lines indicate the location of the broad WR bumps, and a zoom in on the WR bumps is shown in the figure on the right, where a roughly traced continuum guides the eye. For ease of visual comparisons, individual spectra are normalized by the average value of the continuum at \AA\ (which is given in Table \ref{table-opticalcharacteristics}) then offset. 
The spectra of the rest of the sources in the sample are included in the Appendix.
 }
\end{figure}

\subsection{Optical Spectra}

Targets in NGC 2366, NGC 4214, and NGC 4449 were observed with the 4m Mayall Telescope in April 2013; details are presented in Table \ref{table-obs} in the Appendix. Only the optically brightest targets were observed, as the guider was not functioning which led to hand guiding. Each source was observed with a 1\farcs3 $\times$ 5\farcm4 slit aligned at the parallactic angle and seeing varied from 0\farcs7 to 2\farcs5. Use of the R-C CCD spectrograph with the KPC-10A grating provided spectra with a resolution of $\sim 6$ \AA\ over a wavelength range of 3800-8000 \AA (with $\approx$0\farcs69 per pixel). Exposure times vary for each source, yielding a range of SNR per pixel in the continua of 15 - 40. The data are reduced using standard IRAF routines. Spectra were extracted with a 10 pixel window ($=$ 6\farcs9 on sky), except for NGC 4214 - Object 15/16. The spectra of the targets Object 15 and Object 16 in NGC 4214 could not be separated and thus a wider extraction window of 20 pixels was adopted to get the emission from the combined source (Object 15/16). Spectra were obtained for the spectroscopic standard stars Feige 66, Feige 67, Hilter 600, HZ 44, and Wolf 1346 for flux calibrations.

We observed targets  in M 51 and NGC 6946 with the 6.5m MMT in April 2013 using the multi-object spectrograph Hectospec \citep{fab05}; details are presented in Table \ref{table-obs} in the Appendix. Individual targets were observed by individual fibers that subtend a diameter of 1\farcs5 on the sky, pointed at the coordinates of the observed radio continuum source. Target selection was constrained by fiber positioning, resulting in a random sampling. A 270 line mm$^{-1}$ grating blazed at 5200 \AA\ resulted in spectra with $\sim$ 5 \AA\ resolution over a wavelength range of 3700-9000 \AA. A problem with an LED in the instrument led to contamination of many spectra in the 8430-8445 \AA\ region, and we did not include any of the spectra redwards of 8000 \AA\ in our analysis. Since there was no blocking filter, we expect light redwards of 8000 \AA\  to be contaminated by second-order blue light in any event. Optical spectra obtained with Hectospec were reduced using the new version of IDL Hectospec pipeline SPECROAD \citep{mink07}, which produced sky-subtracted, variance-weighted co-added spectra. Using IRAF procedures, flux calibration was done with spectra of the standard star HD 192281. To increase the signal consistently, all MMT spectra were smoothed with a box car of 5 pixels to a resolution of $\sim$ 6-7 \AA\ , resulting in SNR per pixel in the continua of 15 - 50 after smoothing (individual SNR values are listed in Table \ref{table-bumps}).	

\subsection{Optical Archival Imaging}

Broad-band and narrow-band images are available for NGC 2366, NGC 4214, NGC 4449, and M 51 in the {\em Hubble Space Telescope} (HST) archive\footnotemark. These observations, although taken with various instruments for different programs, provide well-resolved comparisons. The sources in our sample in NGC 6946 are not captured in available archival HST images, and we therefore use archival data from the Kitt Peak National Observatory 2.1m Telescope with the CFIM imager originally obtained as ancillary data for the SINGS survey \citep{ken03}, in which similar filters are available with lower spatial resolution. Color images (B, I, and H$\alpha$) of the regions surrounding a subset of the sources targeted in this sample are shown  Figure \ref{fig-rgb_panel1}; the rest of the sources of the sample are shown in the Appendix in Figures \ref{fig-rgb_panel2}-\ref{fig-rgb_panel3}. The V-band observations, used for photometry in this work, are described in the Appendix in Table \ref{table-imaging}. 

\footnotetext{ Based on observations made with the NASA/ESA Hubble Space Telescope, and obtained from the Hubble Legacy Archive, which is a collaboration between the Space Telescope Science Institute (STScI/NASA), the Space  Telescope European Coordinating Facility (ST-ECF/ESA) and the Canadian Astronomy Data Centre (CADC/NRC/CSA). }

\begin{figure}[t!]
\includegraphics[width=\textwidth,angle=0]{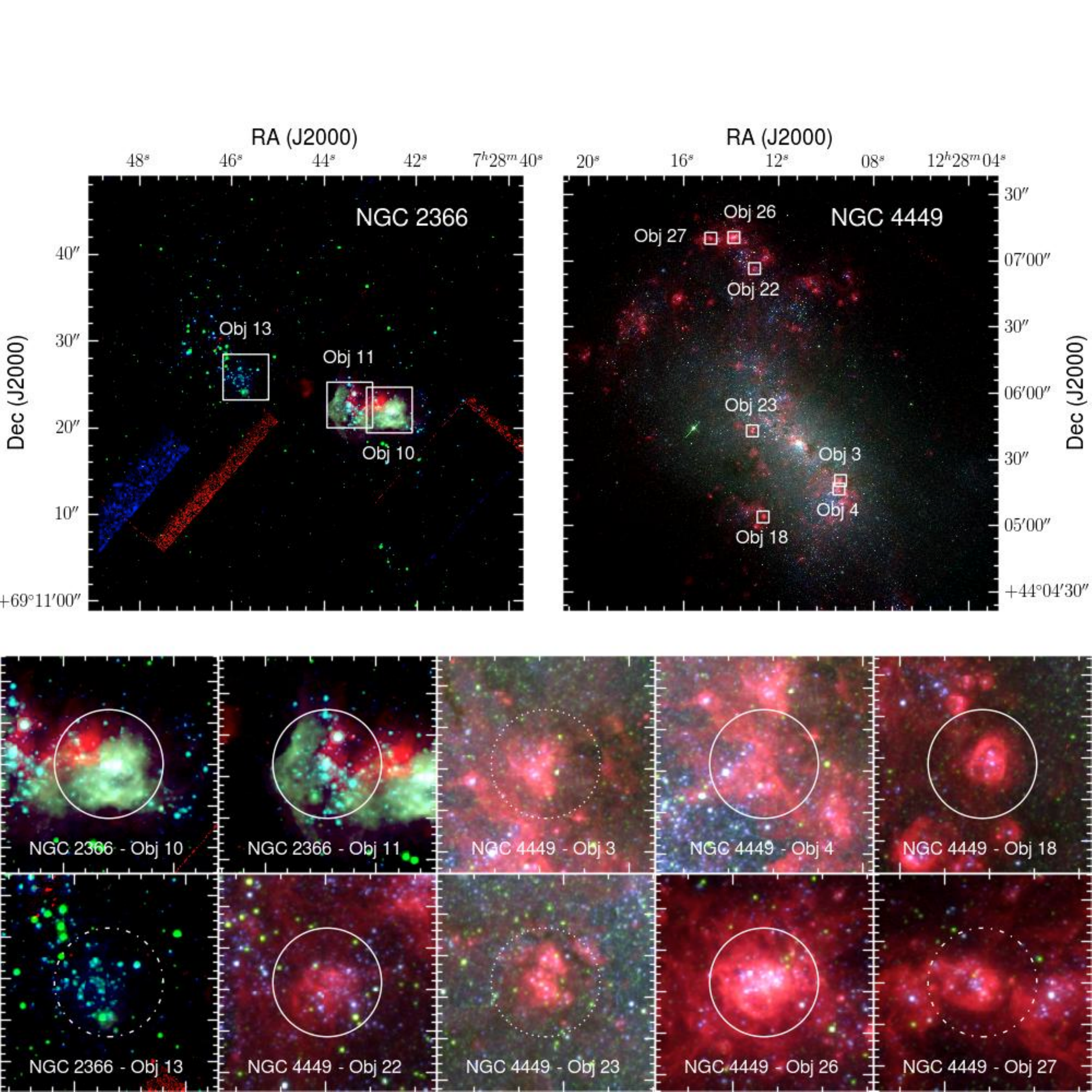}
\caption{\label{fig-rgb_panel1} Archival HST images (H$\alpha$, B, I) of the target galaxies NGC 2366 and NGC 4449. Insets are zoomed in on sample sources, with the corresponding region overlaid on the host galaxy image with a white square for ease of comparison.  White circles (2\farcs5) are overlaid and show the extraction regions used for photometry. The line style of this overlaid circle indicates the source's classification: solid for WR clusters, dot-dash for Candidate clusters, and dotted for Non-WR clusters.
Images of the rest of the images in the sample are included in the Appendix.
 }
\end{figure}

\section{\label{section-wremission} Detecting Wolf-Rayet Emission}

\subsection{\label{section-bump} Identifying the WR Bump}

Using the optical spectra, we determined whether a given source exhibits a significant WR feature, and if it is not present, we estimated limits. For all spectra with any excess emission near $\lambda$ 4686\AA\,  we measured  the equivalent width (EW) with the IRAF SPLOT package after subtracting any superimposed nebular features such as nebular He \textrm{II} 4686 \AA\ or [Fe \textrm{III}] 4658 \AA. The SPLOT task fits with a gaussian, and a single broad bump measurement was used unless the WR bump could be deblended into $\lambda$ 4640/4650 \AA\ and  $\lambda$ 4686 \AA\ components. Uncertainties were determined through IRAF, which performs a Monte-Carlo simulation of 100 trials resulting in an error estimation including the input rms of the spectrum (measured via SPLOT in several line-free regions of the continuum), and then added in quadrature with an additional measurement uncertainty introduced by estimating the continuum. A source was considered to have a significant detection of the WR bump if the SNR$_{\text{bump}}$ is $\geq$ 5 for a single broad bump, or both deblended $\lambda$ 4640/4650 \AA\ and $\lambda$ 4686 \AA\ bumps were SNR$_{\text{bump}}$ $\geq$ 3. 

\begin{figure}[!t]
\centering
\includegraphics[width=0.8\textwidth,angle=0]{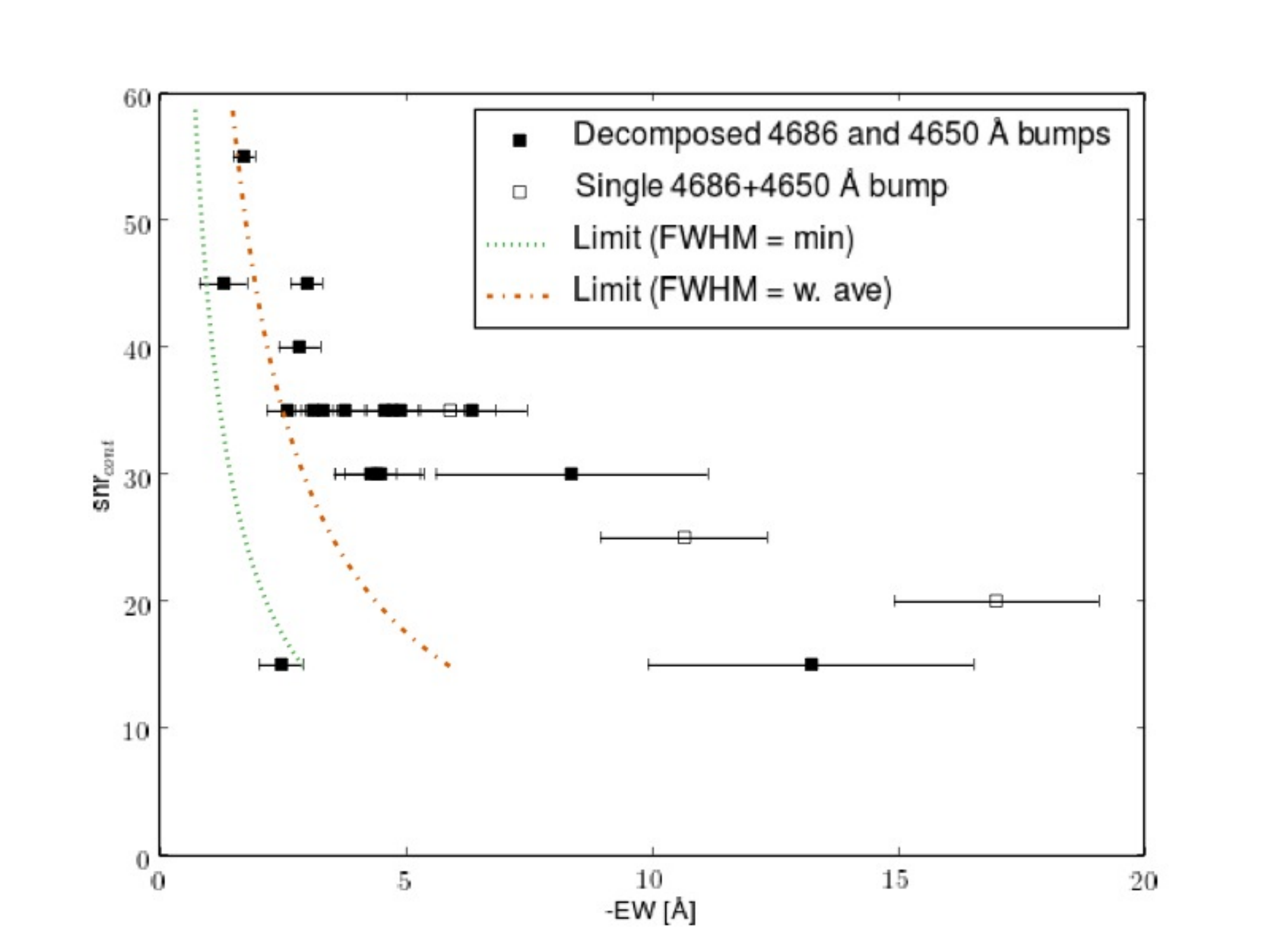}
\caption{\label{fig-significance} A scatter plot showing the sensitivity to identifying a significantly detected WR feature. The colored lines are estimated detection limits adopting an observed FWHM of the WR feature from the sample (as indicated in the legend of either the minimum or weighted average) and requiring a 3$\sigma$ detection.}
\end{figure}

For sources without a clear detection of the WR bump, we found an upper limit of the WR emission using the measured SNR in the optical continuum. This limit can be estimated as EW$_{\text{bump}}$ $<$ $\frac{\text{SNR}_{\text{bump}}}{\text{SNR}_{\text{cont}}} \times (2 \times \text{RES} \times \text{FWHM}_{\text{bump}})^{0.5}$ where RES is the spectral resolution and the label ``bump'' refers to the WR bump. We input the requirement that SNR$_{\text{bump}}$ $\sim$ 3 as above, as well as used a FWHM$_{\text{bump}}$ equal to the minimum value observed in this sample such that FWHM$_{\text{bump}}$ $\sim$ 18 \AA . For comparison, nebular lines were observed to have FWHM $\sim$ 6 \AA\ (the same as the spectral resolution). The observed SNR$_{\text{cont}}$ is plotted versus EW$_{\text{bump}}$  in Figure \ref{fig-significance} and shows both the data points for the WR clusters and the resulting detection limit. 
We also plot the detection limit that would be obtained if the weighted average FWHM for all of the WR clusters was used instead. This comparison helps to illustrate  that if a given source were to have a bump similar to the sources in the sample that are significantly detected, the expected EW would be even larger than the adopted upper limit. The upper limits and WR emission line measurements are presented for the sample in Table \ref{table-bumps}.

\begin{deluxetable}{lllllll}
\tabletypesize{\scriptsize}
\tablewidth{0pt} 
\tablecaption{\label{table-bumps} Classification and the WR Features}
\tablehead{
	 \colhead{Source Name}		&
	 \colhead{snr$_{\text{cont}}$}		&
	 \colhead{F(4650 \AA)/F(H$\beta$)}		&	 
	 \colhead{F(4686 \AA)/F(H$\beta$)}		&	 
	 \colhead{F(5808 \AA)/F(H$\beta$)}		&	 
	 \colhead{-EW(4686 \AA) }		&	 
	 \colhead{Classification}		
}
\startdata

 NGC 2366 - Object 10	 & 45	 & 0.49 (0.25)	 & 0.30 (0.36)	 & ...	 & 1.3 (0.5)	 & WR\\ 
 NGC 2366 - Object 11	 & 35	 & 4.37 (0.18)	 & 2.54 (0.29)	 & 1.95 (0.26)	 & 4.6 (1.5)	 & WR\\ 
 NGC 2366 - Object 13	 & 35	 & ...	 & ...	 & ...	 & $<$1.3	 & Candidate\\ 
 NGC 4214 - Object 13	 & 30	 & 9.52 (0.28)	 & 6.31 (0.30)	 & 6.58 (0.26)	 & 8.4 (2.8)	 & WR\\ 
 NGC 4214 - Object 14	 & 35	 & 3.35 (0.20)	 & 3.17 (0.15)	 & 2.20 (0.24)	 & 6.3 (1.1)	 & WR\\ 
 NGC 4214 - Object 15/16	 & 35	 & 4.22 (0.29)	 & 2.77 (0.26)	 & ...	 & 4.9 (1.3)	 & WR\\ 
 NGC 4449 - Object 3	 & 15	 & ...	 & ...	 & ...	 & $<$2.9	 & Non-WR\\ 
 NGC 4449 - Object 4	 & 30	 & 24.58 (0.24)	 & 33.72 (0.20)	 & 18.09 (0.22)	 & 4.5 (0.9)	 & WR\\ 
 NGC 4449 - Object 18	 & 15	 & 5.04 (0.34)	 & 9.05 (0.24)	 & 0.59 (0.40)	 & 13.2 (3.3)	 & WR\\ 
 NGC 4449 - Object 22	 & 35	 & ...	 & 13.83 (0.14)	 & 25.16 (0.10)	 & 5.9 (0.9)	 & WR\\ 
 NGC 4449 - Object 23	 & 15	 & ...	 & ...	 & ...	 & $<$2.9	 & Non-WR\\ 
 NGC 4449 - Object 26	 & 30	 & 2.62 (0.09)	 & 1.82 (0.10)	 & 2.17 (0.10)	 & 4.3 (0.5)	 & WR\\ 
 NGC 4449 - Object 27	 & 20	 & ...	 & ...	 & ...	 & $<$2.2	 & Candidate\\ 
 NGC 6946 - Object 13	 & 20	 & ...	 & 9.56 (0.10)	 & ...	 & 17.0 (2.1)	 & WR\\ 
 NGC 6946 - Object 37	 & 25	 & ...	 & ...	 & ...	 & $<$1.8	 & Non-WR\\ 
 NGC 6946 - Object 48	 & 25	 & ...	 & 10.91 (0.15)	 & ...	 & 10.6 (1.7)	 & WR\\ 
 NGC 6946 - Object 110	 & 30	 & 3.68 (0.16)	 & 3.36 (0.19)	 & 6.68 (0.17)	 & 4.4 (0.9)	 & WR\\ 
 NGC 6946 - Object 115	 & 15	 & 6.25 (0.17)	 & 3.46 (0.17)	 & 6.92 (0.22)	 & 2.5 (0.5)	 & WR\\ 
 NGC 6946 - Object 117	 & 35	 & 2.44 (0.10)	 & 8.29 (0.08)	 & 6.79 (0.09)	 & 3.7 (0.4)	 & WR\\ 
 M 51 - Object 5	 & 20	 & ...	 & ...	 & ...	 & $<$2.2	 & Candidate\\ 
 M 51 - Object 6	 & 25	 & ...	 & ...	 & ...	 & $<$1.8	 & Non-WR\\ 
 M 51 - Object 11	 & 35	 & ...	 & ...	 & ...	 & $<$1.3	 & Non-WR\\ 
 M 51 - Object 34	 & 30	 & ...	 & ...	 & ...	 & $<$1.5	 & Non-WR\\ 
 M 51 - Object 39	 & 25	 & ...	 & ...	 & ...	 & $<$1.8	 & Non-WR\\ 
 M 51 - Object 44	 & 25	 & ...	 & ...	 & ...	 & $<$1.8	 & Non-WR\\ 
 M 51 - Object 46	 & 55	 & 4.20 (0.10)	 & 2.48 (0.11)	 & 3.26 (0.19)	 & 1.7 (0.2)	 & WR\\ 
 M 51 - Object 57	 & 40	 & 1.63 (0.12)	 & 2.90 (0.13)	 & ...	 & 2.8 (0.4)	 & WR\\ 
 M 51 - Object 60	 & 25	 & ...	 & ...	 & ...	 & $<$1.8	 & Non-WR\\ 
 M 51 - Object 63	 & 30	 & ...	 & ...	 & ...	 & $<$1.5	 & Non-WR\\ 
 M 51 - Object 67	 & 35	 & ...	 & ...	 & ...	 & $<$1.3	 & Non-WR\\ 
 M 51 - Object 73	 & 35	 & 6.82 (0.18)	 & 7.17 (0.14)	 & 2.20 (0.81)	 & 2.6 (0.4)	 & WR\\ 
 M 51 - Object 88	 & 20	 & ...	 & ...	 & ...	 & $<$2.2	 & Non-WR\\ 
 M 51 - Object 90	 & 25	 & ...	 & ...	 & ...	 & $<$1.8	 & Non-WR\\ 
 M 51 - Object 92	 & 30	 & ...	 & ...	 & ...	 & $<$1.5	 & Non-WR\\ 
 M 51 - Object 93	 & 30	 & ...	 & ...	 & ...	 & $<$1.5	 & Non-WR\\ 
 M 51 - Object 94	 & 35	 & 4.80 (0.09)	 & 5.85 (0.09)	 & ...	 & 4.7 (0.5)	 & WR\\ 
 M 51 - Object 96	 & 25	 & ...	 & ...	 & ...	 & $<$1.8	 & Non-WR\\ 
 M 51 - Object 97	 & 20	 & ...	 & ...	 & ...	 & $<$2.2	 & Non-WR\\ 
 M 51 - Object 100	 & 45	 & 6.21 (0.08)	 & 4.92 (0.09)	 & 2.12 (0.19)	 & 3.0 (0.3)	 & WR\\ 
 M 51 - Object 101	 & 35	 & 3.09 (0.11)	 & 4.99 (0.10)	 & 3.75 (0.17)	 & 3.1 (0.4)	 & WR\\ 
 M 51 - Object 103	 & 35	 & 1.63 (0.20)	 & 3.93 (0.11)	 & ...	 & 3.3 (0.4)	 & WR\\ 
 M 51 - Object 105	 & 20	 & ...	 & ...	 & ...	 & $<$2.2	 & Candidate\\ 
 
\enddata
\vspace{-20pt}
\tablecomments{
A table presenting the source name, the snr in the continuum of the spectra, the measured or estimated upper limit of the equivalent width (EW; in units of \AA) of the WR feature at 4686 \AA, and the source classification. If there is insufficient signal to decompose the bump, or for the upper limit estimation, the EW is given for the entire broad bump and listed as 4686 \AA.
}

\tablenotetext{a}{NGC 4214 - Object 3, NGC 4214 - Object 17, and M51 - Object 87 belong in the ``Other'' class, and not listed further .}
\end{deluxetable}

\subsection{\label{section-classification}Classifications According to Observed WR Emission}

\begin{figure}[!t]
\includegraphics[width=\textwidth,angle=0]{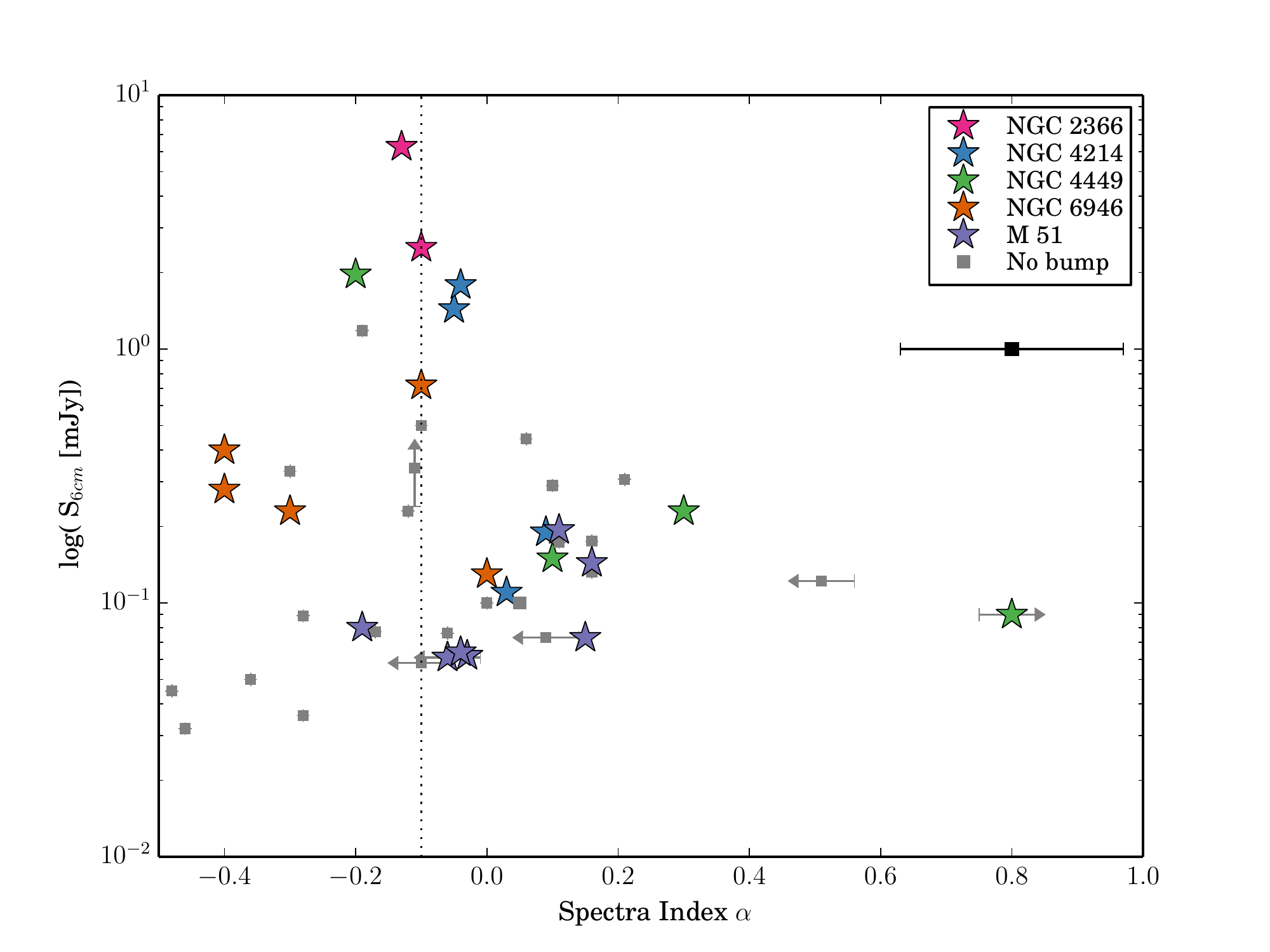}
\caption{\label{fig-radiosample} The radio spectral index $\alpha$ plotted versus the 6cm flux density. WR clusters are the star markers color-coded by host galaxy and `No-bump' sources are the grey squares (both Non-WR clusters and Other classes included here). Typical uncertainties are shown by the black marker below the legend box. It is clear that the `No-bump' and WR clusters sample the same parameter space.}
\end{figure}

We classified the sources in our sample according to the detection of the WR bump. The first category is for sources in which the WR bump was significantly detected. As these sources are emerging massive star clusters that clearly host WR stars, similar to S26, we termed them ``emerging WR clusters'' (frequently referred to as ``WR clusters'' for brevity). For all other sources, we broadly called them ``no-bump'' sources. Any spectra that were clearly different (not H \textrm{II} regions) or were background objects fall into an ``Other'' class. ``Other'' sources are not included further in this work outside of discussing the targeted sample in Figure \ref{fig-radiosample}. The rest of the no-bump sources appeared to be H \textrm{II} regions and we thus assumed they are indeed emerging massive star clusters that we term ``emerging Non-WR'' clusters (similarly, called ``Non-WR clusters'' throughout). Non-WR clusters include a handful of sources we designated as ``Candidate'' sources, which appeared to have marginal WR bump detections by eye  (yet SNR$_{\text{bump}}$  $\leq$ 5).  Applying this classification scheme, we found our sample consists of 21 WR clusters, 17 Non-WR clusters (including 4 Candidates), and 3 Other. The classification of individual sources is given in Table \ref{table-bumps}.

\section{\label{section-environment} Characterizing the Environments}

\subsection{\label{section-photometry} Photometry}

\begin{figure}[!t]
\hspace{-15pt}
\includegraphics*[width=0.55\textwidth,angle=0]{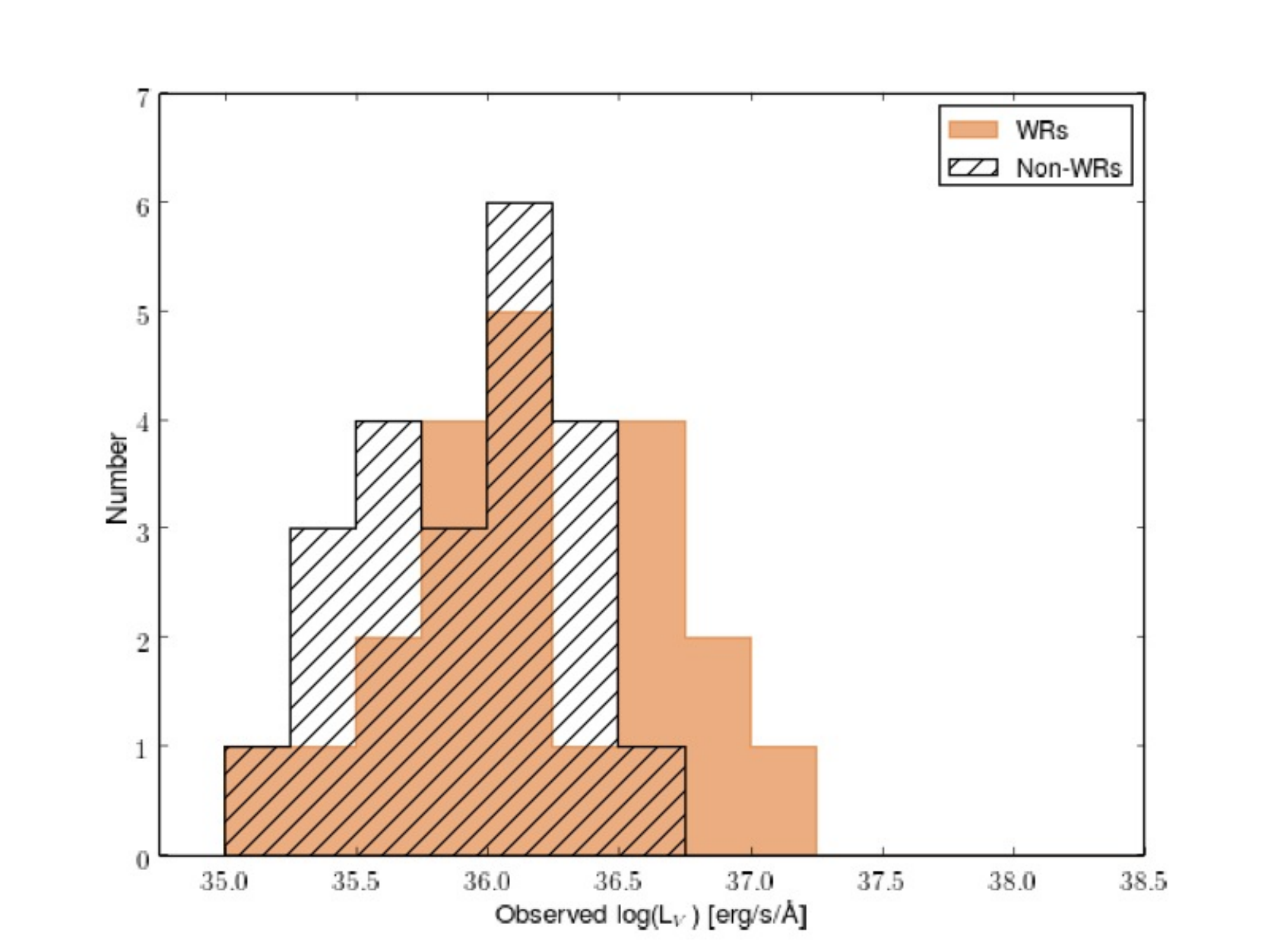}
\hspace{-25pt}
\includegraphics*[width=0.55\textwidth,angle=0]{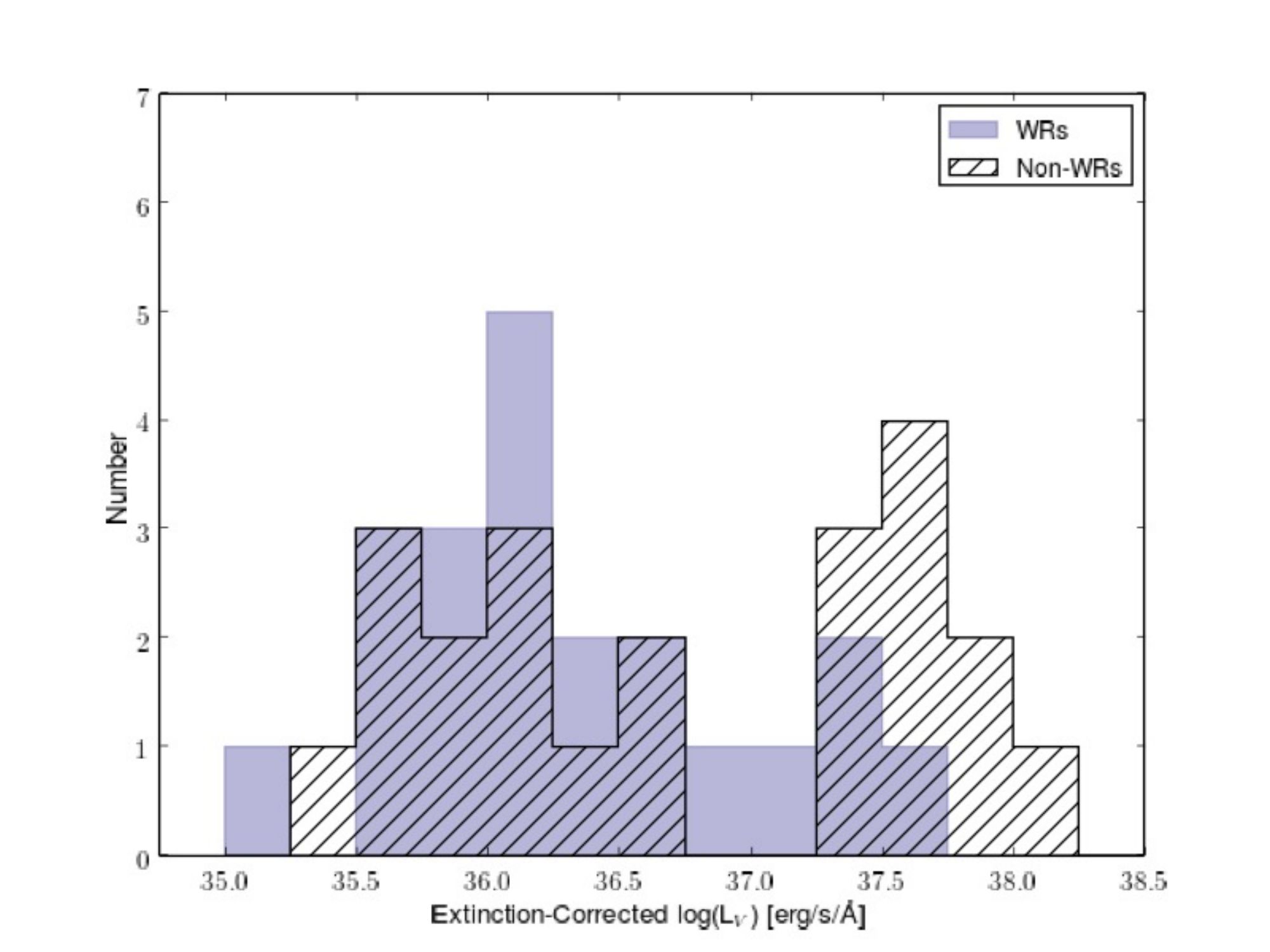}
\caption{\label{fig-vband} Histograms of the V-band photometric luminosity of the sample. Colors indicate the WR cluster class, and the black hatches indicate the Non-WR clusters. Left: the raw (observed) luminosity distribution of both classes is similar. Right: the extinction-corrected (intrinsic) luminosity distribution shows some differences between sources with and without WRs, highlighting that some of the Non-WR clusters exhibit higher extinctions and thus are intrinsically brighter. }
\end{figure}

Optical images were used to obtain a total V-band flux measurement of each source, which helped characterize the sample and to calibrate the optical spectra.  To limit nebular emission contributions, we used a medium-width filter, such as F550M, whenever available. In choosing the extraction apertures, we found that the sources generally reside in complex environments that rarely display a clear boundary, as can be seen in  Figure \ref{fig-rgb_panel1} and in the Appendix in Figures \ref{fig-rgb_panel2}-\ref{fig-rgb_panel3}. We adopted a circular aperture with a radius of 2\farcs5 centered at the coordinates of the target's radio continuum position, except  for NGC 4214 - Object 15/16, which we extracted with a 4\farcs0  radius.  We performed aperture photometry with the IDL procedure SURPHOT \citep{rei08}, where several background annuli were used and the resistant mean and mode were taken to estimate the background value. The uncertainties were dominated by background subtraction and were found by the standard deviation of fluxes calculated using these different background estimates. The raw luminosities are presented in the Appendix in Table \ref{table-opticalcharacteristics}. We found the WR clusters and the Non-WR clusters have similar raw luminosity distributions (see Figure \ref{fig-vband}).

\subsection{\label{section-lines} Nebular Emission Line Measurements and Corrections}

We measured the emission line strengths of the nebular lines in order to determine the interstellar extinction, the optical ionizing flux, and the metallicity. Emission line measurements were obtained using the IRAF SPLOT package and are presented in the Appendix in Tables \ref{table-fluxes_wrs} and \ref{table-fluxes_nonwrs}. Reddening corrections were determined with interstellar extinctions derived from the Balmer line ratios (discussed in Section \ref{section-extinction}). In addition to  the measurement uncertainties (as in Section \ref{section-bump}), uncertainties in the flux were estimated by including the uncertainty in the flux calibration fit and the uncertainty in the reddening correction. High extinctions limited the measurement of the nebular emission lines for many of the Non-WR sources, and thus we did not present measurements for these objects. Zero-point flux corrections, accounting for non-photometric conditions as well as slit corrections,  were made by comparing the V-band flux (see Section \ref{section-photometry}) to the total spectral flux  in the relevant filter bandpass and listed in Table \ref{table-opticalcharacteristics} (in the Appendix). Many sources did not require a flux correction or only minimal adjustment was necessary.

\subsection{Nebular Properties}

\subsubsection{\label{section-extinction} Interstellar Extinction}

Measuring the interstellar extinction for a source provides not only a way to correct the raw flux measurements, but also is a valuable intrinsic property of the environment of the source as well. The interstellar extinction of individual sources were determined using the optical nebular Balmer line ratios of H$\alpha$, H$\beta$, H$\gamma$, and H$\delta$. The observed Balmer decrement was converted to A$_\text{V}$ via the extinction curve for  the super star cluster analog 30 Doradus of the LMC \citep{mis99,fitz85}, and the average weighted A$_\text{V}$ was adopted for each source. The use of a Milky Way extinction curve produces similar results, and as these extinction curves are roughly parallel over the optical regime  \citep{dmp14}, reddening corrections applied to the measured emission lines follow the same wavelength dependence. We found that the extinctions measured with the 30 Doradus extinction curve are, on average, roughly 0.9 times the extinctions derived with the Milky Way extinction curve, and thus the extinctions adopted in this work may be slightly underestimated. However, the results and interpretation of this work are robust to a moderate increase in extinction.  For example, the line ratios remain the same (as the extinction curve for the 30 Dor extinction curve and the Milky Way extinction curve are parallel). The line fluxes, and properties derived from the line fluxes, may be increased by up to 5\% for A$_{\text{V}} \leq$ 0.5, 10\% for A$_{\text{V}} \leq$ 1.0, and 20\% for A$_{\text{V}} \leq$2.0, which would result in higher inferred ionization fluxes, and therefore also larger cluster masses.  However, the conclusions of this work are relatively unaffected by the choice of extinction curve. If the extinction was measured to be negative, the adopted extinction was zero. We used the available Balmer lines in the case that a source was without the complete Balmer series suite. The derived interstellar extinctions are given in Table \ref{table-fundamental}.

\begin{deluxetable}{lllllll}
\tabletypesize{\scriptsize}
\tablewidth{0pt} 
\tablecaption{\label{table-fundamental} Source Properties}
\tablehead{
	 \colhead{Source}		&
	 \colhead{A$_{\text{V}}$}		&
	 \colhead{Age }		&
	 \colhead{Q$_\text{o}$(H$\beta$)}		&
	 \colhead{Q$_\text{o}$(radio)}		&
	\colhead{L$_{V, i}$}		& 
	 \colhead{Mass}		\\
	 \colhead{}		&	    
	\colhead{mag}		&	    
	 \colhead{Myr}		&	    
	 \colhead{10$^{49}$ photons s$^{-1}$ }		&
	 \colhead{10$^{49}$ photons s$^{-1}$ }		&    
  	 \colhead{erg s$^{-1}$ }		&
	 \colhead{M$_\sun$}		\\	    

}		 
\startdata

NGC 2366 - Object 10	 & 	0.0	 & 	2.0 (0.6)	 & 	165.8 (14.2)	 & 	505.5 (50.7)	 & 	1.7e+36 (2.9e+34)	 & 	4.1e+04\\ 
NGC 2366 - Object 11	 & 	0.1	 & 	3.2 (0.7)	 & 	63.3 (7.9)	 & 	210.3 (21.8)	 & 	9.3e+35 (2.3e+34)	 & 	1.2e+04\\ 
NGC 2366 - Object 13	 & 	0.0	 & 	4.2 (0.2)	 & 	9.8 (0.8)	 & 	22.8 (3.0)	 & 	2.3e+35 (3.3e+34)	 & 	5.0e+03\\ 
NGC 4214 - Object 13	 & 	0.0	 & 	3.3 (0.6)	 & 	17.5 (2.2)	 & 	8.3 (2.4)	 & 	3.6e+35 (4.7e+34)	 & 	5.8e+03\\ 
NGC 4214 - Object 14	 & 	0.0	 & 	2.9 (0.5)	 & 	28.9 (4.1)	 & 	105.2 (12.4)	 & 	3.6e+35 (4.5e+34)	 & 	4.8e+03\\ 
NGC 4214 - Object 15/16	 & 	0.0	 & 	2.8 (0.0)	 & 	41.9 (4.2)	 & 	69.5 (8.5)	 & 	9.7e+35 (1.3e+35)	 & 	1.3e+04\\ 
NGC 4449 - Object 3	 & 	0.1	 & 	3.7 (0.5)	 & 	15.2 (5.6)	 & 	42.8 (6.7)	 & 	4.9e+35 (1.6e+35)	 & 	8.6e+03\\ 
NGC 4449 - Object 4	 & 	0.2	 & 	8.0 (0.3)	 & 	12.5 (2.4)	 & 	22.3 (4.5)	 & 	1.8e+36 (1.8e+35)	 & 	5.4e+04\\ 
NGC 4449 - Object 18	 & 	0.0	 & 	3.4 (0.3)	 & 	11.3 (1.1)	 & 	10.4 (1.0)	 & 	1.4e+35 (3.5e+34)	 & 	2.4e+03\\ 
NGC 4449 - Object 22	 & 	0.1	 & 	5.3 (0.2)	 & 	8.8 (1.7)	 & 	14.2 (4.0)	 & 	7.6e+35 (8.5e+34)	 & 	1.4e+04\\ 
NGC 4449 - Object 23	 & 	0.2	 & 	10.0 (0.0)	 & 	0.5 (0.1)	 & 	64.8 (10.1)	 & 	6.1e+35 (2.9e+35)	 & 	3.1e+04\\ 
NGC 4449 - Object 26	 & 	0.1	 & 	3.0 (0.4)	 & 	106.3 (9.2)	 & 	258.8 (26.4)	 & 	1.6e+36 (7.3e+34)	 & 	3.1e+04\\ 
NGC 4449 - Object 27	 & 	0.0	 & 	4.5 (0.1)	 & 	7.8 (7.6)	 & 	37.6 (6.4)	 & 	3.6e+35 (1.2e+34)	 & 	9.6e+03\\ 
NGC 6946 - Object 13	 & 	0.7	 & 	2.8 (0.3)	 & 	14.5 (4.3)	 & 	123.8 (22.3)	 & 	3.3e+37 (7.6e+35)	 & 	6.8e+05\\ 
NGC 6946 - Object 37	 & 	1.2	 & 	4.4 (0.3)	 & 	4.8 (2.5)	 & 	26.9 (6.0)	 & 	2.0e+37 (1.4e+35)	 & 	5.9e+05\\ 
NGC 6946 - Object 48	 & 	0.5	 & 	2.9 (0.7)	 & 	9.6 (2.9)	 & 	38.9 (5.7)	 & 	1.4e+37 (2.7e+35)	 & 	2.8e+05\\ 
NGC 6946 - Object 110	 & 	0.3	 & 	2.3 (0.2)	 & 	22.0 (6.4)	 & 	187.6 (28.0)	 & 	1.9e+37 (8.9e+35)	 & 	4.0e+05\\ 
NGC 6946 - Object 115	 & 	0.1	 & 	3.1 (0.2)	 & 	11.0 (2.0)	 & 	48.3 (13.5)	 & 	6.8e+36 (3.2e+35)	 & 	1.4e+05\\ 
NGC 6946 - Object 117	 & 	0.2	 & 	5.0 (0.1)	 & 	15.1 (3.5)	 & 	25.6 (10.2)	 & 	2.1e+37 (8.0e+35)	 & 	6.1e+05\\ 
M 51 - Object 5	 & 	0.0	 & 	2.9 (0.2)	 & 	17.3 (1.6)	 & 	37.6 (9.5)	 & 	1.1e+36 (2.4e+34)	 & 	2.3e+04\\ 
M 51 - Object 6	 & 	0.8	 & 	4.2 (0.1)	 & 	16.7 (4.0)	 & 	16.5 (6.9)	 & 	5.4e+36 (5.8e+35)	 & 	1.2e+05\\ 
M 51 - Object 11	 & 	0.0	 & 	4.3 (0.2)	 & 	9.8 (0.8)	 & 	31.4 (7.4)	 & 	1.2e+36 (1.4e+35)	 & 	3.2e+04\\ 
M 51 - Object 34	 & 	1.2	 & 	6.4 (0.0)	 & 	11.7 (5.1)	 & 	62.9 (8.1)	 & 	3.9e+37 (1.8e+36)	 & 	1.2e+06\\ 
M 51 - Object 39	 & 	1.2	 & 	6.4 (0.0)	 & 	38.4 (19.4)	 & 	68.1 (8.9)	 & 	6.8e+37 (3.4e+36)	 & 	2.2e+06\\ 
M 51 - Object 44	 & 	0.6	 & 	2.9 (0.8)	 & 	26.9 (9.0)	 & 	89.2 (10.6)	 & 	7.9e+35 (1.1e+35)	 & 	1.6e+04\\ 
M 51 - Object 46	 & 	0.0	 & 	2.9 (0.0)	 & 	103.4 (9.2)	 & 	46.7 (9.5)	 & 	4.0e+36 (3.0e+35)	 & 	8.2e+04\\ 
M 51 - Object 57	 & 	0.3	 & 	2.8 (0.2)	 & 	10.5 (2.5)	 & 	24.8 (5.8)	 & 	4.1e+35 (1.6e+35)	 & 	8.3e+03\\ 
M 51 - Object 60	 & 	1.2	 & 	6.5 (0.0)	 & 	9.1 (1.3)	 & 	23.2 (6.1)	 & 	3.4e+37 (1.6e+36)	 & 	1.1e+06\\ 
M 51 - Object 63	 & 	1.7	 & 	6.5 (0.0)	 & 	38.9 (9.8)	 & 	18.6 (5.5)	 & 	1.4e+38 (2.4e+36)	 & 	4.5e+06\\ 
M 51 - Object 67	 & 	1.8	 & 	6.5 (0.0)	 & 	16.5 (3.2)	 & 	45.9 (6.9)	 & 	7.2e+37 (2.9e+35)	 & 	2.3e+06\\ 
M 51 - Object 73	 & 	0.0	 & 	4.2 (0.1)	 & 	20.7 (8.6)	 & 	29.5 (6.6)	 & 	2.1e+36 (8.6e+34)	 & 	4.8e+04\\ 
M 51 - Object 88	 & 	1.5	 & 	6.3 (0.0)	 & 	10.4 (2.7)	 & 	39.7 (6.9)	 & 	2.0e+37 (4.7e+35)	 & 	1.2e+06\\ 
M 51 - Object 90	 & 	1.8	 & 	6.3 (0.0)	 & 	18.2 (4.8)	 & 	158.3 (17.0)	 & 	2.8e+37 (3.4e+35)	 & 	9.2e+05\\ 
M 51 - Object 92	 & 	0.1	 & 	4.3 (0.3)	 & 	14.9 (3.3)	 & 	39.2 (7.8)	 & 	2.1e+36 (1.2e+35)	 & 	5.9e+04\\ 
M 51 - Object 93	 & 	0.7	 & 	5.0 (0.2)	 & 	11.2 (2.3)	 & 	29.9 (6.4)	 & 	4.8e+36 (6.4e+34)	 & 	1.2e+05\\ 
M 51 - Object 94	 & 	0.0	 & 	2.8 (0.1)	 & 	25.9 (7.9)	 & 	23.1 (4.7)	 & 	1.1e+36 (1.6e+34)	 & 	2.2e+04\\ 
M 51 - Object 96	 & 	1.9	 & 	10.0 (0.4)	 & 	1.4 (0.3)	 & 	25.8 (6.2)	 & 	2.3e+37 (2.4e+35)	 & 	1.7e+06\\ 
M 51 - Object 97	 & 	1.2	 & 	4.2 (0.1)	 & 	77.1 (28.5)	 & 	228.4 (23.7)	 & 	4.4e+37 (9.2e+35)	 & 	1.0e+06\\ 
M 51 - Object 100	 & 	0.0	 & 	3.0 (0.1)	 & 	303.1 (23.6)	 & 	68.9 (8.3)	 & 	4.2e+36 (1.1e+35)	 & 	8.5e+04\\ 
M 51 - Object 101	 & 	0.0	 & 	3.0 (0.1)	 & 	18.3 (1.5)	 & 	17.2 (3.9)	 & 	1.6e+36 (1.5e+35)	 & 	3.2e+04\\ 
M 51 - Object 103	 & 	0.2	 & 	3.0 (0.2)	 & 	12.8 (3.0)	 & 	47.8 (8.5)	 & 	1.1e+36 (2.0e+35)	 & 	2.2e+04\\ 
M 51 - Object 105	 & 	0.7	 & 	2.2 (0.1)	 & 	9.6 (4.0)	 & 	90.2 (11.9)	 & 	5.0e+35 (3.2e+34)	 & 	9.8e+03\\ 

\enddata
\tablecomments{For each emerging massive star cluster in this table we list the following: the interstellar extinction as measured by Balmer lines, the age estimated by
the equivalent width of H$\beta$, the ionizing flux as inferred from H$\beta$ and thermal radio emission, the intrinsic extinction-corrected luminosity, and the stellar mass.}
\end{deluxetable}

\subsubsection{\label{section-nebular} Ionized Gas Density and Temperature}

The ionized gas conditions of WR sources were determined using the nebular emission line ratios. To determine the electron density and temperatures in the ionized gas, we utilized the NEBULAR package in IRAF, which uses the five-level atom model \citep{der87}. The electron density was estimated using the S \textrm{II} line ratio 6716$\lambda$/6731$\lambda$. The electron temperatures were given by: O$^{+}$ electron temperature (T(O \textrm{II})) was measured using the [O \textrm{II}] ratio (3727$\lambda$/(7319$\lambda$+7330$\lambda$)), and O$^{++}$ electron temperature (T(O \textrm{III})) was measured using the [O \textrm{III}] ratio (4959$\lambda$+5007$\lambda$)/4363$\lambda$. The results are presented in Table \ref{table-nebular}; the densities and temperatures of the WR clusters were typical for extragalactic H \textrm{II} regions \citep{hh09}.

\begin{deluxetable}{llllll}
\tabletypesize{\scriptsize}
\tablewidth{0pt} 
\tablecaption{\label{table-nebular} Nebular Properties of the WR Clusters}
\tablehead{
	 \colhead{Source}		&
	 \colhead{n$_{\text{e}}$}		&
	 \colhead{T([O  \textrm{II}])}		&
	 \colhead{T([O  \textrm{III}])}		&
	\colhead{12+$\log$(O/H)}		&
	\colhead{z}		\\ 
	 \colhead{}		&
	 \colhead{cm$^{-3}$}		&
	 \colhead{10$^4$ K}		&
	 \colhead{10$^4$ K}		&	    
	 \colhead{}		&	    
  	 \colhead{}		\\
}		 
\startdata
NGC 2366 - Object 10	 & 	512 (264)	 & 	1.3 (0.3)	 & 	1.6 (0.1)	 & 	7.9 (0.1)	 & 	0.002\\ 
NGC 2366 - Object 11	 & 	32 (300)	 & 	1.7 (0.1)	 & 	1.4 (0.1)	 & 	7.9 (0.1)	 & 	0.003\\ 
NGC 4214 - Object 13	 & 	76 (134)	 & 	1.1 (0.1)	 & 	1.0 (0.1)	 & 	8.4 (0.2)	 & 	0.008\\ 
NGC 4214 - Object 14	 & 	111 (158)	 & 	1.2 (0.2)	 & 	1.1 (0.1)	 & 	8.2 (0.2)	 & 	0.005\\ 
NGC 4214 - Object 15/16	 & 	50 (66)	 & 	1.2 (0.1)	 & 	1.2 (0.1)	 & 	8.2 (0.1)	 & 	0.005\\ 
NGC 4449 - Object 4	 & 	75 (355)	 & 	1.5 (0.3)	 & 	1.9 (0.3)	 & 	7.9 (0.3)	 & 	0.002\\ 
NGC 4449 - Object 18	 & 	50 (27)	 & 	1.0 (0.1)	 & 	1.3 (0.1)	 & 	8.5 (0.2)	 & 	0.009\\ 
NGC 4449 - Object 22	 & 	22 (128)	 & 	1.2 (0.2)	 & 	2.0 (0.3)	 & 	8.3 (0.4)	 & 	0.006\\ 
NGC 4449 - Object 26	 & 	157 (131)	 & 	1.1 (0.1)	 & 	1.0 (0.0)	 & 	8.5 (0.1)	 & 	0.008\\ 
NGC 6946 - Object 13	 & 	49 (270)	 & 	1.5 (0.4)	 & 	0.7 (0.1)	 & 	9.0 (0.2)	 & 	0.032\\ 
NGC 6946 - Object 48	 & 	557 (1070)	 & 	1.9 (0.0)	 & 	4.3 (4.1)	 & 	9.3 (0.2)	 & 	0.062\\ 
NGC 6946 - Object 110	 & 	101 (332)	 & 	1.8 (0.2)	 & 	1.1 (0.1)	 & 	9.1 (0.2)	 & 	0.039\\ 
NGC 6946 - Object 115	 & 	15 (249)	 & 	1.3 (0.3)	 & 	1.7 (0.2)	 & 	9.1 (0.1)	 & 	0.035\\ 
NGC 6946 - Object 117	 & 	496 (771)	 & 	1.1 (0.7)	 & 	2.0 (0.4)	 & 	8.9 (0.1)	 & 	0.023\\ 
M 51 - Object 46	 & 	129 (107)	 & 	1.0 (0.1)	 & 	0.4 (0.0)	 & 	9.1 (0.4)\tablenotemark{a}	 & 	0.040\tablenotemark{a}\\ 
M 51 - Object 57	 & 	152 (285)	 & 	1.0 (0.3)	 & 	1.7 (0.3)	 & 	9.1 (0.1)	 & 	0.037\\ 
M 51 - Object 73	 & 	115 (711)	 & 	0.9 (0.4)	 & 	2.1 (1.2)	 & 	9.2 (0.2)	 & 	0.043\\ 
M 51 - Object 94	 & 	44 (282)	 & 	0.7 (0.1)	 & 	3.0 (1.6)	 & 	9.4 (0.2)	 & 	0.066\\ 
M 51 - Object 100	 & 	149 (90)	 & 	0.9 (0.1)	 & 	2.3 (0.1)	 & 	9.5 (0.1)	 & 	0.092\\ 
M 51 - Object 101	 & 	89 (90)	 & 	0.7 (0.1)	 & 	4.3 (0.7)	 & 	9.3 (0.1)	 & 	0.060\\ 
M 51 - Object 103	 & 	60 (176)	 & 	1.0 (0.2)	 & 	2.7 (0.8)	 & 	9.1 (0.1)	 & 	0.039\\ 

\enddata

\tablecomments{
This table presents the electron density, electron temperatures, and metallicity measured for the WR clusters.}
\tablenotetext{a}{An approximate metallicity adopted for this source.}

\end{deluxetable}

\subsubsection{\label{section-metallicity} Metallicity}

Massive stellar evolution is sensitive to metallicity, in part due to the dependence of mass-loss on metallicity \citep{vink05,cro06,hai15}. This results in a well-known trend for a decrease in the ratio of WN to WCs with decreasing metallicity. Therefore determining the metallicity was crucial for understanding our sources, especially as our sample spanned a wide range of environments. Because certain oxygen lines become faint at high metallicity, different methods were necessary to estimate the metallicity of each source throughout the sample. 

The most accurate method, known as the T$_e$ or ``direct method'', was used for all WR sources in the galaxies NGC 2366, NGC 4214, and NGC 4449 (low to moderate metallicity environments). The T$_e$ method used two distinct temperature zones in the photoionized H \textrm{II} region using the O$^{+}$ and O$^{++}$ electron temperatures \citep{izo94, izo97}. The IRAF package NEBULAR was used to measure ionic abundances, with the appropriate electron temperature.The O$^{+}$ ionic abundance was measured with  ionic abundances of 3727$\lambda$ and the doublet 7319/7330$\lambda$, and O$^{2+}$ ionic abundance was determined using ionic abundances for the lines 4363$\lambda$, 4959$\lambda$, and 5007$\lambda$. The total oxygen abundance was then derived by O/H = O$^{+}$/H$^{+}$+O$^{++}$/H$^{+}$.

For WR sources in galaxies NGC 6946 and M 51, which lied on the so-called ``upper branch'' where  log([N \textrm{II}]/[O \textrm{II}]) $>$ -0.8 \citep{vz98}, a strong line method was adopted. Certain strong nebular emission lines are easily observed and used to find the parameters R3$=$([O \textrm{III}] 4959 \AA\ + [O \textrm{III}] 5007\AA) / H$\beta$ and R2 $=$([O \textrm{II}] 3727 \AA\ + [O \textrm{II}] 3729 \AA)/ H$\beta$. To estimate the metallicity using the R3 and R2 parameters, we adopted the method of \citet{kk04} (often called the KK04 method) in which the oxygen abundance was found iteratively. The metallicity (z) was found using the variables x$=$log(R2+R3) and y$=$log(R3/R2) and the relations $$ \log(q) = \frac{32.81-1.153y^2+z \times (-3.396-0.025y+0.1444y^2)}{4.603-0.3119y-0.163y^2+z \times (-0.48-0.0271y+0.02037 y^2)} $$ and $$ z = 9.72 - 0.777 x -  0.951  x^2 - 0.072 x^3 - 0.811 x^4 - \log(q) \times (0.0737 - 0.0713  x - 0.141 x^2 + 0.0373 x^3 - 0.058  x^4) $$ \citep{kk04}. 

Conversely, the purpose of identifying approximate metallicities of the Non-WR clusters was for later comparisons to STARBURST99 models. Additionally, the necessary nebular emission lines to measure the metallicity were not observed for many of the Non-WR clusters.  As such,  we chose the best match of available metallicity models  as approximately averaged by the WR clusters in the same host galaxy. This approximate metallicity is also adopted for one WR source, M 51 - Object 46, as the derived metallicity was unphysical, which we attribute to relatively weak oxygen line emission.

All derived and adopted metallicities are listed in Table \ref{table-nebular}. For both methods, we converted between metallicity z and 12+log(O/H) by assuming a simple scaling relation and the solar metallicity value 12+log(O/H)$_{\sun}$ $=$ 8.69 \citep{asp09}.

\subsection{\label{section-ionizingflux} Estimating the Ionizing Flux}

Ionizing photons are produced by the massive stars that are harbored in the clusters in our sample.  If we assume that all ionizing (Lyman) photons are absorbed by hydrogen atoms, the so-called Case B approximation, then every ionization results in a recombination that produces a Balmer photon. Thus with the Case B assumption, we can use the measured H$\beta$ flux to infer the ionizing flux. However, this optical light can be obscured, causing the inferred ionizing flux to be underestimated. The ionizing flux can also be inferred from radio free-free emission, which would be unaffected by such extinction. While the optical interstellar extinction was measured in Section \ref{section-extinction} and is an important component to understanding the nebular environment of the source, using nebular lines has the caveat that it only measures extinction towards gas that is not very heavily extincted. As such, characterizing the ionizing flux inferred from both the radio and the optical data provides the opportunity to reveal optically obscured ionized gas, which could not be identified with optical spectra alone; therefore, we calculated the ionizing flux estimated from both radio and optical methods. 

The ionizing flux inferred from the optical emission was estimated through the empirical relations presented in \citet{sv98} using the emission line flux of H$\beta$ as $$ Q_{\text{o, H} \beta} \sim L_{H\beta}/4.76^{-13} \text{ phot} \; \text{s}^{-1}. $$ The ionizing flux  inferred from radio wavelengths can be found by $$ Q_{\text{o, radio}} \geq 6.3 \times 10^{52} (\frac{T_e}{10^4K}) ^{-0.45} (\frac{\nu}{\text{GHz}})^{0.1} \frac{L_{\nu \text{, thermal}}}{10^{27} \text{ erg} \; s^{-1} \;\text{Hz}^{-1}} \;\text{s}^{-1} \text{\citep{con92}.}  $$  As a radio flux measurement at 6cm was available for the entire sample, it was used here although higher frequencies will include less non-thermal contributions if they are available. We input the O$^{++}$ temperatures derived for the WR clusters (see Section \ref{section-nebular}) and an effective temperature of 10$^4$ K, the approximate average of the estimated temperatures, for the Non-WR cluster class. Of note, both the radio and optical inferred ionizing fluxes are typically considered to be lower limits if dust is absorbing ionizing photons or there is ionizing photon leakage. We list the ionizing flux inferred from both the optical and radio data in Table \ref{table-fundamental}.

\subsection{\label{section-fundamental} Fundamental Properties and STARBURST99 Models}

We present here both the fundamental properties of intrinsic luminosity, stellar mass, and age of the sample as well as stellar population synthesis models of  STARBURST99 \citep{sb99, lei14}. These properties provide the basis for much of the analysis in this study, and the age and mass were found using the predictions of the STARBURST99 models at a given metallicity.

The intrinsic V-band luminosities were determined by correcting the observed raw V-band photometry for the optical extinction (Section \ref{section-extinction}); these are presented in Table \ref{table-fundamental}. Figure \ref{fig-vband} shows that extinction-corrected luminosity distributions were not the same for the WR and Non-WR classes, in contrast to the observed luminosity distributions. Several of the Non-WR clusters have high extinctions that result in higher intrinsic luminosity compared to similar observed luminosity.

To estimate the age of each cluster, we used predictions from the stellar population synthesis models of  STARBURST99  \citep{sb99, lei14}, which simulate a simple starburst of a given metallicity.  The presence of thermal radio emission suggests that our sources are quite young, and we therefore assume an instantaneous burst of star formation. We ran STARBURST99 v7.0.0 \citep{lei14} for the four different metallicity tracks (z$=$ 0.004, 0.008, 0.020, and 0.040) that most closely matched the metallicity of the host galaxies of the sample--these are also the metallicities that were adopted for the Non-WR clusters as described in Section \ref{section-metallicity}.  We adopted a Kroupa IMF \citep{kroupa01} with mass limits of 0.1 - 120 M$_{\sun}$, using Geneva evolutionary tracks (single-star tracks) with high mass loss and  Pauldrach (WM-Basic)/Hillier (CMFGEN) atmospheres \citep{pauld98,hill98,hill99,smi02} for each STARBURST99 run.

\begin{figure*}[!t]
\hspace{-15pt}
\includegraphics*[width=0.55\textwidth,angle=0]{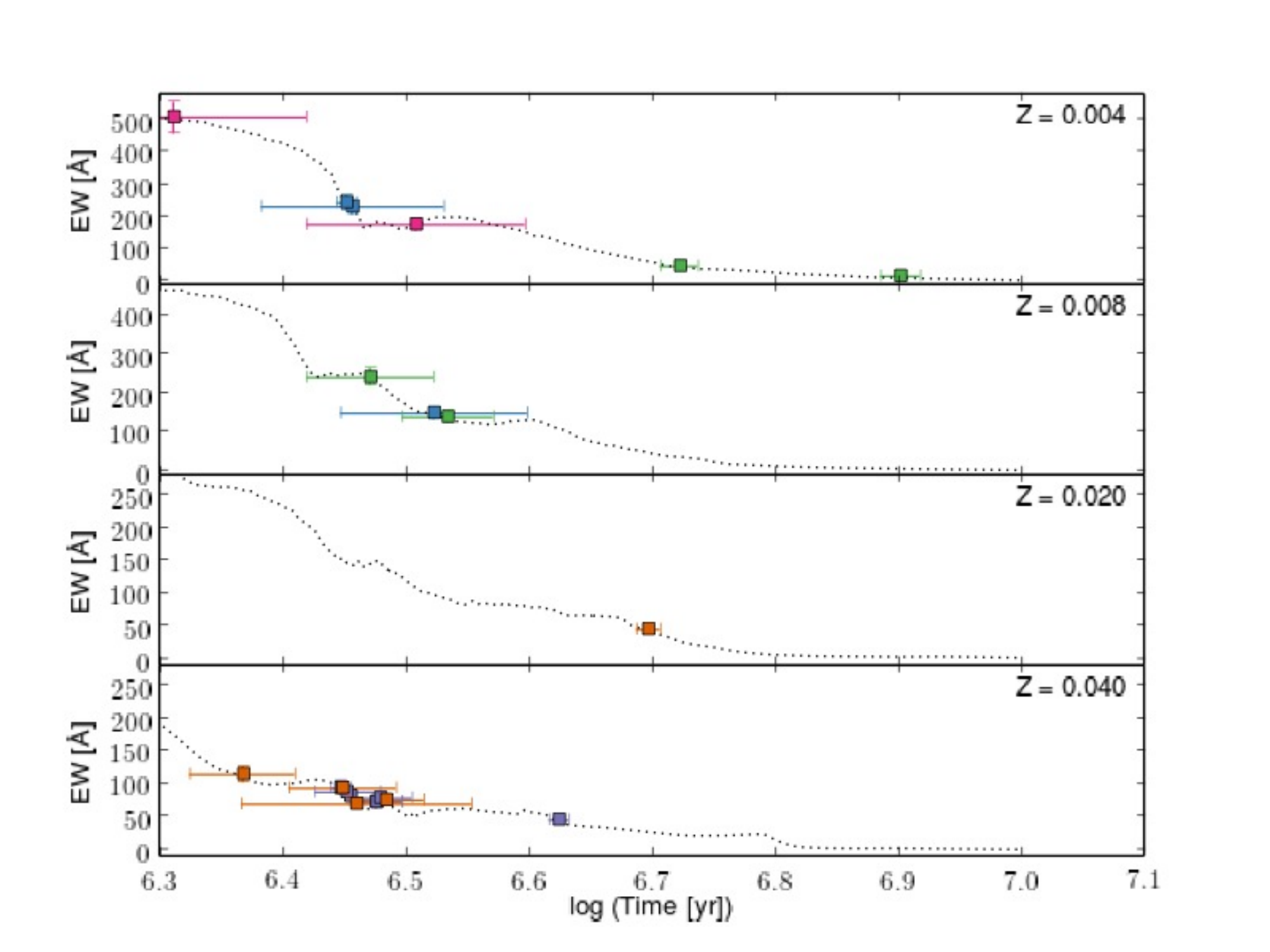}
\hspace{-20pt}
\includegraphics*[width=0.55\textwidth,angle=0]{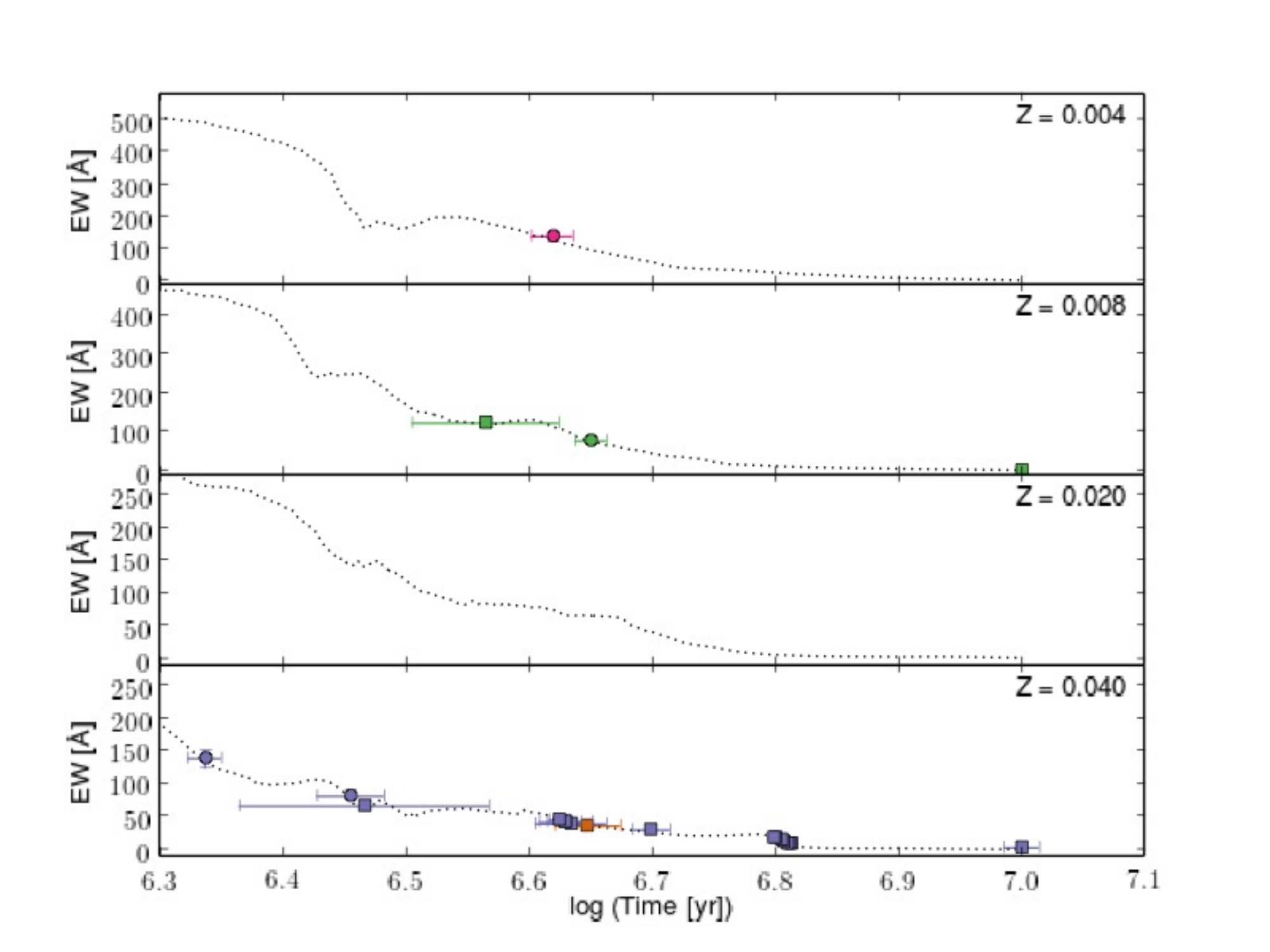}
\caption{\label{fig-ages_sb99} Ages of the sources in the sample (with WRs on the left, and Non-WRs on the right) as estimated by the measured equivalent width (EW) of H$\beta$ and predictions from STARBURST99 models, with the appropriate metallicity track plotted as separate panels. Sources are color-coded by host galaxy, as in Figure \ref{fig-radiosample}.}
\end{figure*}

The age of each source was estimated by comparing the measured equivalent width of H$\beta$ to that predicted by the STARBURST99 model with the appropriate metallicity track. Uncertainties were from the uncertainty of the measured equivalent width. In Figure \ref{fig-ages_sb99}, we plot the resulting ages against the STARBURST99 predictions. Throughout this work, the appropriate STARBURST99 model for each individual source was identified with this estimated age and metallicity track. Because it is more realistic to assume that there may be binary stars within each cluster, it is important to note that the use of binary evolution tracks would systematically increase the age estimates made here \citep[see Figure 6 in][]{es09}.

\begin{figure}[t]
\includegraphics[width=0.8\textwidth,angle=0]{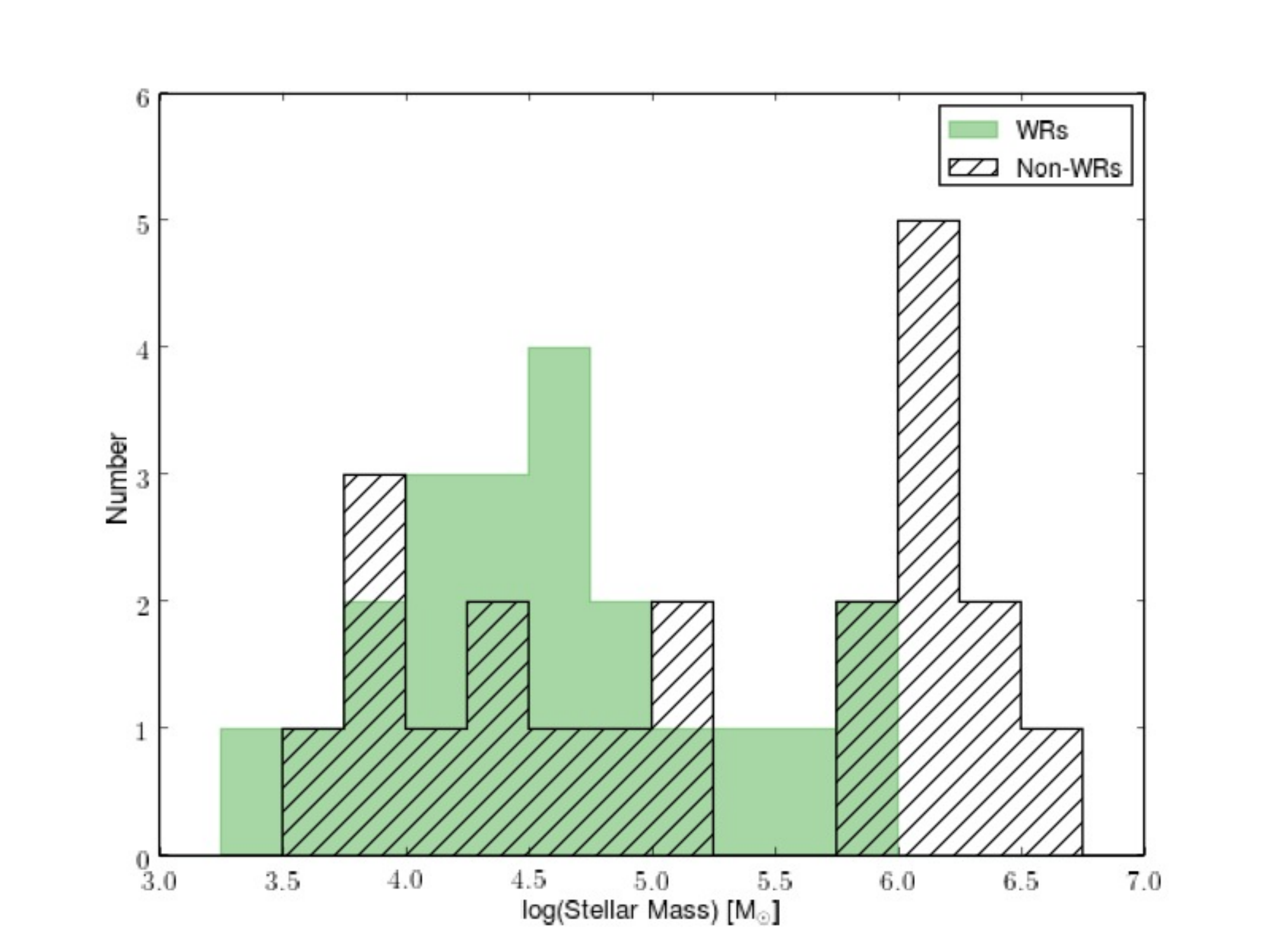}
\caption{\label{fig-mass} The distributions of the stellar cluster mass, obtained by scaling the V-band photometry to STARBURST99 models, of the sample.}
\end{figure}

The stellar mass of each cluster was then estimated by scaling the intrinsic V-band luminosity to the predicted luminosity from STARBURST99 models. The predicted luminosity was obtained by passing the synthetic spectra produced by STARBURST99, of the appropriate metallicity and age for each source, through the bandpass describing the observed V-band image (i.e. F550M or the KP 2.1m V-band filters). The stellar masses of the clusters are presented in Table \ref{table-fundamental}. Figure \ref{fig-mass} shows the cluster mass distributions of the sample, the shape of which somewhat reflect the distribution of the intrinsic luminosities. Overall, the estimated cluster masses fall in the general range for massive star clusters to super star clusters ($>$10$^4$ M$_\sun$), with most WR clusters at typical massive star cluster masses.

\subsection{\label{section-populations} Determining the Massive Star Populations}

Assessing the massive star populations in individual sources enables the comparison to both simulations and to observations of other regions.  We estimated the number of WR stars and O-stars in each emerging WR cluster using the optical spectral observations by following the methods of \citet{sv98} and \citet{gus00}. The resulting WR and O-star populations, as well as population ratios, are given in Table \ref{table-pops} for the emerging WR clusters. 

The WR bumps are used here not only confirm the presence of WR stars, but to estimate the WR populations directly. The number of WR stars was found from $N_{\text{WR}}=L_{\text{WR}}/L_{\text{o,WR}}$, where $L_{\text{o,WR}}$ was the typical luminosity of the WR feature produced by a single WR star and $L_{\text{WR}}$ was the observed luminosity of the same feature from the source. The WC subtypes dominate the C  \textrm{IV} $\lambda$ 5808 \AA\ emission bump. We therefore found the WC populations using the approximation of $N_{\text{WC}}=L(5808 \text{ \AA})/L_{\text{WC4}}(5808 \text{ \AA})$, assuming that WC4 stars are representative and have a typical luminosity at 5808 \AA\ is 3.0$\times$10$^{36}$ ergs s$^{-1}$ as measured in the LMC \citep{sv98}. 

The blue WR bump is produced by both WN and WC subtypes. Therefore, we estimated the population of WN stars from the blue WR bump after subtracting off the WCs' contribution. The WCs' contribution to the blue bump feature can be estimated using the luminosity of the red bump with the coefficient $k=L_{\text{WC4}} (4650  \text{ \AA})/ L_{\text{WC4}} (5808 \text{ \AA})$ \citep{gus00}; we adopted the value of k$=$1.71 $\pm$ 0.53 \citep{sv98, gus00}. The WN population was then found from the remaining observed emission in this WR feature and by assuming that a WN7 is representative and have a luminosity of 2.0 $\times$ 10$^{36}$ erg s$^{-1}$ in the blue bump \citep[4650 $+$ 4686 \AA;][]{gus00} as in the LMC to Milky Way environments. While O supergiants (Of-type stars) also produce He \textrm{II} $\lambda$ 4686 \AA\ emission, it is quite weak compared to that of WNs. Very massive unevolved stars will resemble hydrogen-rich WN stars; the WR stars in the R136 supercluster are such objects \citep{mh98}. However, the presence of the C  \textrm{IV} $\lambda$ 5808 \AA\ feature in some of our sources shows that, in those cases at least, we are dealing with an evolved massive star population, as this line must be due to WC stars.

Similarly to determining the WR populations, the ionizing flux inferred from the optical Balmer lines can be used to approximate the population of O-stars. The number of O7V stars can be simply estimated by subtracting  the ionizing flux produced by the WR stars and assuming an ionizing flux produced by a typical O7V star. We adopt ionizing flux values of $Q_{\text{o,O7V}} \sim 10^{48.75}$ s$^{-1}$ from an O7V star \citep{mart05} and $Q_{\text{o,WR}} \sim 10^{49}$ s$^{-1}$ for a WR star \citep{gus00,sch99}. To estimate the total O-star population, it is necessary to account for the fact that different subtypes of O-stars produce different ionizing flux values. We account for the different O-star subtypes, which are produced by an IMF, with the parameter $\eta_{\text{o}} = N_\text{O7V}/N_\text{O}$. Thus, the number of O-stars is then found by $$N_{\text{O}}=(Q_{\text{o}} - N_{\text{WR}} Q_{\text{o,WR}})/(\eta_o Q_{\text{o,O7V}})$$ \citep{gus00}. We estimate O-star population uncertainties by accounting for the measured flux uncertainty in, estimated uncertainties on $\eta_{\text{o}}$, and uncertainties for the subtracted WR populations.

\begin{deluxetable}{llllll|lll}
\tabletypesize{\scriptsize}
\tablewidth{0pt} 
\tablecaption{\label{table-pops} Massive Star Populations in the WR Clusters}
\tablehead{
	 \colhead{Source}		&
	 \colhead{O7V}		&
	 \colhead{O}		&
	 \colhead{WN}		&
	 \colhead{WC}		&
	 \colhead{WR$_{\text{total}}$}		&
	 \colhead{WC/WN}		&
	 \colhead{WR/O7V}		& 
	 \colhead{WR/O}		\\

}		 
\startdata
NGC 2366 - Object 10	 & 	272 (26)	 & 	239 (23)	 & 	3 (2)	 & 	...	 & 	3 (2)	 & 	...	 & 	0.011 (0.008)	 & 	0.013 (0.008)\\ 
NGC 2366 - Object 11	 & 	92 (17)	 & 	75 (16)	 & 	6 (5)	 & 	2 (1)	 & 	8 (5)	 & 	0.333 (0.139)	 & 	0.087 (0.066)	 & 	0.107 (0.070)\\ 
NGC 4214 - Object 13	 & 	24 (6)	 & 	23 (6)	 & 	2 (2)	 & 	2 (1)	 & 	4 (2)	 & 	1.000 (0.500)	 & 	0.167 (0.118)	 & 	0.174 (0.098)\\ 
NGC 4214 - Object 14	 & 	46 (8)	 & 	51 (14)	 & 	2 (2)	 & 	1 (1)	 & 	3 (2)	 & 	0.500 (0.500)	 & 	0.065 (0.051)	 & 	0.059 (0.042)\\ 
NGC 4214 - Object 15/16	 & 	62 (13)	 & 	68 (38)	 & 	7 (6)	 & 	...	 & 	7 (6)	 & 	...	 & 	0.113 (0.110)	 & 	0.103 (0.105)\\ 
NGC 4449 - Object 4	 & 	3 (16)	 & 	...	 & 	10 (9)	 & 	4 (2)	 & 	14 (9)	 & 	0.400 (0.180)	 & 	4.667 (11.187)	 & 	... \\ 
NGC 4449 - Object 18	 & 	14 (5)	 & 	15 (8)	 & 	4 (3)	 & 	0 (1)	 & 	4 (3)	 & 	0.030 (0.029)	 & 	0.286 (0.274)	 & 	0.267 (0.245)\\ 
NGC 4449 - Object 22	 & 	9 (4)	 & 	46 (22)	 & 	...	 & 	4 (2)	 & 	4 (2)	 & 	...	 & 	0.444 (0.370)	 & 	0.087 (0.060)\\ 
NGC 4449 - Object 26	 & 	179 (17)	 & 	142 (32)	 & 	2 (2)	 & 	4 (1)	 & 	6 (2)	 & 	2.000 (0.500)	 & 	0.034 (0.015)	 & 	0.042 (0.017)\\ 
NGC 6946 - Object 13	 & 	10 (11)	 & 	21 (24)	 & 	5 (4)	 & 	...	 & 	5 (4)	 & 	...	 & 	0.500 (0.660)	 & 	0.238 (0.332)\\ 
NGC 6946 - Object 48	 & 	10 (8)	 & 	18 (17)	 & 	4 (3)	 & 	...	 & 	4 (3)	 & 	...	 & 	0.400 (0.467)	 & 	0.222 (0.268)\\ 
NGC 6946 - Object 110	 & 	34 (12)	 & 	42 (15)	 & 	...	 & 	3 (1)	 & 	3 (1)	 & 	...	 & 	0.088 (0.060)	 & 	0.071 (0.035)\\ 
NGC 6946 - Object 115	 & 	13 (4)	 & 	18 (5)	 & 	...	 & 	1 (1)	 & 	1 (1)	 & 	...	 & 	0.077 (0.088)	 & 	0.056 (0.058)\\ 
NGC 6946 - Object 117	 & 	22 (6)	 & 	22 (7)	 & 	...	 & 	2 (1)	 & 	2 (1)	 & 	...	 & 	0.091 (0.066)	 & 	0.091 (0.054)\\ 
M 51 - Object 46	 & 	169 (18)	 & 	317 (127)	 & 	3 (3)	 & 	5 (2)	 & 	8 (4)	 & 	1.667 (0.667)	 & 	0.047 (0.028)	 & 	0.025 (0.016)\\ 
M 51 - Object 57	 & 	16 (5)	 & 	30 (15)	 & 	2 (1)	 & 	...	 & 	2 (1)	 & 	...	 & 	0.125 (0.094)	 & 	0.067 (0.047)\\ 
M 51 - Object 73	 & 	26 (17)	 & 	26 (8)	 & 	5 (5)	 & 	1 (1)	 & 	6 (5)	 & 	0.200 (0.200)	 & 	0.231 (0.268)	 & 	0.231 (0.205)\\ 
M 51 - Object 94	 & 	34 (17)	 & 	69 (42)	 & 	7 (5)	 & 	...	 & 	7 (5)	 & 	...	 & 	0.206 (0.207)	 & 	0.101 (0.095)\\ 
M 51 - Object 100	 & 	424 (86)	 & 	590 (128)	 & 	54 (42)	 & 	10 (4)	 & 	65 (42)	 & 	0.185 (0.058)	 & 	0.153 (0.121)	 & 	0.110 (0.075)\\ 
M 51 - Object 101	 & 	29 (3)	 & 	41 (5)	 & 	1 (1)	 & 	1 (1)	 & 	2 (1)	 & 	1.000 (1.000)	 & 	0.069 (0.041)	 & 	0.049 (0.025)\\ 
M 51 - Object 103	 & 	19 (6)	 & 	26 (9)	 & 	2 (2)	 & 	... & 	2 (2)	 & 	...	 & 	0.105 (0.121)	 & 	0.077 (0.081)\\ 

\enddata
\tablecomments{By using the strength of the measured WR bump and the H$\beta$ flux, we estimate the massive star populations in the individual WR clusters, as well as calculate the population ratios (e.g. WC/WN and WR/O). We present both population estimates for O7V stars, directly estimated from the H$\beta$ flux after subtracting the WR contributions, as well as for all O-stars, which requires an additional estimation of $\eta_o$ to account for the IMF across the O-star populations.}
\end{deluxetable}

\section{\label{section-results} Results}

\subsection{\label{section-commonality} The Commonality of WR Stars in Radio-selected Emerging Clusters}
 
Definitions of ``emerging'' and ``embedded'' can vary, we use the following for the purposes of this paper based on observables: fully embedded--no optical emission is observed; partially embedded--the optical extinction A$_{\text{v}} > 1$ or the ionizing flux inferred from the radio compared to the optical such that $ Q_{\text{o, radio}} / Q_{\text{o, H} \beta}  >1$; emerging --  not fully emerged and can be partially embedded or have less extinction; and fully emerged--no nebular emission is observed. All of the WR and Non-WR clusters in our study are therefore considered to be emerging, and 39 of the 45 sources are partially embedded.

\begin{figure}[!t]
\includegraphics[width=0.8\textwidth]{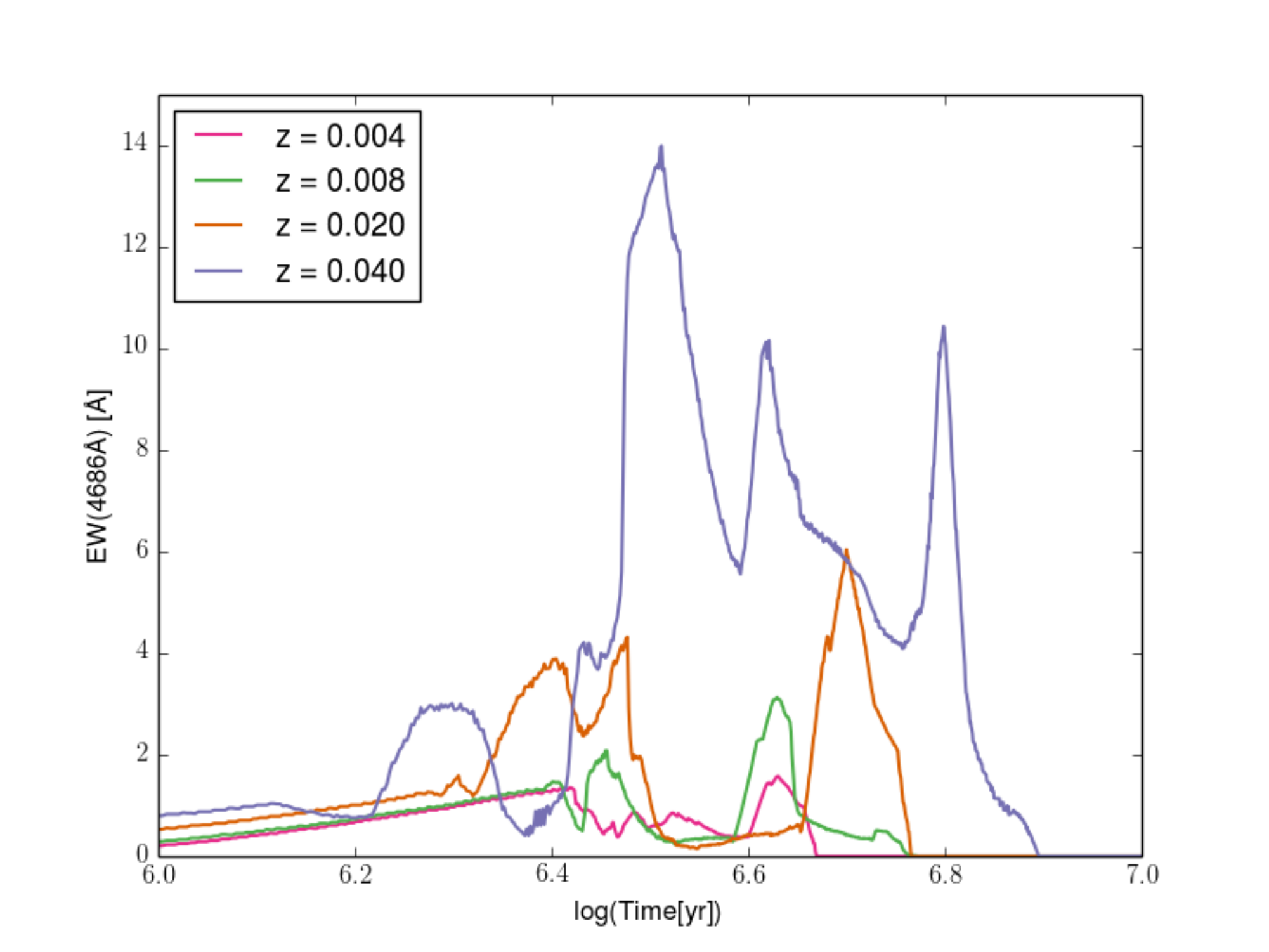}
\caption{\label{fig-sb99-ew}  STARBURST99 \citep{sb99} predictions for the behavior of the equivalent width of the WR emission at 4686 \AA\ for all relevant metallicity tracks. It is clear that the WR emission is not constant over the WR phase. }
\end{figure}

The emerging massive star clusters of our sample span a wide range of estimated ages; these ages are more diverse than we expected based on previous work on  thermal radio emission timescales. Thus, it is important to investigate the temporal behavior of the WR emission across those ages to understand if these emerging massive star clusters could or should be in the WR phase at some level. We turn to the predictions of the STARBURST99 models, and find that the WR bump can be produced throughout much of the time that is spanned by the sources in our sample. However, the strength of the WR bump varies greatly with time, which we show in Figure \ref{fig-sb99-ew} by plotting the STARBURST99 predicted EW of the WR bump at 4686 \AA\ versus time. In fact, if we consider only the behavior of the strength of the WR emission over time, which rises and falls--appearing to turn on and off, it is  clear that one should not expect to detect WR emission throughout the duration of the WR phase for an entire cluster. Many clusters could happen to be observed at a time  in between WR emission peaks. The minimum EW observed could be altered by the mass of the cluster, metallicity, and other star formation parameters, and could even result in no WR emission in many cases, especially where stochasticity is at play.

These  STARBURST99 predictions above show that the EW of the WR bump can vary in time, and therefore over the ages spanned by the emerging massive star clusters in our sample, a detection of the WR bump could be quite rare. For instance for a 10$^6$ M$_\sun$ mass cluster at a metallicity of z=0.020, the EW of the WR bump is expected to be above a detection limit of 2\AA\ only $\sim$35\% of the time from $\sim$ 1 to 6 Myr. In spite of these expectations, we have found $\sim$ 50\% of the sources in our radio-selected sample show significant detections of the WR bump in their optical spectra. Unfortunately due to the different observing constraints imposed when obtaining optical spectra with different instruments (namely fiber placement using the MMT versus hand-guiding at the Mayall Telescope),  the completeness of this sample cannot be reliably evaluated. Regardless, we do not think that this high percentage is due to preferentially selecting or observing sources with WR stars; the source selection was based on targets chosen from radio continuum studies and the quality of obtained optical spectra of these targets.  We show that all types of sources in the sample, whether a WR bump is detected or not, span roughly the same parameter space of the radio properties from which they were chosen, as shown by Figure \ref{fig-radiosample} where we plot the flux density at 6cm F$_{\lambda, 6cm}$ versus the radio spectral index $\alpha$. We adopted a SNR requirement of 15 in the observed spectra, rather than an optical brightness cutoff,  that allowed for different exposure times and observing conditions. Sources with or without a bump did not appear to be more limited by this choice. Figure \ref{fig-vband} shows the distribution of the observed raw V-band luminosities of the WR clusters is roughly the same as for the Non-WR clusters, although the brightest sources do have WR detections. Because the sample does not appear to be biased towards WR detections due to our source selection process, the sheer number of emerging WR clusters observed in this sample is quite meaningful. The high percentage of WR detections for sources in our sample indicates that WR stars may commonly be present in massive star clusters that are emerging. Overall, this work has shown that it can no longer be assumed that all massive star clusters emerge before their massive star habitants begin to evolve off of the main sequence.

\subsection{\label{section-comparepops} Comparing the Massive Star Populations}

\begin{figure}[!t]
\includegraphics[width=\textwidth,angle=0]{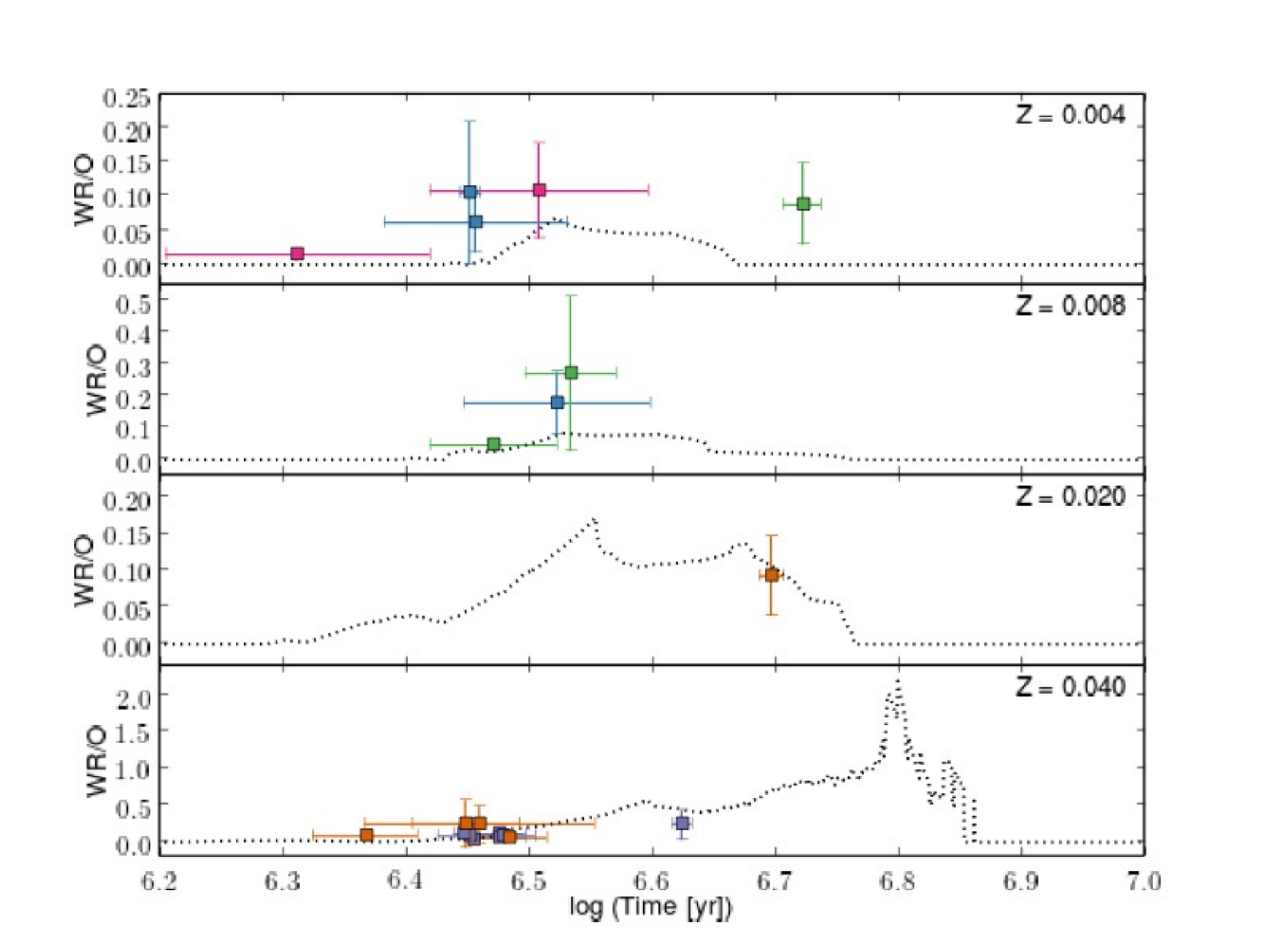}
\caption{\label{fig-pops_sb99} The estimated ratio of the number of WRs to O-stars plotted with STARBURST99 predictions (dotted line). Data points are for the emerging WR clusters, colors are the same as in Figure \ref{fig-radiosample}. NGC 4449 - Object 4 is not shown, which has an observed WR/O $\sim$ 5, and is discussed in Section \ref{section-comparepops}. The estimated populations ratios of sources in our sample are generally consistent with the STARBURST99 models, occasionally somewhat higher.
}
\end{figure}

\begin{figure}[t]
\includegraphics[width=\textwidth,angle=0]{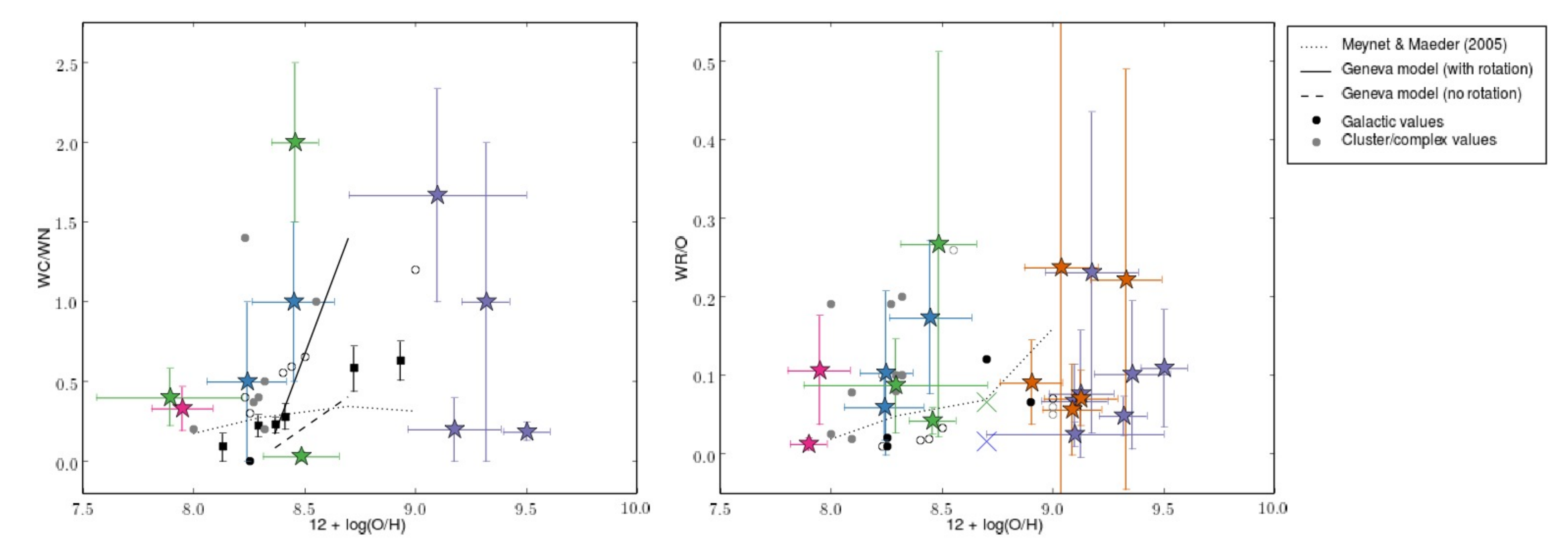}
\caption{\label{fig-pops} Populations ratios estimated for our sources in comparison to their metallicity (left: WC/WN, right: WR/O). Observations of other regions, both individual star-forming regions and averaged galactic areas, are overplotted; references can be found in \citet{sok15}. NGC 4449 - Object 4 is not shown, which has an observed WR/O $\sim$ 5. Uncertainties clutter the plot, but it is clear that the population ratios are similar to predictions and other regions.}
\end{figure}

Plotting the ratio of the massive star populations, meaning the number of WCs compared to WNs or WR stars to O-stars,  is an informative tool that can test evolutionary models but also put the observations in context to understand individual sources or general trends. We plotted the observed WR/O ratio versus time in comparison to STARBURST99 models for our sample (Figure \ref{fig-pops_sb99}). This plot is useful to see what WR/O populations we could expect for a coeval population that fully samples the IMF. Our emerging WR clusters were consistent with these predictions, although occasionally somewhat higher.  As noted in Section \ref{section-fundamental}, the STARBURST99 models include only single stars. If binary evolution is considered, the expected WR/O ratio is higher at any given metallicity \citep[see Figure 5 in][]{eit08}. As our clusters realistically include some binaries, it is no surprise that the WR/O ratio thus is occasionally higher than that expected from STARBURST99. We also plotted the WR/O ratio versus metallicity (Figure \ref{fig-pops}) to compare to observations of other star-forming regions, which are overplotted. Uncertainties dominated  this plot, but we again found that the WR/O estimates for our sources appear typical. One source, NGC 4449 - Object 4, was quite unusual compared to the rest of the sample. With an estimated WR/O $\sim$ 5 and age of $\sim 8$ Myr, this source was not shown in either of these plots. The optical spectra show much weaker nebular lines than any other source (see Fig. \ref{fig-fullspec1}). Perhaps this source can be explained by few O-stars remaining un-evolved and much of the ionized gas leaking out, possibly due to increased feedback within this specific cluster.

Shown in Figure \ref{fig-pops}, the WC/WN ratio typically decreases with decreasing metallicity. Discussed in Section \ref{section-metallicity}, this is a direct result of changes in the WR winds in these environments. Thus, the WC/WN ratio is regarded as one of the best observables to compare to stellar evolution models, but is also useful to check our population estimates. The best agreement between models and observations has been at low metallicity, but there has been much improvement at the higher metallicity with the most recent evolutionary models, particularly from the Geneva group that are shown at the solid and dashed lines in the plot. We found that our sources display normal WC/WN values compared to the other plotted regions and in line with the model predictions, although they are fairly loosely constrained in our sample.

\subsection{Evaluating Whether the Sources are still Embedded}

\begin{figure}[h]
\includegraphics[width=\textwidth]{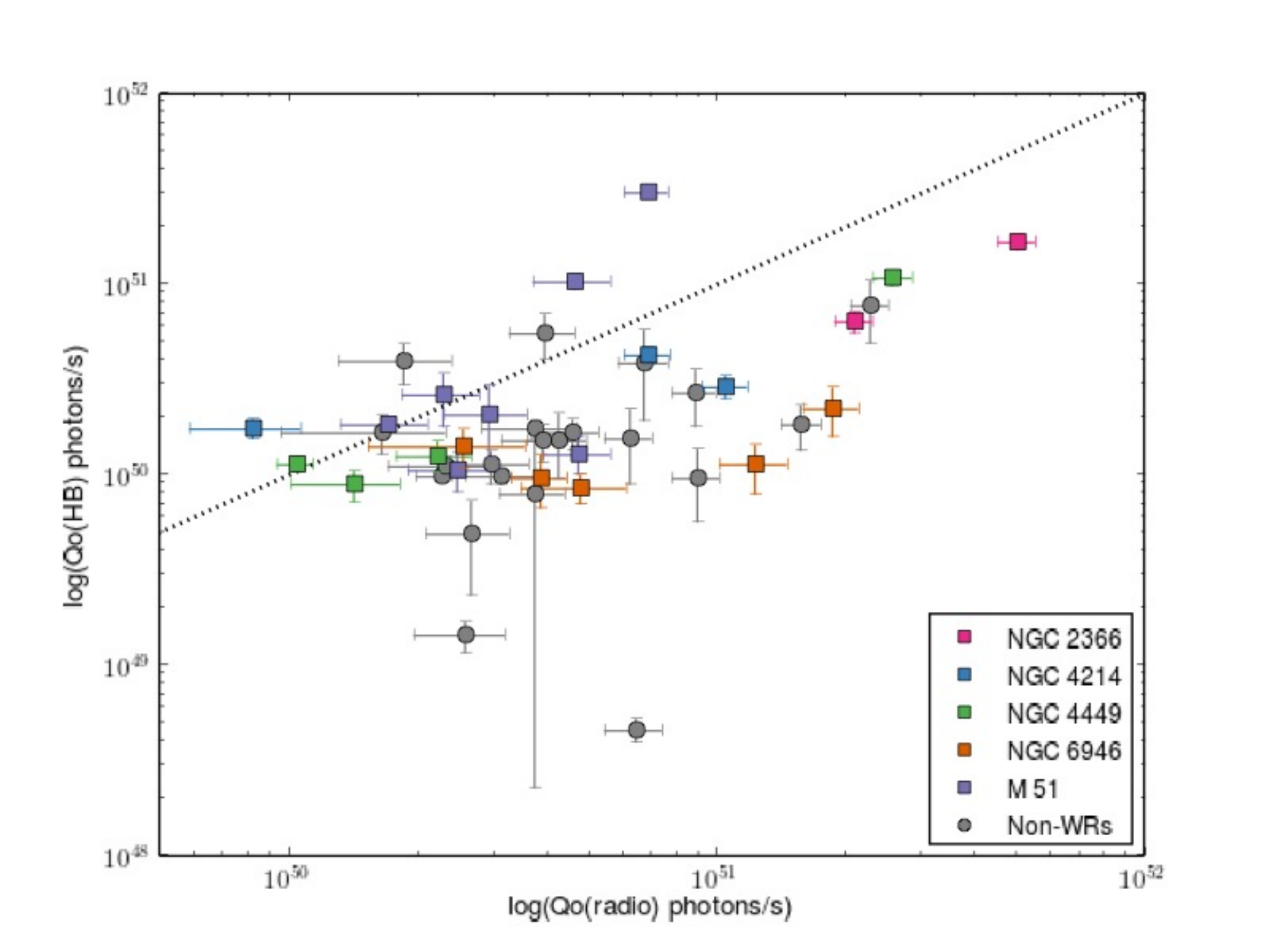}

\caption{\label{fig-ionizingflux} A plot comparing the ionized flux inferred from thermal radio emission to that inferred from optical nebular H$\beta$ emission. The colors are the same as Figure \ref{fig-radiosample}. Most sources show a higher ionizing flux inferred from the radio than from the optical observations, suggestive that the sources are still partially embedded. }
\end{figure}

It is evident that the  ionizing flux values estimated from the radio emission are much higher than from the optical. Plotting the radio inferred ionizing flux versus the optically inferred ionizing flux on log-log scale in Figure \ref{fig-ionizingflux}, the majority of the sample lies below a 1:1 ratio. While the radio flux density at 6 cm may have some non-thermal contributions and thus imply a higher ionizing flux, many of the sources have such large radio inferred ionizing fluxes that major non-thermal contributions ($>$ 50\% or more) would be necessary to explain the observed trend. This observed trend suggests that most of these massive star clusters are still partially embedded, as more ionizing flux is inferred from the radio, and thus are not fully emerged. 

\subsection{The Excitation of the Sample}
 
\begin{figure}[!t]
\includegraphics[width=0.7\textwidth,angle=0]{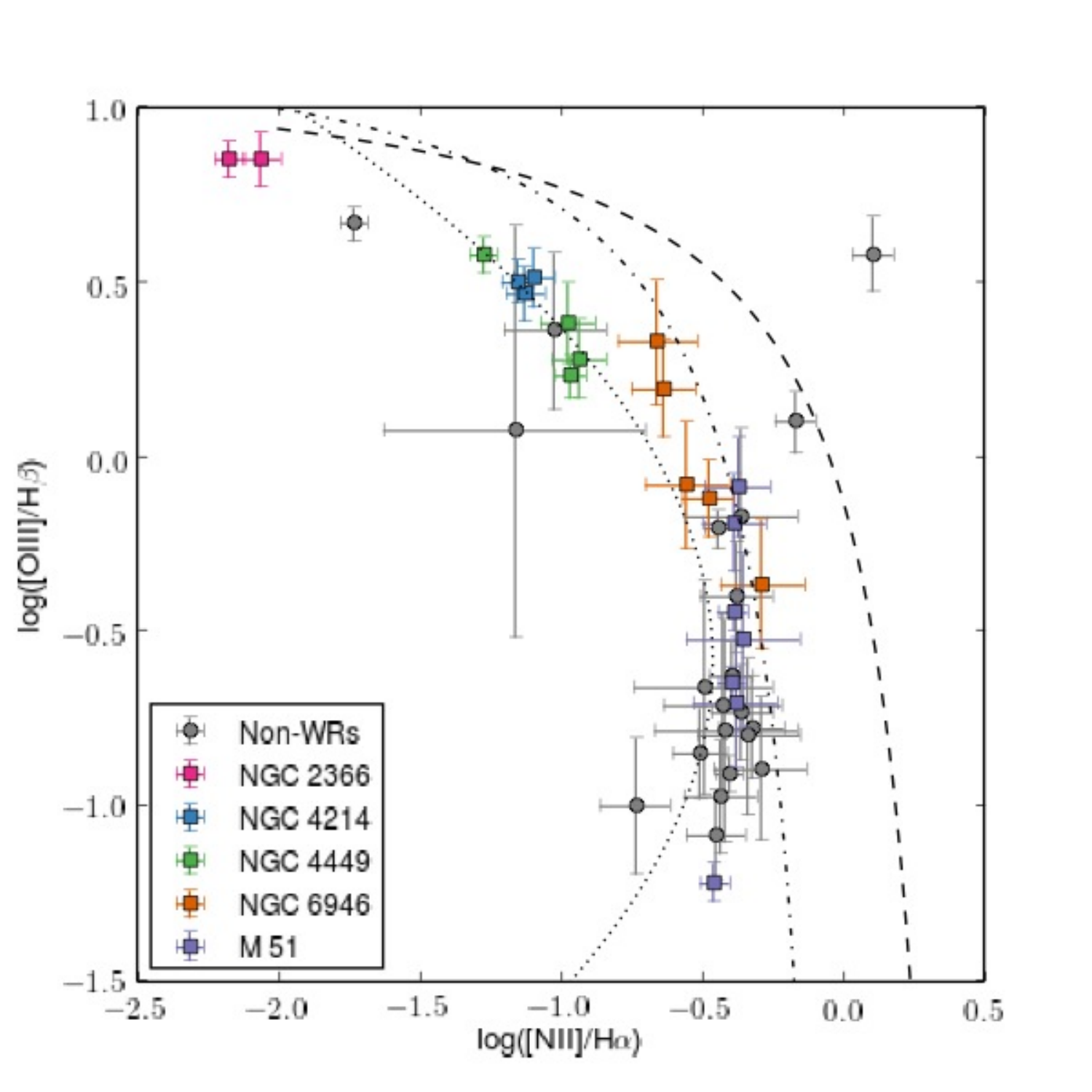}
\caption{\label{fig-bpt} The BPT diagram \citep{bpt} used to evaluate the excitation mechanisms. We see that the full sample spans across the diagram. Generally, the sample  is also above where average star-forming galaxies lie shown by the dotted line \citep{bpc08} and borders the theoretical and empirical limits that are produced  by star formation along that are shown as dashed and dot-dash lines \citep{kew01, kauf03}.}
\end{figure}

To gauge how extreme the star formation environments of the emerging massive star clusters may be, we utilize the Baldwin, Philips, \& Terlevich (BPT) diagram as an optical diagnostic \citep{bpt}. As shown in Figure \ref{fig-bpt}, this is typically a plot of the ratio of the nebular lines of [N \textrm{II}]/H$\alpha$ versus [O \textrm{III}]/H$\beta$ and is used to evaluate the mechanisms exciting the nebular emission. This is useful to identify if nebular lines are being excited by star formation alone; the theoretical and empirical limits \citep{kew01,kauf03} are overplotted as dashed and dot-dash lines.  Incontestably, most star-forming galaxies lie well below this star formation limit, as shown by the dotted line that shows the average of Sloan Digital Sky Survey galaxies \citep{bpc08}. If sources lie above or to the right of these limits, there must be some other contribution providing excitation, such as from shocks or an AGN. Therefore, the BPT diagram is most commonly used to distinguish between star-forming galaxies and active galactic nuclei (AGN), although it also can indicate the strength of the ionization parameter U and the excitation parameter q for a given metallicity.
 
The emerging massive star clusters in our sample span the BPT diagram, reflective of the range of metallicities. There does not appear to be any obvious differences in the location of sources with and without WR stars. Most sources lie above the average star-forming galaxy track and appear to border the star formation limit. The direction of this displacement indicates that the emerging massive star clusters have higher ionization and/or excitation parameters than typical H \textrm{II} regions, especially our sources in the spiral galaxies M 51 and NGC 6946 with higher metallicity (right side of the plot).
 
The sources on the left side of this plot are also interesting. Low values of [N \textrm{II}]/H$\alpha$ ($<$ -1.0) host low metallicity sources, which often severe as analogs for understanding star formation at high redshift. Our sources  in NGC 4214 and NGC 4449 roughly lie with the average star-forming galaxies in the BPT diagram at these metallicities. This is also the same area as Green Pea galaxies, which are local extreme galaxies known for high ionization parameters and high [O \textrm{III}]/[O \textrm{II}] ratios \citep{jo13},  conditions similar to high redshift star-forming galaxies that may be responsible for reionization \citep{no14}.  Furthermore, most published BPT diagrams do not extend to ratios $<$ -1.50 and the models shown may not be valid here, yet all of our sources in NGC 2366 fall in this category, which has not yet been well characterized and likely very similar to high redshift objects. Thus, these sources at low metallicity may represent modes of star formation similar to that in the early Universe.

\subsection{\label{section-nonwrs} A Different Population: Non-WR Clusters}

\begin{figure}[!t]
\includegraphics[width=0.95\textwidth,angle=0]{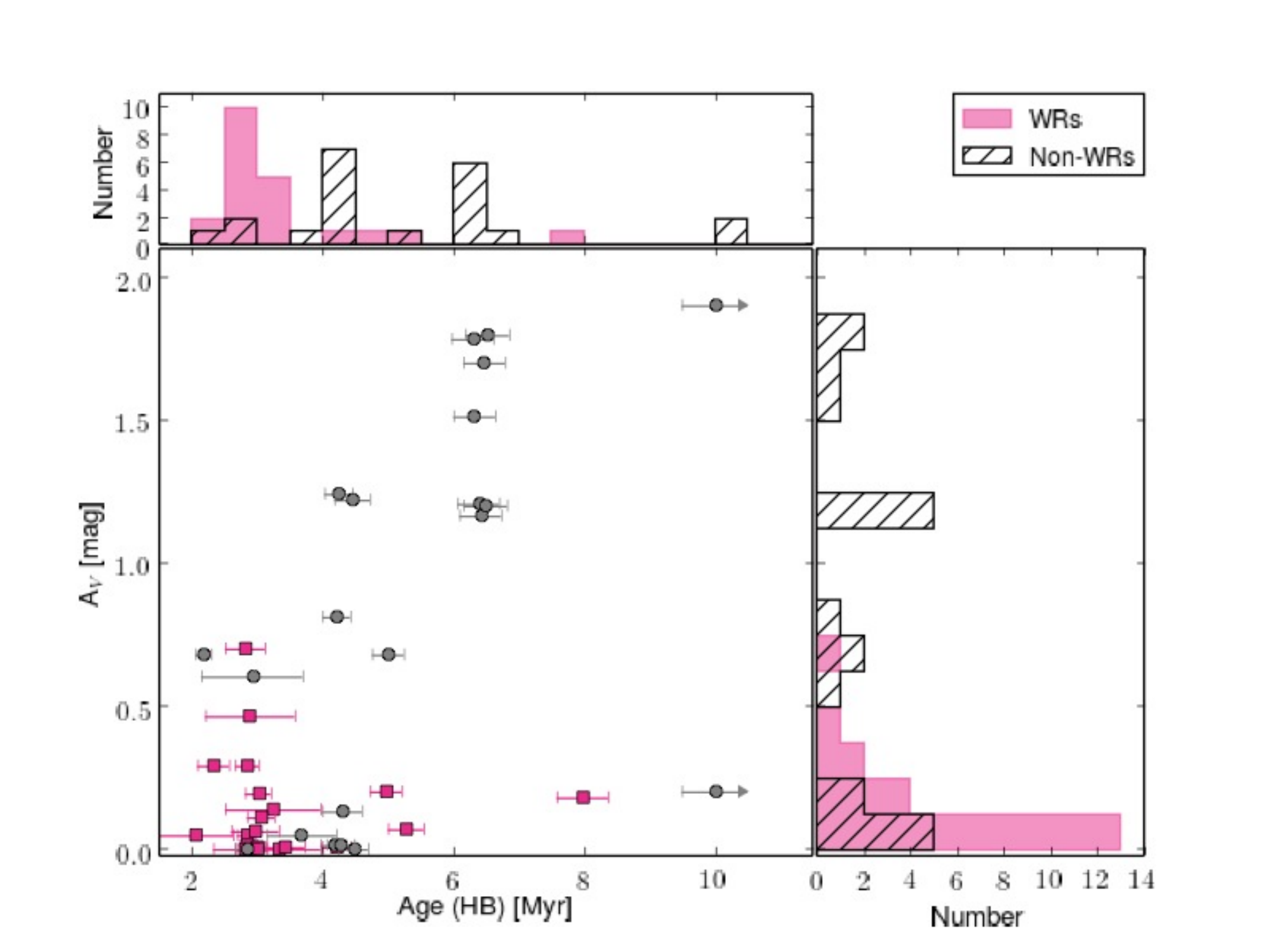}
\caption{\label{fig-age_extinction} A look into the ages and extinctions of the sources in the sample. The top/right panel shows a histogram of ages/extinctions and the distribution of these properties are markedly different between the two classes of emerging massive star clusters (WR clusters in pink, non-WR clusters in grey). The main figure shows these properties in comparison as extinction versus age. We found that the most highly extincted sources do not have detected WR features and tend to have larger ages. }
\end{figure}

\begin{figure}[!t]
\includegraphics[width=0.9\textwidth,angle=0]{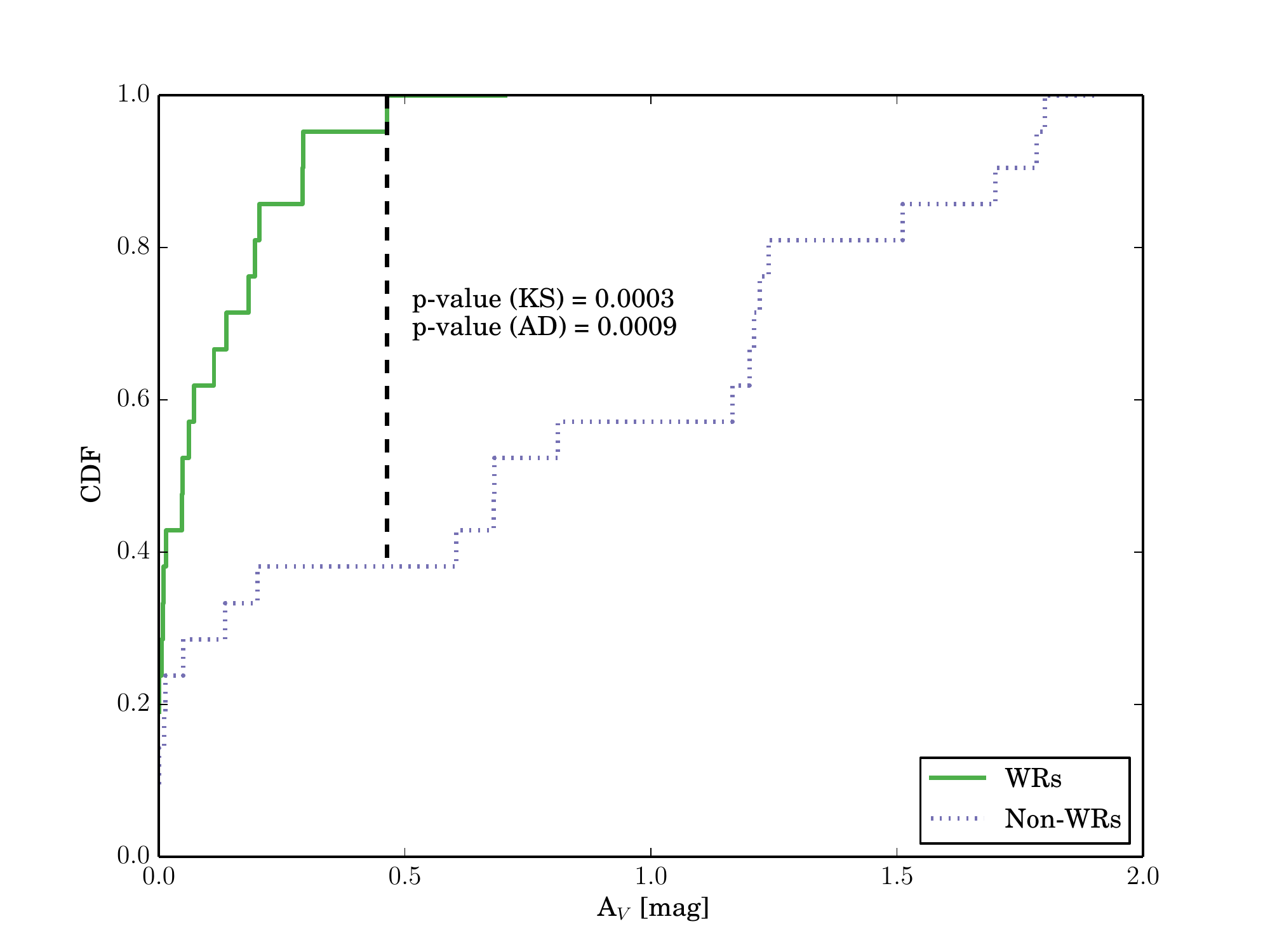}
\caption{\label{fig-ks} A plot showing the cumulative distribution functions (CDF) of the extinctions of the WR and Non-WR clusters. We performed the Kolmogorov-Smirnov (KS) test and the Anderson-Darling (AD) test on the distributions of the ages and extinctions. For the extinction distributions, we found p-values of 0.0003 (KS test) and 0.0009 (AD test), showing that the extinction distributions of the two classes are statistically different. The age distributions are similarly statistically different.}
\end{figure}

We have presented many similarities between the WR clusters and the Non-WR clusters thus far, other than the fact that only the WR clusters exhibit the WR bump. However, we have found that the Non-WR clusters are also distinct in their extinctions and ages. We plot the age versus extinction in Figure  \ref{fig-age_extinction}.  As the upper and side panels show, the distributions of both of these properties are markedly different between these classes. Moreover, when both properties are examined (the main plot), the sources with the highest extinctions are evidently older and do not contain detectable WR features. To further examine this trend, we performed the Kolmogorov-Smirnov (KS) test and the Anderson-Darling (AD) test on the distribution of the extinctions (see Figure \ref{fig-ks}).

The KS and AD tests can be used to find confidence limits, where an output p-value less than 0.05 would reject the null hypothesis that the samples come from the same distribution. We find the extinction of the sources in the two classes (WR and Non-WR clusters) are statistically different with a p-value of 0.0003 from the KS test and a p-value of 0.0009 from the AD test, indicating that do not come from the same underlying extinction distribution.  Similarly, the ages of the WR and Non-WR clusters do not come from the same underlying distribution, with p-values of 0.001 (KS test) and 0.006 (AD test). This plot shown in Figure \ref{fig-ks} indicates that there is a population of Non-WR clusters that is quite different from the rest of the emerging massive star clusters. This anomalous  population of Non-WR clusters tend to be older, extincted, and do not show signs of WR stars.

Although the KS and AD tests indicate different populations, the massive star clusters in our sample are very messy environments that make proving any evolutionary scenario rather difficult.The subset of Non-WR clusters presented here that are 1) old, 2) have high extinction, and 3) do not currently have WR stars may form the most curious result from this project. Most of these anomalous Non-WR clusters have stellar masses that should be able to produce WR stars, especially as many are at high metallicities. Given the mass and metallicity of the clusters, it is challenging to understand why {\em all} of these Non-WR clusters exhibit these three characteristics and appear distinct from the rest of the sample (as in Figure \ref{fig-age_extinction}). If these Non-WR clusters are in fact old, highly-extincted, WR-free massive star clusters, then what could they be? The most obvious explanation may be that they did host WR stars at an earlier time and they no longer exist. However, in this case we would expect that these previous WR stars must have exploded as supernova, which is why they are no longer producing WR bumps. Supernova remnants produce radio emission with a negative spectral index that can dominate over the thermal radio emission.  Supernova remnants can inject high energy particles that will give rise to synchrotron emission for up to 1 Gyr \citep{hee15}, and thus the thermal radio-selection of our sources makes this scenario unlikely. Additionally, the clusters would have needed to be in unusually dense environments to remain partially embedded if supernova explosions had occurred. Thus, we may not consider this scenario viable.

However, the stellar death throes of massive stars are not yet a solved problem--so perhaps the endpoints of massive stellar evolution are the answer here. While the common assumption is that they will explode as supernova, there is a possibility that massive stars instead may directly collapse into a black hole. \citet{fryer99} found that stars with masses greater than 40 $M_\sun$ may form a black hole without a supernova explosion (although they may undergo gamma ray bursts if rotating). An archival search for supernova progenitors found none came from massive stars above 18 M$_\sun$, which can explained by either the WR phase that is too faint just before the point of core-collapse or that massive stars produce failed supernova and black holes instead \citep{smartt15}. There has additionally been an identification of one possible progenitor of a failed supernova as the disappearance of a 25-30 M$_\sun$ yellow supergiant \citep{rey15}.  Thus the subsample of old, highly-extincted Non-WR clusters can be explained if most stars that are massive enough to become WR stars do not explode as supernova, but instead directly and quietly collapse into black holes. In the case that these clusters had WR stars but they have all died, there would be no indications of either WR stars or supernova in our subsample of Non-WR clusters. One caveat to this scenario is the age: while these Non-WR clusters are past the onset of the WR phase, the majority are less than 8 Myr old and thus not all of the massive stars should have perished and the stars from initial masses of $\sim$ 25 M$_\sun$ could be in the WR phase \citep{es09}. To offer other possible explanations, we discuss here how any of the three derived characteristics (age, extinction, WR detection) could be inaccurate, and the implications (of course combinations are possible as well):

\begin{enumerate} 
\item  Could the anomalous Non-WR clusters be {\em young} instead?: We estimated the cluster ages using  the observed equivalent width of H$\beta$ and STARBURST99 predictions. In addition to uncertainties in the method, there are a few potential scenarios that may result in an overestimation of ages. 
\begin{enumerate}
\item If ionizing photons were to leak out of the H \textrm{II} region, the cluster could appear older than it actually is. However, the equivalent width would need to be reduced from roughly 100-50 \AA\ to the observed $\sim$10 \AA\ in many cases. It is difficult to justify why only the high extinction sources would have leaked such large amounts of their ionizing photons.
\item Similarly, the right distribution of dust in or surrounding the clusters may be responsible for low equivalent width measurements and the resulting derived older ages. If there is a clumpy screen of dust, then the stellar continuum can suffer less extinction due to dust than nebular emission lines \citep{cal94}, and thus this extinction may reduce the equivalent width of the Balmer lines. However, in the scenario described by \citet{cal94}, the optical continuum is produced by an older stellar population than the emission lines. Thus for this scenario to be considered a feasible explanation, there would need to be both a dust shield and distinct spatial distributions between the WRs (and young ionizing stars) versus the older stars producing the observed continuum without the WR bump. 

\item Another possibility is that the Non-WR clusters are not producing massive enough stars to produce the equivalent widths of H$\beta$ that are expected for young massive star clusters, and thus appearing older than they are. This can occur through several mechanisms. Stochasticity may be effecting these massive star clusters, which becomes important for clusters with masses $\sim10^5$ M$_\sun$ and less \citep{fl10}. Thus, the cluster may not produce stars with stellar masses that fully sample the upper end of the IMF. Additionally, an irregular bottom heavy IMF may be able to justify how a population of stars can be formed without massive stars that become WRs. If these sources were made up of several star clusters, rather than a single massive star cluster that should fully sample the IMF, then the combined stellar populations may only be composed of stars with low mass. At extragalactic distances, the possibility of several densely-packed low mass clusters is hard to rule out.

\end{enumerate}

\item Could the anomalous  Non-WR clusters have low internal {\em extinction} instead?: This could be explained by foreground extinction in the host galaxy that results in measurements of high extinction for these sources. As many of these Non-WR clusters lie in the spiral arms of M 51, foreground extinction from the host galaxy likely does contribute on some level. However, if the measured high values are solely due to foreground extinction from the host galaxy rather than internal extinction in the clusters themselves, then why does this foreground extinction only effect the old clusters without WR detections requires a resolution. 
\item  Could the anomalous Non-WR clusters actually have {\em WR stars} that were not detected by our observations?:
While the Non-WR clusters have high extinctions, and it might seem appealing to assume that these extinctions are diminishing the WR signal, this cannot explain non-detections of the WR bump. Since the WR bump is due to stellar features, the WR bump and the continuum should be affected by extinction in the same way. The non-detections could result instead from various scenarios when additional light from other massive stars washes out the  WR bump, such as the following possibilities: 
\begin{enumerate}
\item One scenario is if the WR stars are weak-lined WR stars. The weakest WR stars can have an $\lambda$ 4686\AA\ equivalent width a factor of one hundred less than other WR stars \citep{cro07}. However, weaker lined WR stars are more prevalent at lower metallicities \citep{cro06}, and as the majority of our Non-WR clusters are found in high metallicity environments, this explanation is unlikely. 
\item Another possibility is aperture-like effects; \citet{kpv13} showed that aperture size can effect a potential WR detection, where the WR signal can be unknowingly wiped out depending on the distribution of WR stars and aperture choice. In our observations, the WR signal could be diluted by including additional light from beyond the cluster. Yet, our sources in each galaxy were treated uniformly, and thus this effect should be present in the entire sample and therefore does not reasonably explain why the Non-WR clusters are different.
\item If the star formation were not coeval, then a population of newly formed O-stars could wash out the WR bump produced by still present WR stars formed in an earlier wave of star formation. Yet, the ages of the Non-WR clusters, which are estimated from nebular observations that reflect the most recent star formation, are typically past the onset of the WR phase instead of before. Thus the estimated ages of our sample are not in line with this scenario.
\item If the IMF of the cluster differed from the typical Salpeter IMF, there could be more O-stars produced relative to WR stars than expected (or less massive stars in total). If this were the case, then the detection of the WR bump would be more difficult because the light from additional O-stars may dilute the WR feature. For instance, a slightly steeper IMF was observed in M31 that could suggest that the number of massive stars ($>8$ M$_\sun$) would be 25\% less than expected with a Kroupa IMF \citep{wei15}. However, little variation has been observed in the IMF universally \citep{bcm10,mas11}, making this explanation difficult to justify. 
\item As shown by the STARBURST99 predictions shown in Figure \ref{fig-sb99-ew}, the strength of the integrated WR features can vary over time. Thus, some of the sources without WR detections could have been observed during one of the low points (appearing ``off'') of the duty cycle of the WR features. However, there is no reason that this should correlate with age or extinction, as in Figure \ref{fig-age_extinction}.

\end{enumerate}

\end{enumerate}

\section{\label{section-conclusions} Conclusions and Discussion}

To investigate the potential evolutionary role of WR stars in clearing surrounding natal material from massive star clusters, we have obtained optical spectra of 42 emerging massive star clusters to search for WR detections. Targets were identified from sources that exhibit thermal radio emission in radio continuum studies of the star-forming galaxies NGC 2366, NGC 4214, NGC 4449, NGC 6946, and M 51. The observed properties of the sources in the sample indicate that these massive star clusters are intense regions of star-formation that have not yet completely cleared their natal material. Most sources in the sample, with or without WR stars, exhibit less ionizing flux inferred from the optical emission lines than would be inferred from the thermal radio emission; this suggests that some light is being blocked and thus that these sources are still partially embedded. We also find that the observed nebular line ratios tend to border what can be produced by star formation alone, shown by the BPT diagram in Figure \ref{fig-bpt}, suggesting the presence of a hard radiation field.

We found that 50\% of the sample exhibited significant detections of the WR bump, and have constrained their nebular environments. The observed ages and high number of WR detections show that the thermal radio emission does not dissipate before WR stars start to appear. Thus, we have not observed a period during which there is thermal radio emission concurrent with optical emission before the WR phase for the cluster begins, as may have been expected.
 
Overall, we find that our observations of the emerging WR clusters are consistent with the hypothesis that WR stars may be contributing to the removal of natal material during the cluster emergence. In particular, we find the interstellar extinction in the emerging WR clusters is  lower than most of the non-WR clusters. Moreover, the observed differences between the ages and extinctions of the WR cluster and the Non-WR cluster classes have important evolutionary implications. Naively, if it is assumed that WR stars are partly responsible for clearing a cluster, then clusters without WR stars should have higher extinctions, which is exactly what we see. Comparing both the ages and extinctions of the sources, we found that some emerging massive star clusters appear to remain embedded for longer, and that these clusters do not show detections of the WR bump;  many Non-WR clusters with high extinction are $\sim$ 5 Myr older than most WR clusters  (see Figure \ref{fig-age_extinction}). Thus if sources without WR stars stay embedded longer than sources with WR stars,  the WR stars could actually be helping these clusters to emerge. This may suggest that WR stars make the emerging process more efficient or accelerated, and possibly necessary in some cases, for clusters to clear obscuring natal material.

While we do offer alternative scenarios to the observed ages, extinctions, or WR detections and it is clear that many unknowns may hamper our interpretation of this data, our observations of relatively cleared out WR clusters and extincted, old clusters that do not exhibit WR features are quite compelling and indicative of the importance of the WR stars. 
Thus, further optical spectral observations of additional massive star clusters exhibiting thermal radio emission are needed. An expanded sample will greatly expand our understanding of emerging massive star clusters and super star clusters and provide further constraints to some of the suggested scenarios. We suspect that our radio selection process is an important component in this study, as it identifies clusters that still have gaseous material but are not subject to extinction. This method is different than typical observed collections of extragalactic H \textrm{II} regions, which are mostly found through optical brightness criteria, and may explain why sources like the Non-WR clusters have not been previously scrutinized. While  radio continuum surveys to obtain high enough sensitivities to detect these extragalactic H \textrm{II} regions are expensive, the new capabilities of the Karl G. Jansky Very Large Array have increased  continuum sensitivities by more than an order of magnitude. New radio continuum studies are already coming out, such as the Star Formation in Radio Survey that has resolved thermal radio sources \citep{murphy12}, and show there is hope for  this field to rapidly advance.

To truly confirm the role of the WR stars in how massive star clusters emerge,  complex simulations are needed that not only model massive star clusters but also specifically include the WR phase and incorporate different feedback mechanisms. Current technology cannot reach these massive and super star cluster mass scales with the resolutions needed that also include feedback processes, and it is not reliable to simply scale from lower mass systems \citep{bk15}. Fortunately,  we are already witnessing advancements in capabilities. For instance, super star cluster mass scales have been reached  by recent simulation of radiation feedback on a nascent super star cluster \citep{so15}. Thus, with increased radio telescope sensitivities and the continued improvements towards complex simulations, we will soon be able to  disentangle the physical effects of WR stars in massive star cluster evolution.

\acknowledgments

We thank the anonymous referee for improving the quality of this work. This research is supported by NSF grant 1413231 (PI: K. Johnson). K.R.S. gratefully acknowledges support provided  by Sigma Xi Grants-In-Aid of Research and observing support from NOAO for her Ph.D thesis. K.E.J. additionally acknowledges support provided by the David and Lucile Packard Foundation. Special thanks to Sean Moran for running the Hectospec pipeline.

Observations at the 6.5m MMT at Fred Lawrence Whipple Observatory are supported by the F. H. Levinson Fund of the Silicon Valley Community Foundation. Based on observations at Kitt Peak National Observatory, National Optical Astronomy Observatory (NOAO Prop. ID: 2013A-0317; PI: K. Sokal), which is operated by the Association of Universities for Research in Astronomy (AURA) under cooperative agreement with the National Science Foundation. This paper uses data products produced by the OIR Telescope Data Center, supported by the Smithsonian Astrophysical Observatory. This research has utilized NASA's Astrophysics Data System Service. 

\bibliography{ms_ref_preprint}
\label{references}

\clearpage

\input{appendix_preprint}

\end{document}

%% file: appendix_preprint.tex
\appendix

\setcounter{table}{0}
\renewcommand{\thetable}{A\arabic{table}}

\setcounter{figure}{0}
\renewcommand{\thefigure}{A\arabic{figure}}

\section{Appendix: Source Spectra and Images }

\begin{figure*}[h!]
\includegraphics*[width=9cm,angle=0]{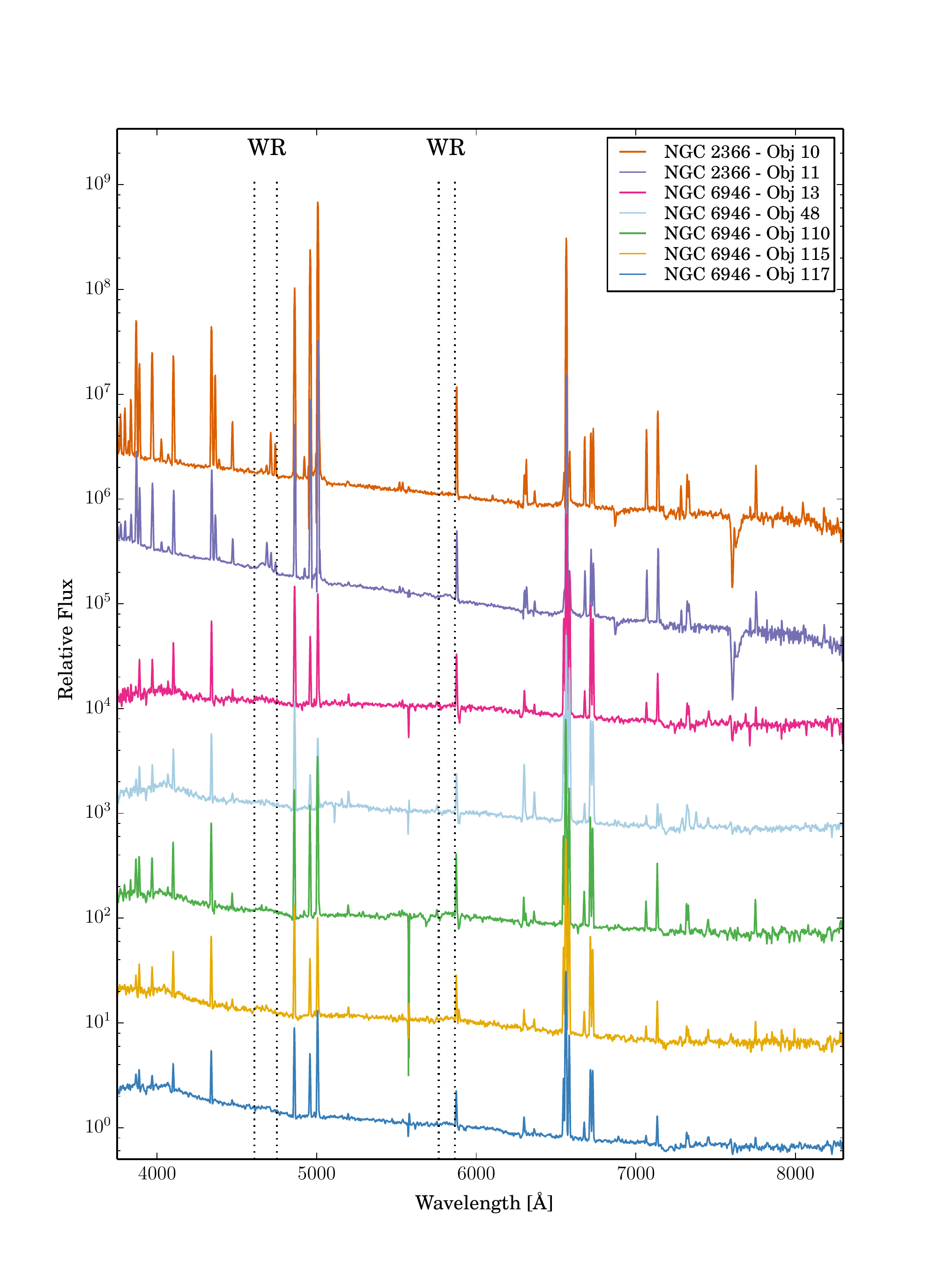}
\hspace{-25pt}
\includegraphics*[width=9cm,angle=0]{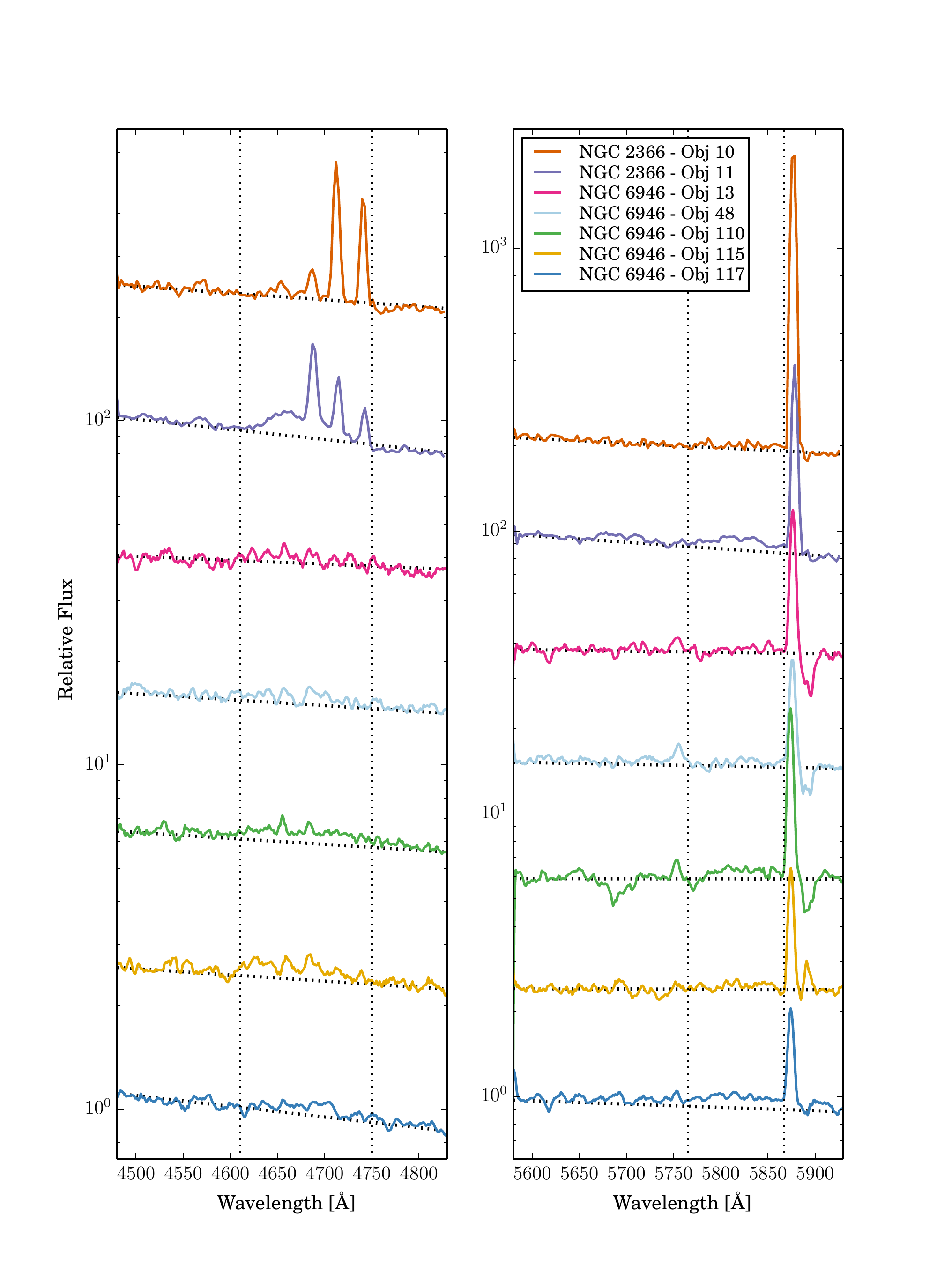}
\caption{\label{fig-fullspec2} Optical spectra observed with the 4m Mayall Telescope at KPNO and the 6.5m MMT of another subset of the WR clusters, otherwise the same as Figure \ref{fig-fullspec1}.}
\end{figure*}

\begin{figure*}[h!]
\includegraphics*[width=9cm,angle=0]{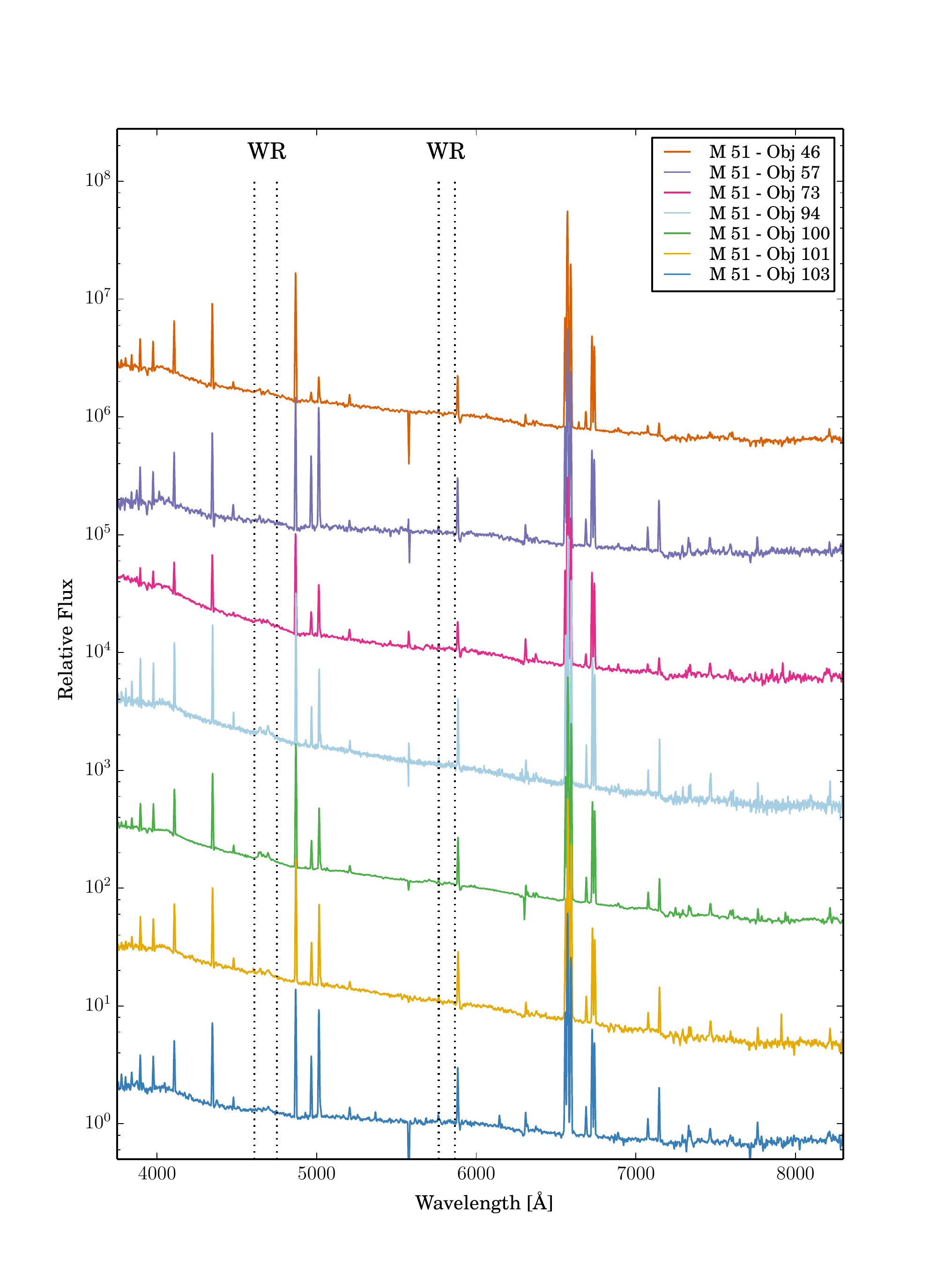}
\hspace{-25pt}
\includegraphics*[width=9cm,angle=0]{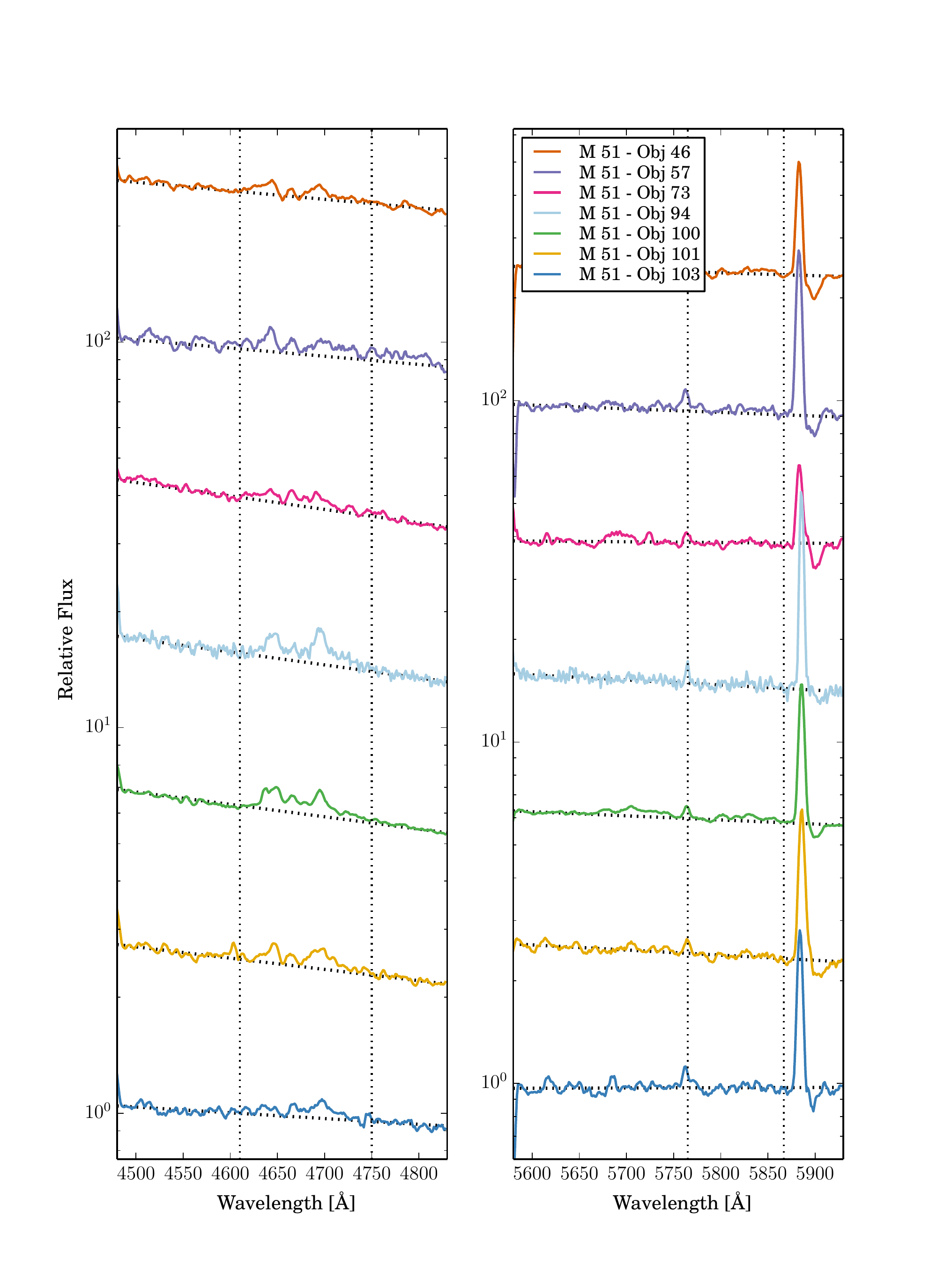}
\caption{\label{fig-fullspec3} Optical spectra observed with the 6.5m MMT of the rest of the WR clusters, otherwise the same as Figure \ref{fig-fullspec1}.}
\end{figure*}

\begin{figure*}[h!]
\includegraphics*[width=9cm,angle=0]{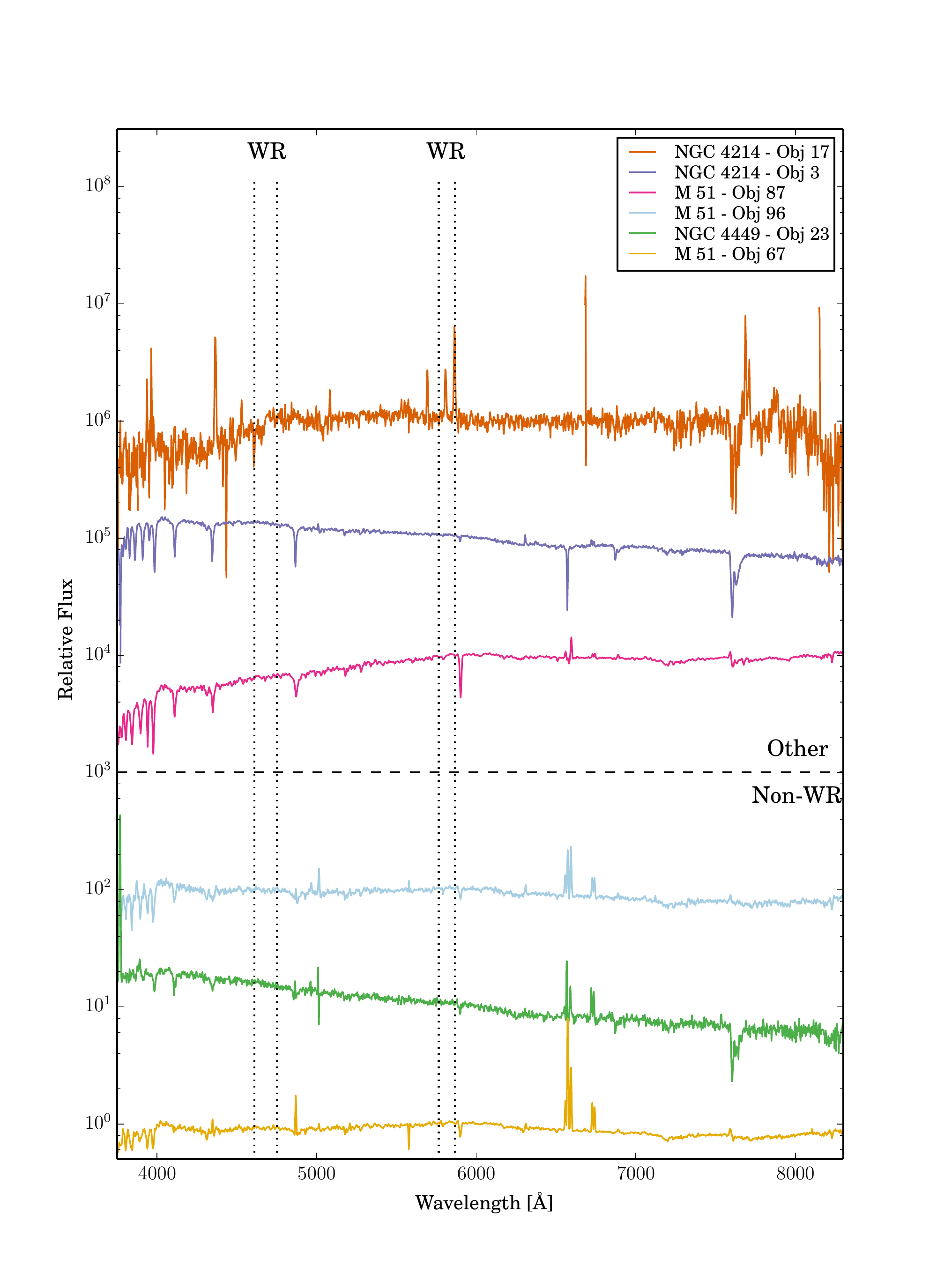}
\hspace{-25pt}
\includegraphics*[width=9cm,angle=0]{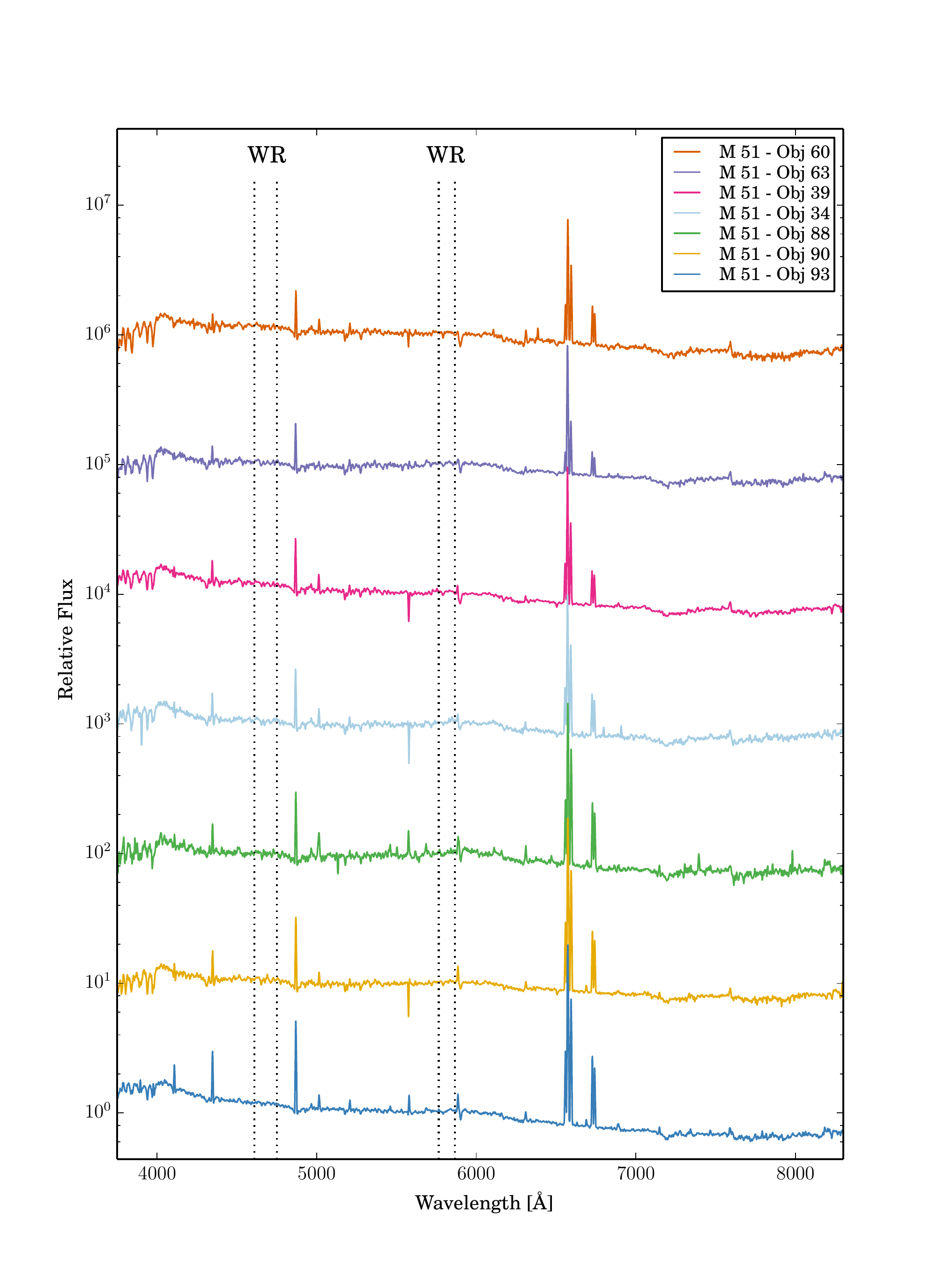}
\caption{\label{fig-nonwrspec} Optical spectra observed with the 4m Mayall Telescope at KPNO and the 6.5m MMT of a subset of the `no-bump' sources, otherwise the same as Figure \ref{fig-fullspec1} without the zoom-in panels of WR feature regions.}
\end{figure*}

\begin{figure*}[h!]
\includegraphics*[width=9cm,angle=0]{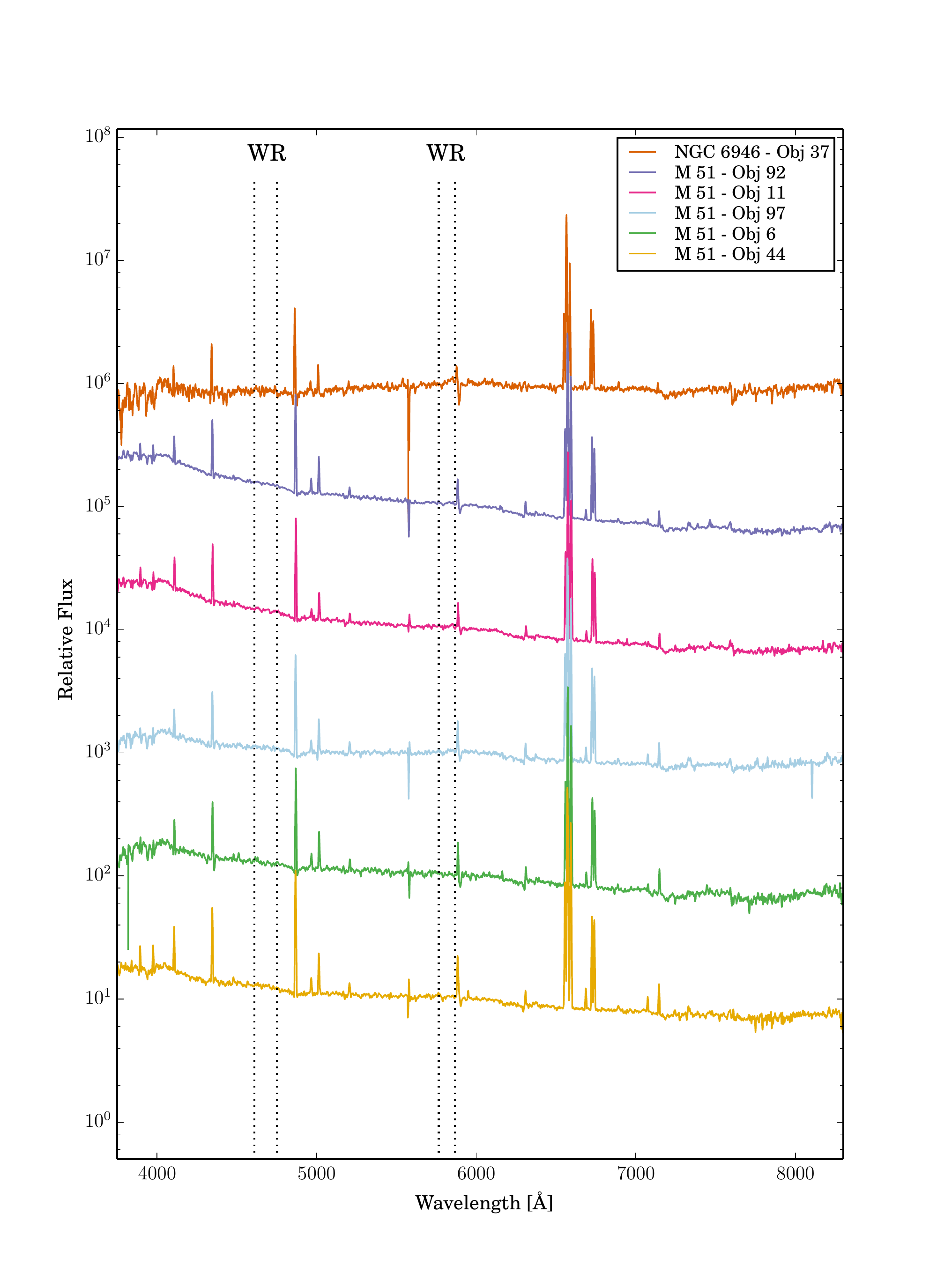}
\hspace{-25pt}
\includegraphics*[width=9cm,angle=0]{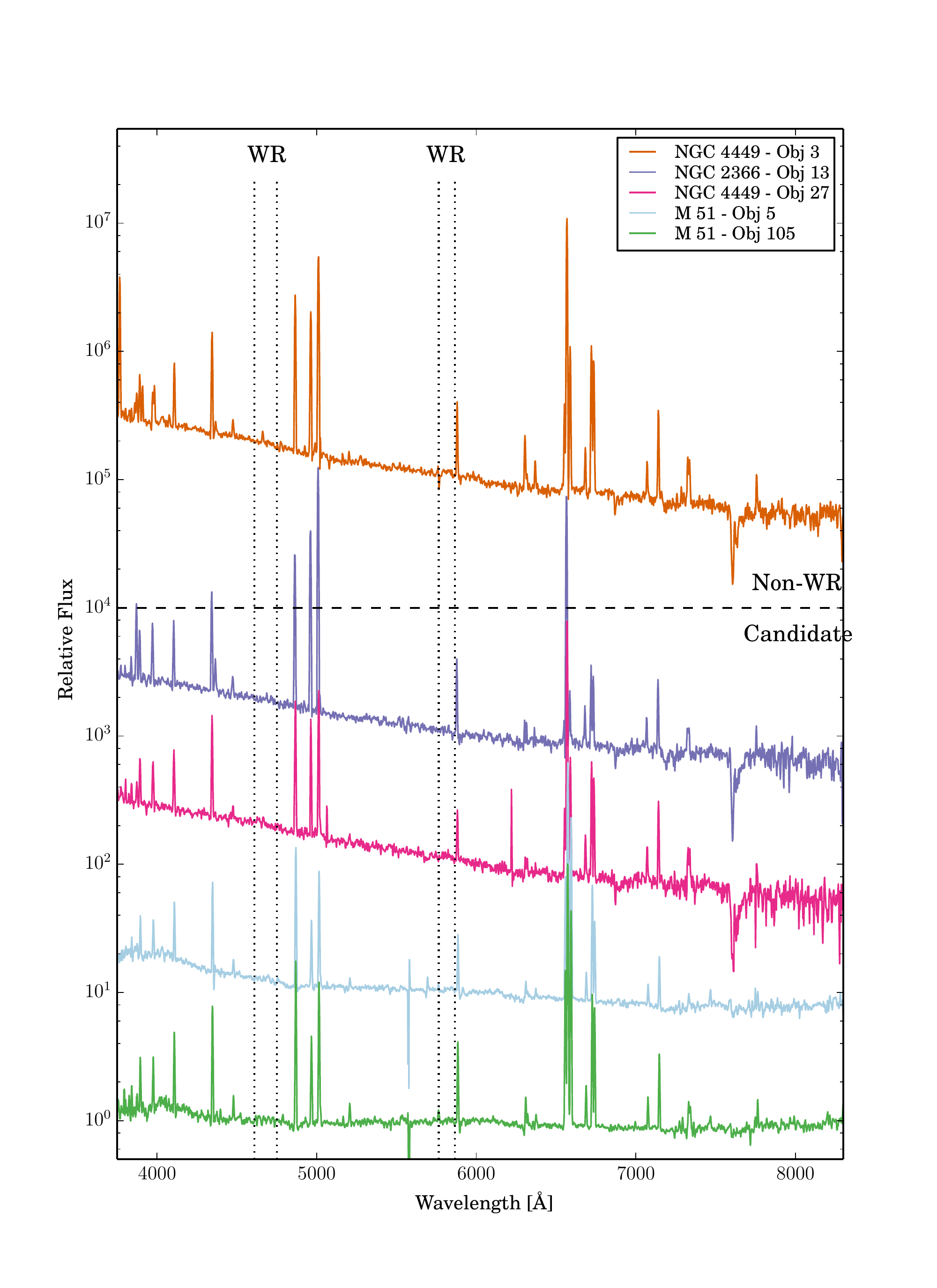}
\caption{\label{fig-candidatespec} Optical spectra observed with the 6.5m MMT of the rest of the `no-bump' sources, otherwise the same as Figure \ref{fig-nonwrspec}.}
\end{figure*}

\begin{figure}[h!]
\includegraphics[width=\textwidth,angle=0]{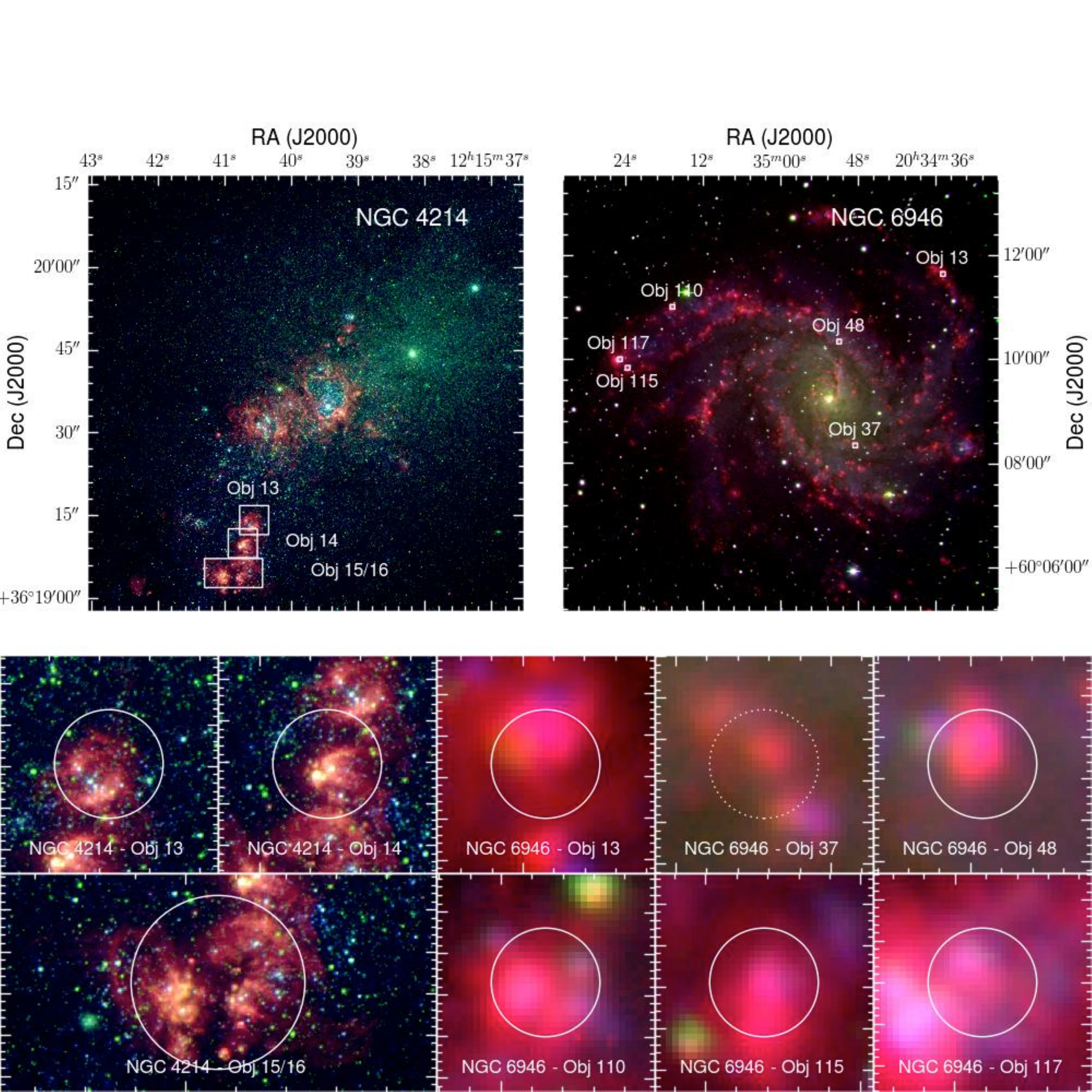}
\caption{\label{fig-rgb_panel2} Archival HST and KP 2.1m images (H$\alpha$, B, I) of the target galaxies NGC 4214 and NGC 6946. Insets and overlaid regions are the same as \ref{fig-rgb_panel1} with the exception that the extracted region and corresponding overlaid circle on source  NGC 4214 - Object 15/16 is 4.0''.}
\end{figure}

\begin{figure}[h!]
\includegraphics[width=\textwidth,angle=0]{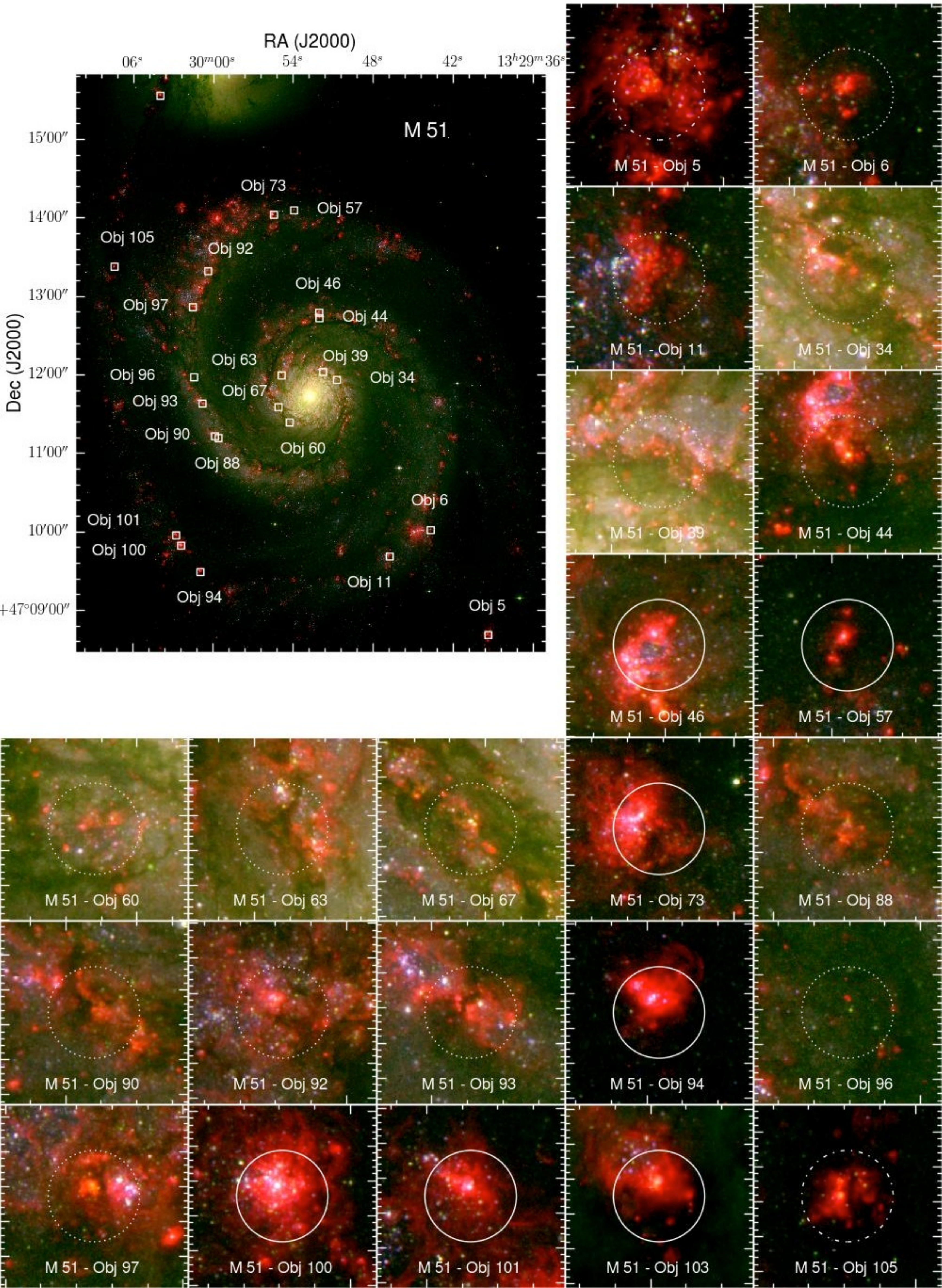}
\caption{\label{fig-rgb_panel3} Archival HST images (H$\alpha$, B, I) of the target galaxy M51. Insets and overlaid regions are the same as Figure \ref{fig-rgb_panel1}. }
\end{figure}

We present the optical spectra obtained for this study of the sources in our sample and archival images of these interesting objects. In addition to the spectra shown in Figure \ref{fig-fullspec1}, additional spectra for the WR class are shown throughout Figures \ref{fig-fullspec2}-\ref{fig-fullspec3}, and the full sample throughout Figures \ref{fig-fullspec2}-\ref{fig-candidatespec}. In the same fashion, an example image showing a subset of these regions was presented in Figure \ref{fig-rgb_panel1};the rest of the sample is presented here in Figures \ref{fig-rgb_panel2}-\ref{fig-rgb_panel3}. We also include tables describing the observations and various measured properties of the sources, such as line fluxes.

\begin{deluxetable}{llllll}
\tabletypesize{\scriptsize}
\tablewidth{0pc}
\tablecaption{\label{table-obs} Optical Spectral Observations} 
\tablehead{
\colhead{Source}      &
\colhead{Right Ascension $\alpha$ }  &
\colhead{Declination $\delta$ }   &  
\colhead{Exposure Time [s]}      &
\colhead{Date Observed\tablenotemark{a}}      
}
\startdata
NGC 2366 - Object 10	&	7:28:42.58	&	+69:11:22.0 	& 	2400 	&	2013-04-12	\\
NGC 2366 - Object 11	&	7:28:43.71	&	+69:11:22.4	&	1800 			&	2013-04-15\tablenotemark{b} \\
NGC 2366 - Object 13	&	7:28:45.69	&	+69:11:25.8	&	1800 			&	2013-04-13	\\
	&			&			&	1800			&	2013-04-15	\\
NGC 4214 - Object 3	&	12:15:38.18	&	+36:19:44.9	&	1200 		&	2013-04-12	\\
	&			&			&	900		&	2013-04-14	\\
	&			&			&	900		&	2013-04-15	\\
NGC 4214 - Object 13	&	12:15:40.56	&	+36:19:14.1	&	2100 	&	2013-04-12\tablenotemark{c}	\\
	&			&			&	900		&	2013-04-14	\\
	&			&			&	1800 		&	2013-04-15\tablenotemark{d}	\\	
NGC 4214 - Object 14	&	12:15:40.73	&	+36:19:09.9	& 1800		&	2013-04-14	\\
	&			&			&	900		& 	2013-04-15	\\
NGC 4214 - Object 15/16\tablenotemark{e} 	&	12:15:40.87\tablenotemark{e}	&	+36:19:04.4\tablenotemark{e}	&	1800		&	2013-04-12	\\
	&			&			&	2700 		&	2013-04-14	\\
	&			&			&	600		&	2013-04-15	\\
NGC 4214 - Object 17	&	12:15:41.36	&	+36:21:14.1	&	2700		&	2013-04-14	\\
	&			&			&	1800 		&	2013-04-15	\\
NGC 4449 - Object 3	&	12:28:09.37	&	+44:05:20.2	&	1200	&	2013-04-13	\\
	&			&			&	2700 	&	2013-04-14	\\
	&			&			&	900	&	2013-04-15	\\
NGC 4449 - Object 4	&	12:28:09.44	&	+44:05:16.3	&	1800 	&	2013-04-13	\\
NGC 4449 - Object 18	&	12:28:12.63	&	+44:05:03.7	&	1800 	&	2013-04-13	\\
	&			&			&	1800	&	2013-04-14	\\
NGC 4449 - Object 22	&	12:28:12.99	&	+44:06:56.3	& 1800 	&	2013-04-13	\\
	&			&			&	1800 	&	2013-04-14	\\
NGC 4449 - Object 23	&	12:28:13.08	&	+44:05:42.7	&	1800 	&	2013-04-13	\\
NGC 4449 - Object 26	&	12:28:13.86	&	+44:07:10.4	&	1800 	&	2013-04-13	\\
NGC 4449 - Object 27	&	12:28:14.83	&	+44:07:10.0	&	1800 	&	2013-04-13	\\
	&			&			&	1800 	&	2013-04-15	\\
	
NGC 6946 - Object  13	&	20:34:34.85 	&	+60:11:38.6 	&	3600.	&	2013-04-08	\\
NGC 6946 - Object 37	&	20:34:48.43 	&	+60:08:20.8	&	3600.	&	2013-04-07	\\
NGC 6946 - Object 48	&	20:34:50.92	&	+60:10:20.7	&	3600.	&	2013-04-07	\\
NGC 6946 - Object 110	&	20:35:16.71	&	+60:11:00.6	&	3600.   &	2013-04-07	\\
	&			&			&	3600.	&	2013-04-08	\\
NGC 6946 - Object 115	&	20:35:23.62	&	+60:09:50.2	&	3600.	&	2013-04-08	\\
NGC 6946 - Object 117	&	20:35:24.82	&	+60:10:00.0	&	3600.	&	2013-04-07	\\
M 51 - Object 5	&	13:29:39.36	&	+47:08:40.7	&	6300.	&	2013-04-02	\\
	&			&			&	6300.	&	2013-04-06	\\
M 51 - Object 6	&	13:29:43.67	&	+47:10:01.0	&	6300.	&	2013-04-02	\\
	&			&			&	6300.	&	2013-04-06	\\	
M 51 - Object 11	&	13:29:46.74	&	+47:09:40.8	&	6300.	&	2013-04-02	\\
	&			&			&	12600. 	&	2013-04-06	\\
M 51 - Object 34	&	13:29:50.70	&	+47:11:55.9	&	6300.	&	2013-04-06	\\
M 51 - Object 39	&	13:29:51.73 	&	+47:12:01.9	&	6300.	&	2013-04-02	\\
	&			&			&	6300.	&	2013-04-06	\\
M 51 - Object 44	&	13:29:52.01 	&	+47:12:42.9	&	6300.	&	2013-04-02	\\
M 51 - Object 46	&	13:29:52.03	&	+47:12:47.2	&	6300.	&	2013-04-06	\\
M 51 - Object 57	&	13:29:53.91	&	+47:14:05.4	&	6300.	&	2013-04-06	\\
M 51 - Object 60	&	13:29:54.24	&	+47:11:23.2	&	6300.	&	2013-04-02	\\
M 51 - Object 63	&	13:29:54.84	&	+47:11:59.2	&	6300.	&	2013-04-02	\\
	&			&			&	6300.	&	2013-04-06	\\
M 51 - Object 67	&	13:29:55.08	&	+47:11:35.0	&	6300.	&	2013-04-06	\\
M 51 - Object 73	&	13:29:55.42	&	+47:14:02.1	&	6300.	&	2013-04-02	\\
	&			&			&	6300.	&	2013-04-06	\\
M 51 - Object 87	&	13:29:59.53	&	+47:15:58.3	&	6300.	&	2013-04-06	\\
M 51 - Object 88	&	13:29:59.58 	&	+47:11:11.3	&	6300.	&	2013-04-06	\\
M 51 - Object 90	&	13:29:59.84	&	+47:11:12.7	&	6300.	&	2013-04-02	\\
	&			&			&	6300.	&	2013-04-06	\\
M 51 - Object 92	&	13:30:00.36	&	+47:13:18.9	&	6300.	&	2013-04-02	\\
	&			&			&	12600. 	&	2013-04-06	\\	
M 51 - Object 93	&	13:30:00.78	&	+47:11:37.7	&	6300.	&	2013-04-02	\\
	&			&			&	6300.	&	2013-04-06	\\
M 51 - Object 94	&	13:30:00.93	&	+47:09:28.9	&	6300.	&	2013-04-02	\\
	&			&			&	6300.	&	2013-04-06	\\
M 51 - Object 96	&	13:30:01.41	&	+47:11:57.8	&	6300.	&	2013-04-06	\\
M 51 - Object 97	&	13:30:01.50	&	+47:12:51.4	&	6300.	&	2013-04-02	\\
	&			&			&	6300.	&	2013-04-06	\\
M 51 - Object 100	&	13:30:02.38	&	+47:09:49.1	&	6300.	&	2013-04-06	\\
M 51 - Object 101	&	13:30:02.75	&	+47:09:56.9	&	6300.	&	2013-04-02	\\
M 51 - Object 103	&	13:30:03.95	&	+47:15:33.0	&	6300.	&	2013-04-02	\\
	&			&			&	6300.	&	2013-04-06	\\
M 51 - Object 105	&	13:30:07.38 	&	+47:13:22.3	&	6300.	&	2013-04-02	\\
	&			&			&	12600. 	&	2013-04-06	\\

\enddata
\tablecomments{This table presents the observations obtaining the optical spectra, by source, with the 4m Mayall Telescope at KPNO and the 6.5m MMT. We determine flux corrections (see Section \ref{section-lines}) to account for any non-ideal observing conditions.}

\tablenotetext{a} {yyyy-mm-dd}
\tablenotetext{b} {at wind limit}
\tablenotetext{c} {some clouds}
\tablenotetext{d} {bad seeing}
\tablenotetext{e} {The targets NGC 4214 - Obj 15 and NGC 4214 - Obj 16 could not be separated. Therefore, the combined spectrum is used and the averaged positions of Object 15 (12:15:40.76 +36:19:04.5) and Object 16 (12:15:40.98 +36:19:04.3) is presented.}
\end{deluxetable}

\begin{deluxetable}{llllllll}
\tabletypesize{\scriptsize}
\tablewidth{0pt} 
\tablecaption{\label{table-imaging} Archival V-Band  Imaging Observations}
\tablehead{
	 \colhead{Galaxy}			&
	 \colhead{Telescope}				&
	 \colhead{Filter}			&
	 \colhead{Description}		&	 
	 \colhead{Instrument}		&	 
	 \colhead{Date Observed}		&
	 \colhead{Proposal ID}		&	 
	 \colhead{PI}		
}
\startdata

NGC 2366 	&	HST	&	F547M	&	V	&	WFPC2		& 1996-01-08	&	6096	 & L. Drissen	 \\	

NGC 4214	&	HST	&	F547M	&	V	&	WFC3(UVIS)	& 2009-12-23	&	11360 &	R. O'Connell \\

NGC 4449	&	HST	& 	F550M	&	V	&	ACS(WFC)	& 2005-11-18	&	10522	& D. Calzetti	\\

NGC 6946	&	KP\tablenotemark{a}	& 	V	&	V	&	CFIM(t2ka)	& 2001-11-08	&	...	& K. Gordon\tablenotemark{b}		\\

M 51			&	HST	& 	F555W	&	wide-V	&	ACS(WFC)	& 2005-01-20	&	10452	& S. Beckwith		\\
\enddata

\tablecomments{Observing information for archival optical images with  medium and broad V-band filters.
}
\tablenotetext{a}{KP$=$ Kitt Peak National Observatory 2.1m Telescope. }
\tablenotetext{b}{Primary observer.}

\end{deluxetable}

\begin{deluxetable}{lllll}
\tabletypesize{\scriptsize}
\tablewidth{0pt} 
\tablecaption{\label{table-opticalcharacteristics} Optical Characteristics of the Sample}
\tablehead{
	 \colhead{Source}		&
	 \colhead{F$_{\text{cont, 6000 \AA}}$}		&
	 \colhead{Flux Correction }		&
	 \colhead{-EW(H$\beta$)}		&
	\colhead{L$_{V, o}$}		\\ 
	 \colhead{}		&
	 \colhead{ergs cm$^{-2}$ s$^{-1}$ }		&
	 \colhead{}		&	    
	 \colhead{\AA}		&	    
  	 \colhead{erg s$^{-1}$ }		\\
}		 
\startdata
NGC 2366 - Object 10	 & 	7.88e-16	 & 	1.0	 & 	505.0 (50.5)	 & 	1.48e+36 (2.908e+34)\\ 
NGC 2366 - Object 11	 & 	6.23e-16	 & 	1.0	 & 	176.1 (17.6)	 & 	6.74e+35 (2.348e+34)\\ 
NGC 2366 - Object 13	 & 	1.70e-16	 & 	1.0	 & 	134.9 (13.5)	 & 	2.27e+35 (3.291e+34)\\ 
NGC 4214 - Object 13	 & 	3.45e-16	 & 	1.0	 & 	148.0 (14.8)	 & 	3.58e+35 (4.652e+34)\\ 
NGC 4214 - Object 14	 & 	3.81e-16	 & 	1.0	 & 	226.9 (22.7)	 & 	3.61e+35 (4.547e+34)\\ 
NGC 4214 - Object 15/16	 & 	6.74e-16	 & 	1.1	 & 	241.7 (24.2)	 & 	8.69e+35 (1.254e+35)\\ 
NGC 4449 - Object 3	 & 	2.01e-16	 & 	1.2	 & 	123.5 (12.3)	 & 	4.41e+35 (1.644e+35)\\ 
NGC 4449 - Object 4	 & 	1.43e-15	 & 	1.0	 & 	13.0 (1.3)	 & 	1.21e+36 (1.837e+35)\\ 
NGC 4449 - Object 18	 & 	1.28e-16	 & 	1.0	 & 	138.0 (13.8)	 & 	1.38e+35 (3.473e+34)\\ 
NGC 4449 - Object 22	 & 	2.90e-16	 & 	1.0	 & 	45.1 (4.5)	 & 	6.45e+35 (8.496e+34)\\ 
NGC 4449 - Object 23	 & 	3.49e-16	 & 	1.0	 & 	2.5 (0.3)	 & 	3.87e+35 (2.918e+35)\\ 
NGC 4449 - Object 26	 & 	6.20e-16	 & 	1.4	 & 	241.2 (24.1)	 & 	1.38e+36 (7.294e+34)\\ 
NGC 4449 - Object 27	 & 	1.65e-16	 & 	1.5	 & 	78.1 (7.8)	 & 	3.59e+35 (1.175e+34)\\ 
NGC 6946 - Object 13	 & 	7.34e-17	 & 	1.0	 & 	92.6 (9.3)	 & 	6.46e+36 (7.571e+35)\\ 
NGC 6946 - Object 37	 & 	6.52e-17	 & 	1.0	 & 	36.4 (3.7)	 & 	1.23e+36 (1.354e+35)\\ 
NGC 6946 - Object 48	 & 	1.07e-16	 & 	1.0	 & 	69.5 (7.0)	 & 	4.67e+36 (2.705e+35)\\ 
NGC 6946 - Object 110	 & 	1.97e-16	 & 	2.9	 & 	114.2 (11.4)	 & 	9.75e+36 (8.923e+35)\\ 
NGC 6946 - Object 115	 & 	1.22e-16	 & 	1.6	 & 	73.2 (7.3)	 & 	5.29e+36 (3.233e+35)\\ 
NGC 6946 - Object 117	 & 	2.77e-16	 & 	2.4	 & 	43.9 (4.4)	 & 	1.33e+37 (7.997e+35)\\ 
M 51 - Object 5	 & 	1.36e-16	 & 	3.4	 & 	81.9 (8.2)	 & 	1.12e+36 (2.362e+34)\\ 
M 51 - Object 6	 & 	1.07e-16	 & 	2.1	 & 	43.8 (4.4)	 & 	8.33e+35 (5.775e+35)\\ 
M 51 - Object 11	 & 	1.40e-16	 & 	2.2	 & 	40.1 (4.0)	 & 	1.12e+36 (1.359e+35)\\ 
M 51 - Object 34	 & 	2.28e-16	 & 	1.6	 & 	13.1 (1.3)	 & 	2.38e+36 (1.789e+36)\\ 
M 51 - Object 39	 & 	6.21e-16	 & 	2.8	 & 	11.3 (1.2)	 & 	4.65e+36 (3.351e+36)\\ 
M 51 - Object 44	 & 	1.44e-16	 & 	1.5	 & 	65.5 (6.6)	 & 	1.98e+35 (1.081e+35)\\ 
M 51 - Object 46	 & 	6.60e-16	 & 	4.5	 & 	79.3 (7.9)	 & 	3.89e+36 (2.997e+35)\\ 
M 51 - Object 57	 & 	5.89e-17	 & 	1.0	 & 	87.1 (8.7)	 & 	2.07e+35 (1.553e+35)\\ 
M 51 - Object 60	 & 	2.32e-16	 & 	1.2	 & 	8.8 (0.9)	 & 	2.16e+36 (1.620e+36)\\ 
M 51 - Object 63	 & 	4.12e-16	 & 	3.4	 & 	9.2 (0.9)	 & 	2.84e+36 (2.411e+36)\\ 
M 51 - Object 67	 & 	1.98e-16	 & 	1.0	 & 	7.6 (0.8)	 & 	1.15e+36 (2.859e+35)\\ 
M 51 - Object 73	 & 	2.29e-16	 & 	2.7	 & 	43.6 (4.4)	 & 	2.03e+36 (8.617e+34)\\ 
M 51 - Object 88	 & 	4.35e-16	 & 	5.3	 & 	17.1 (1.7)	 & 	6.23e+35 (4.672e+35)\\ 
M 51 - Object 90	 & 	8.71e-17	 & 	1.0	 & 	18.0 (1.8)	 & 	4.60e+35 (3.448e+35)\\ 
M 51 - Object 92	 & 	1.88e-16	 & 	1.6	 & 	38.5 (3.9)	 & 	1.57e+36 (1.169e+35)\\ 
M 51 - Object 93	 & 	1.33e-16	 & 	1.1	 & 	27.9 (2.8)	 & 	9.97e+35 (6.381e+34)\\ 
M 51 - Object 94	 & 	1.12e-16	 & 	1.1	 & 	94.2 (9.4)	 & 	1.10e+36 (1.592e+34)\\ 
M 51 - Object 96	 & 	8.07e-17	 & 	1.0	 & 	1.4 (0.2)	 & 	2.91e+35 (2.413e+35)\\ 
M 51 - Object 97	 & 	3.44e-16	 & 	4.1	 & 	41.0 (4.1)	 & 	2.52e+36 (9.240e+35)\\ 
M 51 - Object 100	 & 	1.95e-15	 & 	6.0	 & 	70.9 (7.1)	 & 	4.15e+36 (1.101e+35)\\ 
M 51 - Object 101	 & 	1.11e-16	 & 	1.0	 & 	71.0 (7.1)	 & 	1.59e+36 (1.463e+35)\\ 
M 51 - Object 103	 & 	8.33e-17	 & 	1.5	 & 	78.5 (7.9)	 & 	6.92e+35 (1.984e+35)\\ 
M 51 - Object 105	 & 	2.80e-17	 & 	1.0	 & 	137.7 (13.8)	 & 	1.04e+35 (3.161e+34)\\ 

\enddata
\tablecomments{ A table presenting various optical characteristics used throughout this paper. Specifically, the spectra shown in Figures \ref{fig-fullspec1} - \ref{fig-candidatespec} are normalized to the average continuum values presented in this table, as well as the following for ``Other'' sources (not listed above): M 51 - Object 87:  F$_{\text{cont, 6000 \AA}}$ $=$ 6.33e-15 ergs cm$^{-2}$ s$^{-1}$, 
NGC 4214 - Object 3: F$_{\text{cont, 6000 \AA}}$ $=$ 1.07e-15 ergs cm$^{-2}$ s$^{-1}$, and NGC 4214 - Object 17: F$_{\text{cont, 6000 \AA}}$ $=$ 6.44e-17 ergs cm$^{-2}$ s$^{-1}$. The Flux Correction is determined from the V-band photometry and is discussed in Section \ref{section-lines}.}
\end{deluxetable}

\begin{deluxetable}{llllllllllllll|l}
\tabletypesize{\scriptsize}
\tablewidth{0pt} 
\tablecaption{\label{table-fluxes_wrs} Emission Line Fluxes for WR Clusters\tablenotemark{a}}
\tablehead{
	 \colhead{Source}		&
	 \colhead{ [O  \textrm{II}]}		&
	 \colhead{H$\delta$}		&
	 \colhead{ H$\gamma$ }		&
	 \colhead{[O  \textrm{III}]}		&	    
	 \colhead{ [O  \textrm{III}] }		&	 
  	 \colhead{ [O  \textrm{III}] }		&	 
	\colhead{ [N  \textrm{II}]}		&	 
  	\colhead{H $\alpha$ }		&	   
	\colhead{  [N  \textrm{II}]}		&	 		 
	\colhead{[S  \textrm{II}] }		&	 
 	\colhead{ [S  \textrm{II}] }		&	 
 	\colhead{ [O  \textrm{II}] }		&	    
	\colhead{ [O  \textrm{II}] }	&
	 \colhead{ H $\beta$ \tablenotemark{a}}			\\
	 \colhead{}		&
	 \colhead{3727 \AA}		&
	 \colhead{4102 \AA}		&
	 \colhead{4341 \AA}		&
	 \colhead{4363 \AA}		&	         
	 \colhead{4959 \AA}		&	 
  	 \colhead{5007 \AA}		&	 
	\colhead{6548 \AA}		&	 
  	\colhead{6563 \AA}		&	   
	\colhead{6584 \AA}		&	 		 
	\colhead{6717 \AA}		&	 
 	\colhead{6732 \AA}		&	 
 	\colhead{7319 \AA}		&	    
	\colhead{7330 \AA}		&
	\colhead{4861 \AA}		\\
}		 
\rotate
\startdata

N2366 - Obj 10	 & 	45 (4)	 & 	23 (2)	 & 	46 (4)	 & 	15 (1)	 & 	238 (20)	 & 	705 (60)	 & 	0.8 (0.1)	 & 	292 (24)	 & 	2.0 (0.2)	 & 	3.6 (0.3)	 & 	3.4 (0.3)	 & 	1.0 (0.1)	 & 	0.8 (0.1)	 & 	590 (50) \\ 
N2366 - Obj 11	 & 	69 (10)	 & 	23 (3)	 & 	39 (5)	 & 	12 (2)	 & 	205 (25)	 & 	707 (87)	 & 	1.1 (0.2)	 & 	305 (33)	 & 	2.6 (0.3)	 & 	5.2 (0.6)	 & 	3.9 (0.5)	 & 	1.1 (0.3)	 & 	1.5 (0.5)	 & 	225 (28) \\ 
N4214 - Obj 13	 & 	277 (41)	 & 	22 (3)	 & 	46 (6)	 & 	2.0 (0.4)	 & 	97 (12)	 & 	292 (36)	 & 	7.2 (0.8)	 & 	274 (30)	 & 	20 (2)	 & 	19 (2)	 & 	14 (2)	 & 	3.0 (0.5)	 & 	2.6 (0.5)	 & 	80.8 (10.2)\\ 
N4214 - Obj 14	 & 	219 (37)	 & 	23 (4)	 & 	45 (7)	 & 	2.8 (0.5)	 & 	108 (15)	 & 	326 (45)	 & 	6.1 (0.7)	 & 	245 (29)	 & 	20 (2)	 & 	18 (2)	 & 	14 (2)	 & 	2.9 (0.4)	 & 	2.8 (0.4)	 & 	133 (19)\\ 
N4214 - Obj 15/16	 & 	248 (27)	 & 	30 (3)	 & 	46 (5)	 & 	3.8 (0.4)	 & 	145 (14)	 & 	316 (31)	 & 	6.1 (0.6)	 & 	296 (27)	 & 	21 (2)	 & 	20 (2)	 & 	14 (1)	 & 	2.8 (0.3)	 & 	3.2 (0.4)	 & 	194 (19)\\ 
N4449 - Obj 4	 & 	327 (77)	 & 	16 (4)	 & 	34 (7)	 & 	7 (2)	 & 	92 (17)	 & 	240 (45)	 & 	13 (2)	 & 	312 (49)	 & 	33 (5)	 & 	34 (6)	 & 	26 (4)	 & 	11 (3)	 & 	... 	 & 	34.1 (6.6)\\ 
N4449 - Obj 18	 & 	316 (36)	 & 	26 (3)	 & 	48 (5)	 & 	2.2 (0.3)	 & 	54 (5)	 & 	170 (17)	 & 	10.2 (1.0)	 & 	268 (25)	 & 	29 (3)	 & 	27 (3)	 & 	18 (2)	 & 	2.8 (0.5)	 & 	2.8 (0.6)	 & 	30.7 (3.1) \\ 
N4449 - Obj 22	 & 	417 (96)	 & 	29 (6)	 & 	57 (12)	 & 	6 (1)	 & 	69 (13)	 & 	190 (35)	 & 	10 (2)	 & 	304 (47)	 & 	35 (5)	 & 	51 (8)	 & 	36 (5)	 & 	6 (1)	 & 	3.6 (0.7)	 & 	24.0 (4.5)\\ 
N4449 - Obj 26	 & 	204 (19)	 & 	24 (2)	 & 	46 (4)	 & 	2.1 (0.2)	 & 	126 (11)	 & 	376 (32)	 & 	5.3 (0.4)	 & 	294 (24)	 & 	16 (1)	 & 	16 (1)	 & 	12 (1)	 & 	2.4 (0.2)	 & 	2.1 (0.2)	 & 	289 (25) \\ 
N6946 - Obj 13	 & 	195 (73)	 & 	22 (7)	 & 	43 (14)	 & 	... 	 & 	28 (8)	 & 	84 (24)	 & 	41 (10)	 & 	457 (109)	 & 	128 (31)	 & 	57 (13)	 & 	41 (10)	 & 	3.4 (0.8)	 & 	2.8 (0.7)	 & 	14.6 (4.4) \\ 
N6946 - Obj 48	 & 	129 (49)	 & 	23 (8)	 & 	42 (14)	 & 	3 (1)	 & 	13 (4)	 & 	43 (13)	 & 	75 (18)	 & 	452 (110)	 & 	232 (56)	 & 	67 (16)	 & 	68 (16)	 & 	7 (2)	 & 	5 (1)	 & 	12.6 (3.9) \\ 
N6946 - Obj 110	 & 	197 (71)	 & 	24 (8)	 & 	43 (14)	 & 	1.6 (0.6)	 & 	69 (20)	 & 	212 (60)	 & 	35 (8)	 & 	486 (113)	 & 	107 (25)	 & 	52 (12)	 & 	40 (9)	 & 	4.2 (1.0)	 & 	4 (1)	 & 	28.9 (8.4) \\ 
N6946 - Obj 115	 & 	179 (39)	 & 	25 (5)	 & 	44 (8)	 & 	1.9 (0.5)	 & 	25 (4)	 & 	76 (13)	 & 	43 (6)	 & 	480 (71)	 & 	159 (23)	 & 	51 (7)	 & 	37 (5)	 & 	2.7 (0.7)	 & 	2.0 (0.6)	 & 	11.0 (2.0) \\ 
N6946 - Obj 117	 & 	218 (62)	 & 	23 (6)	 & 	46 (12)	 & 	5 (1)	 & 	51 (12)	 & 	156 (35)	 & 	30 (6)	 & 	395 (74)	 & 	91 (17)	 & 	39 (7)	 & 	36 (7)	 & 	3.3 (0.6)	 & 	3.5 (0.7)	 & 	18.3 (4.2)\\ 
M 51 - Obj 46	 & 	49 (5)	 & 	26 (2)	 & 	47 (4)	 & 	... 	 & 	2.0 (0.2)	 & 	6.0 (0.5)	 & 	43 (4)	 & 	369 (31)	 & 	129 (11)	 & 	28 (2)	 & 	22 (2)	 & 	0.9 (0.2)	 & 	... 	 & 	70.6 (6.3) \\ 
M 51 - Obj 57	 & 	157 (46)	 & 	24 (6)	 & 	45 (12)	 & 	2.0 (0.5)	 & 	27 (6)	 & 	82 (19)	 & 	56 (11)	 & 	413 (79)	 & 	174 (33)	 & 	33 (6)	 & 	27 (5)	 & 	1.9 (0.5)	 & 	1.1 (0.3)	 & 	7.18 (1.70)\\ 
M 51 - Obj 73	 & 	216 (112)	 & 	29 (14)	 & 	50 (23)	 & 	1.0 (0.7)	 & 	9 (4)	 & 	30 (12)	 & 	49 (16)	 & 	343 (112)	 & 	152 (50)	 & 	47 (15)	 & 	37 (12)	 & 	4 (1)	 & 	... 	 & 	14.1 (5.9) \\ 
M 51 - Obj 94	 & 	160 (61)	 & 	28 (10)	 & 	49 (16)	 & 	1.1 (0.4)	 & 	7 (2)	 & 	20 (6)	 & 	47 (11)	 & 	333 (81)	 & 	138 (34)	 & 	28 (7)	 & 	20 (5)	 & 	0.7 (0.3)	 & 	0.6 (0.3)	 & 	17.7 (5.4)\\ 
M 51 - Obj 100	 & 	109 (9)	 & 	26 (2)	 & 	46 (4)	 & 	0.9 (0.2)	 & 	7.2 (0.6)	 & 	22 (2)	 & 	54 (4)	 & 	401 (31)	 & 	161 (12)	 & 	31 (2)	 & 	26 (2)	 & 	0.9 (0.3)	 & 	0.7 (0.2)	 & 	207 (16) \\ 
M 51 - Obj 101	 & 	148 (13)	 & 	26 (2)	 & 	47 (4)	 & 	2.8 (0.3)	 & 	11.8 (1.0)	 & 	36 (3)	 & 	48 (4)	 & 	350 (28)	 & 	142 (11)	 & 	24 (2)	 & 	18 (1)	 & 	0.6 (0.3)	 & 	0.6 (0.2)	 & 	12.5 (1.0) \\ 
M 51 - Obj 103	 & 	181 (51)	 & 	24 (6)	 & 	45 (11)	 & 	3.1 (0.8)	 & 	20 (5)	 & 	64 (15)	 & 	66 (12)	 & 	484 (90)	 & 	200 (37)	 & 	44 (8)	 & 	32 (6)	 & 	1.9 (0.4)	 & 	1.4 (0.4)	 & 	8.78 (2.02) \\

\enddata

\tablenotetext{a}{The extinction corrected flux of measured emission lines in comparison to H$\beta$, given in the last column with units of 10$^{-15}$ ergs cm$^{-2}$ s$^{-1}$.}
\end{deluxetable}

\begin{deluxetable}{lllllll|l}
\tabletypesize{\scriptsize}
\tablewidth{0pt} 
\tablecaption{\label{table-fluxes_nonwrs} Emission Line Fluxes for Non-WR Clusters\tablenotemark{a}}
\tablehead{
	 \colhead{Source}		&
	 \colhead{H$\delta$}		&
	 \colhead{H$\gamma$ }		&
  	 \colhead{[O  \textrm{III}] }		&	 
	\colhead{ [N  \textrm{II}]}		&	 
  	\colhead{H $\alpha$ }		&	   
	\colhead{[N  \textrm{II}]}			&
	\colhead{H $\beta$ \tablenotemark{a}}		\\	 
	 \colhead{}		&
	 \colhead{4102 \AA }		&
	 \colhead{4341 \AA}		&	    
  	 \colhead{5007 \AA}		&	 
	\colhead{6548 \AA}		&	 
  	\colhead{6563 \AA}		&	   	    
	\colhead{6584 \AA}		&
	 \colhead{4861 \AA}		\\
}		 
\rotate
\startdata

NGC 2366 - Object 13	 & 	22 (2)	 & 	44 (3)	 & 	464 (36)	 & 	1.5 (0.2)	 & 	287 (22)	 & 	5.3 (0.4)	 & 	34.7 (2.7) \\ 
NGC 4449 - Object 3	 & 	23 (10)	 & 	48 (20)	 & 	229 (83)	 & 	11 (3)	 & 	402 (118)	 & 	38 (11)	 & 	41.4 (15.3) \\ 
NGC 4449 - Object 23	 & 	... 	 & 	... 	 & 	126 (18)	 & 	53 (7)	 & 	310 (37)	 & 	210 (25)	 & 	1.24 (0.176) \\ 
NGC 4449 - Object 27	 & 	32 (36)	 & 	66 (71)	 & 	119 (113)	 & 	10 (7)	 & 	467 (355)	 & 	32 (24)	 & 	21.3 (20.7) \\ 
NGC 6946 - Object 37	 & 	16 (10)	 & 	45 (26)	 & 	16 (8)	 & 	55 (23)	 & 	449 (185)	 & 	172 (71)	 & 	6.35 (3.32) \\ 
M 51 - Object 5	 & 	26 (3)	 & 	46 (5)	 & 	62 (6)	 & 	57 (5)	 & 	485 (43)	 & 	176 (16)	 & 	11.8 (1.1) \\ 
M 51 - Object 6	 & 	21 (6)	 & 	43 (11)	 & 	17 (4)	 & 	68 (13)	 & 	440 (85)	 & 	208 (40)	 & 	11.4 (2.7) \\ 
M 51 - Object 11	 & 	25 (2)	 & 	47 (4)	 & 	12 (1)	 & 	51 (4)	 & 	398 (32)	 & 	157 (13)	 & 	6.67 (0.56) \\ 
M 51 - Object 34	 & 	14 (7)	 & 	40 (19)	 & 	19 (8)	 & 	48 (16)	 & 	388 (133)	 & 	146 (50)	 & 	10.6 (4.6) \\ 
M 51 - Object 39	 & 	12 (7)	 & 	39 (22)	 & 	22 (11)	 & 	39 (16)	 & 	383 (153)	 & 	123 (49)	 & 	26.2 (13.3) \\ 
M 51 - Object 44	 & 	23 (9)	 & 	42 (16)	 & 	13 (4)	 & 	75 (20)	 & 	476 (126)	 & 	242 (64)	 & 	18.4 (6.1) \\ 
M 51 - Object 60	 & 	... 	 & 	... 	 & 	24 (4)	 & 	51 (6)	 & 	405 (49)	 & 	162 (19)	 & 	7.47 (1.07) \\ 
M 51 - Object 63	 & 	... 	 & 	37 (10)	 & 	10 (4)	 & 	17 (4)	 & 	325 (65)	 & 	60 (12)	 & 	26.6 (6.7) \\ 
M 51 - Object 67	 & 	... 	 & 	41 (10)	 & 	14 (3)	 & 	38 (6)	 & 	387 (61)	 & 	120 (19)	 & 	11.3 (2.2) \\ 
M 51 - Object 88	 & 	17 (6)	 & 	39 (11)	 & 	40 (10)	 & 	52 (11)	 & 	382 (80)	 & 	159 (33)	 & 	37.5 (9.8) \\ 
M 51 - Object 90	 & 	18 (6)	 & 	42 (12)	 & 	11 (3)	 & 	43 (9)	 & 	367 (78)	 & 	135 (28)	 & 	12.4 (3.3)\\ 
M 51 - Object 92	 & 	20 (5)	 & 	46 (11)	 & 	19 (4)	 & 	50 (9)	 & 	362 (64)	 & 	158 (28)	 & 	10.2 (2.2) \\ 
M 51 - Object 93	 & 	22 (5)	 & 	42 (9)	 & 	8 (2)	 & 	49 (8)	 & 	423 (71)	 & 	150 (25)	 & 	7.64 (1.57) \\ 
M 51 - Object 96	 & 	... 	 & 	... 	 & 	380 (55)	 & 	161 (20)	 & 	365 (44)	 & 	464 (56)	 & 	0.977 (0.189) \\ 
M 51 - Object 97	 & 	20 (9)	 & 	43 (18)	 & 	16 (6)	 & 	71 (21)	 & 	488 (143)	 & 	225 (66)	 & 	52.7 (19.4) \\ 
M 51 - Object 105	 & 	22 (11)	 & 	42 (19)	 & 	67 (27)	 & 	75 (25)	 & 	533 (175)	 & 	231 (75)	 & 	6.57 (2.73) \\ 

\enddata

\tablenotetext{a}{The extinction corrected flux of measured emission lines in comparison to H$\beta$, given in the last column with units of 10$^{-15}$ ergs cm$^{-2}$ s$^{-1}$.}

\end{deluxetable}